\documentclass[fleqn,usenatbib]{mnras}

\usepackage{newtxtext,newtxmath}

\usepackage[T1]{fontenc}
\usepackage[utf8]{inputenc}

\DeclareRobustCommand{\VAN}[3]{#2}
\let\VANthebibliography\thebibliography
\def\thebibliography{\DeclareRobustCommand{\VAN}[3]{##3}\VANthebibliography}

\usepackage{graphicx}	
\usepackage{amsmath}	
\usepackage{physics}    
\usepackage{ulem}       
\usepackage{bigints}

\renewcommand{\vec}[1]{\boldsymbol{#1}}

\newcommand{\equ}[1]{eq.~(\ref{eq:#1})}
\newcommand{\equs}[1]{eqs.~(\ref{eq:#1})}
\newcommand{\equm}[1]{(\ref{eq:#1})}

\newcommand{\Equs}[1]{Eqs.~(\ref{eq:#1})}
\newcommand{\equnp}[1]{eq.~\ref{eq:#1}}
\newcommand{\equsnp}[1]{eqs.~\ref{eq:#1}}
\newcommand{\equmnp}[1]{\ref{eq:#1}}
\newcommand{\se}[1]{Section~\ref{sec:#1}}
\newcommand{\ses}[1]{Sections~\ref{sec:#1}}
\newcommand{\sem}[1]{\ref{sec:#1}}
\newcommand{\fig}[1]{Fig.~\ref{fig:#1}}
\newcommand{\figs}[1]{Figs.~\ref{fig:#1}}
\newcommand{\figss}[1]{\ref{fig:#1}}
\newcommand{\Fig}[1]{Figure~\ref{fig:#1}}

\newcommand{\be}{\begin{equation}}
\newcommand{\ee}{\end{equation}}
\newcommand{\bea}{\begin{eqnarray}}
\newcommand{\eea}{\end{eqnarray}}

\newcommand{\no}{\noindent}

\newcommand{\msun}{{\rm M}_\odot}

\newcommand{\ifm}[1]{\relax\ifmmode#1\else$\mathsurround=0pt #1$\fi}
\newcommand{\kms}{\ifmmode\,{\rm km}\,{\rm s}^{-1}\else km$\,$s$^{-1}$\fi}

\newcommand{\Mpc}{\,{\rm Mpc}}
\newcommand{\kpc}{\,{\rm kpc}}
\newcommand{\pc}{\,{\rm pc}}

\newcommand{\yr}{\,{\rm yr}}

\newcommand{\Myr}{\,{\rm Myr}}
\newcommand{\K}{\,{\rm K}}

\newcommand{\ltsima}{$\; \buildrel < \over \sim \;$}
\newcommand{\lsim}{\lower.5ex\hbox{\ltsima}}
\newcommand{\gtsima}{$\; \buildrel > \over \sim \;$}
\newcommand{\gsim}{\lower.5ex\hbox{\gtsima}}

\def\la{\langle}

\def\cmc{\,{\rm cm}^{-3}}

\def\M*{M_{\rm *}}
\def\Mv{M_{\rm v}}
\def\Rv{R_{\rm v}}

\def\Kdegree{{\rm K}}

\def\Pi{\varpi_{_{\rm I}}}

\newcommand{\hypa}{\texttt{IPMSim}}

\usepackage{xcolor}

\title[High-$z$ Filament Cross-Sections]{The Structure and Dynamics of Massive High-$z$ Cosmic-Web Filaments: Three Radial Zones in Filament Cross-Sections}

\author[Y. S. Lu et al.]{Yue Samuel Lu,$^{1,2}$\thanks{E-mail: yul232@ucsd.edu}
Nir Mandelker,$^{3,4}$
S. Peng Oh,$^{2}$
Avishai Dekel,$^{3,5}$
Frank C. van den Bosch,$^{6}$\newauthor%
Volker Springel,$^{7}$
Daisuke Nagai,$^{8}$
Freeke van de Voort$^{9}$
\vspace{0.1cm}\\%
$^{1}$Department of Astronomy and Astrophysics, University of California, San Diego, La Jolla, CA 92093, USA\\%
$^{2}$Physics Department, University of California, Santa Barbara, Broida Hall, Santa Barbara, CA 93106, USA\\%
$^{3}$Centre for Astrophysics and Planetary Science, Racah Institute of Physics, The Hebrew University, Jerusalem, 91904, Israel\\%
$^{4}$Kavli Institute for Theoretical Physics, Kohn Hall, Santa Barbara, CA 93106, USA\\%
$^{5}$Santa Cruz Institute for Particle Physics, University of California, Santa Cruz, CA 95064, USA\\%
$^{6}$Department of Astronomy, Yale University, PO Box 208101, New Haven, CT, USA\\%
$^{7}$Max Planck Institute for Astrophysics, Karl-Schwarzschild-Stra{\ss}e 1, D-85748 Garching, Germany\\%
$^{8}$Department of Physics, Yale University, New Haven, CT, 06520, USA\\%
$^{9}$Cardiff Hub for Astrophysics Research and Technology, School of Physics and Astronomy, Cardiff University, Queen’s Buildings, The Parade, \\%
Cardiff CF24 3AA, UK\\%
}

\date{Accepted XXX. Received YYY; in original form ZZZ}

\pubyear{2023}

\begin{document}
\label{firstpage}
\pagerange{\pageref{firstpage}--\pageref{lastpage}}
\maketitle

\begin{abstract}
We analyse the internal structure and dynamics of cosmic-web filaments that connect massive high-$z$ haloes. Our analysis is based on a high-resolution \texttt{AREPO} cosmological simulation zooming-in on a volume encompassing three ${\rm Mpc}$-scale filaments feeding three massive haloes of $\sim 10^{12}\,\text{M}_\odot$ at $z \sim 4$, embedded in a large-scale sheet. Each filament is surrounded by a cylindrical accretion shock of radius $r_{\rm shock} \sim 50 \,{\rm kpc}$. The post-shock gas is in virial equilibrium with the potential well set by an isothermal dark-matter filament. 
The filament line-mass is $\sim 9\times 10^8\,\text{M}_\odot\,{\rm kpc}^{-1}$, the gas fraction within $r_{\rm shock}$ is the universal baryon fraction, and the virial temperature is $\sim 7\times 10^5 {\rm K}$. These all match
expectations from analytical models for filament properties as a function of halo-mass and redshift. 
The filament cross-section has three radial zones. In the outer ``thermal'' (\textbf{T}) zone, $r \geq 0.65 \, r_{\rm shock}$, 
inward gravity and ram-pressure forces are over-balanced by outwards thermal pressure forces, decelerating the inflowing gas expanding the shock outward. In the intermediate ``vortex'' (\textbf{V}) zone, $0.25 \leq r/  r_{\rm shock} \leq 0.65$, the velocity field is dominated by a quadrupolar vortex structure due to offset inflow along the sheet through the post-shock gas. The outwards force is dominated by centrifugal forces associated with these vortices, with additional contributions from 
global rotation and thermal pressure. The shear and turbulent forces associated with the vortices act inward. The inner ``stream'' (\textbf{S}) zone, $r < 0.25 \, r_{\rm shock}$, is a dense isothermal core, $T\sim 3 \times 10^4 \, {\rm K}$ and $n_{\rm H}\sim 0.01 \,{\rm cm^{-3}}$, defining the cold streams that feed galaxies. The core is formed by an isobaric cooling flow and is associated with a decrease in outwards forces, though it exhibits both inflows and outflows. 

\end{abstract}




\section{Introduction}
\label{sec:intro}

In the last two decades, our understanding of galaxy formation in a cosmological context has undergone a paradigm shift. Galaxies, as we now know, do not form in isolation, but rather are connected through the ``cosmic-web'', which dominates the matter distribution on $\Mpc$ scales in both dark matter (DM) and baryons. This intricate structure has been predicted theoretically \citep{Zeldovich1970,bbks1986,Bond.etal.96,Springel.etal.05}, and can be seen in the distribution of galaxies in observational surveys such as 2dFGRS \citep{Colless2001}, SDSS \citep{tegmark04}, and the 2MASS redshift survey \citep{Huchra05}. Perhaps the most prominent feature of the cosmic-web is a network of \textit{intergalactic filaments}, which contain roughly half of the mass in the Universe \citep{Zeldovich1970,bbks1986,Cautun.etal.14,Eckert.etal.2015,Libeskind.etal.18}. 

\smallskip
The filaments connect the most massive galaxies at any given epoch, which are located at the nodes of the cosmic-web, while more typical galaxies are located along the filaments (see \fig{slices} below). At high redshift, $z\sim (2-6)$ near the peak of cosmic star- and galaxy-formation, these filaments manifest as streams of cold, dense gas ($T\sim 10^4\,\Kdegree$, $n\sim 10^{-2}\,{\rm cm}^{-3}$), predicted to be the main mode of accretion onto high-$z$ galaxies, feeding them directly from the cosmic-web. In massive haloes with virial mass $\Mv \gsim 10^{12}\,\msun$ above the critical mass for the formation of a stable virial shock \citep[][]{Rees77,White78,Birnboim2003shocks,Fielding2017CGM,Stern2021virial}, the streams are predicted to penetrate the hot circumgalactic medium (CGM), maintaining temperatures of $T_{\rm s}\gsim 10^4\Kdegree$ and reaching the central galaxies in roughly a virial crossing time \citep[][]{Dekel.Birnboim.06, dekel2009cold}. This gas supply sustains the large observed star formation rates (SFRs) of $\sim 100\,\msun\yr^{-1}$ without 
galaxy mergers \citep[e.g.,][]{ForsterSchreiber.etal.06, ForsterSchreiber.etal.09, Genzel.etal.06, Genzel.etal.08, Elmegreen.etal.07, Shapiro.etal.08, Stark.etal.08, Wisnioski.etal.15}. 

\smallskip
Though their narrow size and low density (compared to the central galaxies) make the streams difficult to directly detect observationally, numerous observational studies of the CGM and the intergalactic medium (IGM) around massive high-$z$ galaxies, both in absorption \citep[][]{Fumagalli.etal.11, Fumagalli.etal.17, Goerdt.etal.12, vandeVoort.etal.12, Bouche.etal.13, Bouche.etal.16, Prochaska.etal.14} and in emission \citep[e.g.,][]{Steidel.etal.00, Matsuda.etal.06, Matsuda.etal.11, Cantalupo.etal.14, Martin.etal.14a, Martin.etal.14b, Martin.etal.19, Hennawi.etal.15, Farina.etal.17, Cai.etal.2017, Lusso.etal.19, Umehata.etal.19, Daddi.etal.20, Emonts.etal.2023, zhang.etal.2023}, reveal large quantities of cold gas with spatial and kinematic properties consistent with predictions for cosmic filaments and cold streams. Cold streams are ubiquitous in cosmological simulations \citep[e.g.,][]{Keres.etal.05, Ocvirk.etal.08, dekel2009cold, Ceverino.etal.10, vandeVoort.etal.11, Harford2011isothermal}, 
where they are found to be the primary source of gas accretion onto both the halo and the central galaxy \citep[][]{dekel2009cold, Dekel.etal.13}. 

\smallskip
Besides supplying galaxies with cold gas to fuel ongoing star formation, additional effects of filaments and streams on galaxies and DM haloes have been extensively explored theoretically. These include halo spin, galaxy angular momentum growth, and disk formation \citep{Pichon.etal.11, Stewart.etal.11, Stewart.etal.13, Kimm.etal.11, Codis.etal.12, Danovich.etal.12, Danovich.etal.15, Laigle.etal.15, Tillson.etal.15, Gonzales.etal.17}, the shape and alignment of galaxies and haloes \citep{Chen.etal.15, Codis.etal.15, Codis.etal.15b, Tomassetti.etal.16,Ganeshaiah.etal.18,Pandya.etal.19}, driving and sustaining turbulence and disk instabilities \citep{Bournaud.Elmegreen.09,Dekel.etal.09b,Ceverino.etal.10,Genel.etal.12,Ginzburg.etal.22,Forbes.etal.22}, and the emission and absorption signatures of the CGM \citep{Goerdt.etal.10,Goerdt.etal.12,Fumagalli.etal.11,vandeVoort.etal.12,mandelker2020KHI4,mandelker2020LABs}. Gravitational instabilities in and fragmentation of filaments and streams have also been speculated to contribute to the formation of stars and even globular clusters outside of galaxies \citep{mandelker2018cold,Aung.etal.19,Bennett.Sijacki.20}.

\smallskip
Despite their prominence in the large-scale structure of the cosmic-web, and their clear importance for galaxy formation, the structure and properties of filaments are incredibly poorly constrained. Even a question as fundamental as what prevents the filaments from collapsing under their own self-gravity remains open. Previous work has suggested that this could be due to thermal pressure gradients \citep[][]{Harford2011isothermal, Klar.Mucket.2012, ramsoy2021rivers}, coherent filament rotation \citep[][]{birnboim2016stability, mandelker2018cold, WangP.etal.21, xia2021intergalactic}, helical vortices \citep[][]{Codis.etal.12, Laigle.etal.15}, turbulence \citep[][]{mandelker2018cold} or magnetic pressure. However, all of these works relied either on highly simplified model assumptions or on 
low resolution simulations that did not resolve the internal structure of intergalactic filaments. Works that have measured filament properties in large-volume cosmological simulations \citep[e.g.,][]{Cautun.etal.14,Laigle.etal.15,Libeskind.etal.18,Ganeshaiah.etal.18,Ganeshaiah.etal.19,Ganeshaiah.etal.20,Uhlemann.etal.20,Song.etal.20, Galaraga2021, Galaraga2022} commonly involve DM-only N-body simulations or low-resolution hydrodynamic simulations that do not resolve the inner structure of filaments or their support against gravity. While cosmological `zoom-in' simulations achieve much higher resolution in individual galaxies, and recently also in the CGM \citep{vandeVoort.etal.19,Hummels.etal.19,Peeples.etal.19, Suresh.etal.2019, Bennett.Sijacki.20}, cosmic-web filaments in the IGM typically lie outside the high-resolution region and are still poorly resolved. Inferring detailed filament properties from observations is extremely challenging, particularly in the IGM, owing to their low density, narrow size, and uncertain chemical and ionisation compositions. The properties of filaments as a function of halo mass, redshift, and environment thus remain poorly constrained.

\smallskip
A recent effort to study the properties of intergalactic filaments at high-$z$ using cosmological zoom-in simulations with relatively high resolution was made by \citet{ramsoy2021rivers}. These authors used a cosmological zoom-in simulation of a Milky Way (MW)-mass halo, $\Mv\sim 5\times 10^{11}\msun$ at $z=0$, taken from the \texttt{NUT} suite \citep{Powell.etal.11} and run with the adaptive mesh refinement (AMR) code \texttt{RAMSES} \citep{Teyssier.02}. They studied the properties of the main filament that feeds the central halo in the redshift range $z\sim 3.5-8$, focusing primarily on $z\sim 4$ when the filament extended $\lsim 200\kpc$ on either side of the halo and did not seem to connect to any other massive object. The spatial resolution of the analysed filament was $\sim 1.2\kpc$, with small patches around haloes embedded in the filament reaching a resolution of $\sim 0.6\kpc$. These authors found the filament to be well described by a model of an isothermal, self-gravitating, infinite cylinder \citep{ostriker64}, embedded inside an isothermal self-gravitating sheet that dominated the mass distribution at $r\gsim (15-20)\kpc$ from the filament axis. This was found to be true for both the gas and the dark matter, whose density profiles had very similar shapes and widths. An accretion shock was identified around the filament by a sharp increase in gas temperature. However, the post-shock gas was found to cool rapidly, and the shock width was limited to a few kpc. The azimuthally-averaged radial profile of the gas temperature increased by $\lsim 60\%$ between the central value and the peak of the shock. While significant vorticity was identified within the filament, consistent with previous results \citep{Laigle.etal.15}, the filament was found to be primarily supported by thermal pressure. 

\smallskip
The analysis and results of \citet{ramsoy2021rivers} are novel and extremely detailed. However, since their work focused on a single filament feeding an isolated high-$z$ progenitor of a MW-mass halo, it is unclear whether their results describe filaments feeding massive, $\Mv\gsim 10^{12}\msun$, haloes as cold streams at high-$z$, as these are much more massive than the studied system. Additional studies of filaments feeding different mass haloes, in different environments, and at different redshifts are needed to draw more general conclusions about the filament population. 
Finally, while the resolution of $\sim 1\kpc$ in intergalactic filaments is quite good and better than most state-of-the-art simulations, it is unclear whether this is enough to resolve the detailed internal dynamics of filaments given the filament radius of $\sim 15\kpc$, and in particular the support due to turbulence and/or vorticity (for discussions in the context of support of the CGM see, e.g., \citealp{Bennett.Sijacki.20,Lochhaas.etal.22}). 

\smallskip
In this work, we seek to study filaments around more massive haloes, in less isolated regions, and at higher resolution. We use a novel suite of cosmological simulations first introduced in \citet{mandelker2019shattering} and \citet{mandelker2021thermal}, which zoom in on a large patch of the IGM in between two massive haloes, which at $z\sim 2$ are $\Mv\sim 5\times 10^{12}$ M$_{\odot}$ each and connected by a $\sim$ Mpc-scale cosmic filament. We hereafter refer to this suite of simulations as \hypa{}, where ``IPM'' refers to the ``intra-pancake medium'' of multiphase gas within cosmological sheets (or ``Zel'dovich pancakes,'' see \citealp{mandelker2021thermal}; \citealp{Pasha.etal.22}). The large filament at $z\sim 2$ was formed by the merger of several smaller filaments at $z \sim 2.5$. Our analysis focuses on $z\sim 4$, when the system contained several well-defined filaments that connect three haloes with $\Mv\sim 10^{12}\,\msun$ each (see \fig{slices} below). This simulation employs the resolution characteristic of a standard ``zoom-in'' simulation of a single halo within the entire IGM region between the two protoclusters. In the highest resolution version of the simulation used in this work (see \se{Sims}), the typical cell size in the IGM at $z\sim 4$ is $\sim 1\kpc$ near the midplane of the sheet and $\sim 300\pc$ near the cores of the filaments. 
A detailed convergence study presented in \citet{mandelker2021thermal} suggests that this appears sufficient to resolve the formation of multiphase gas in turbulent IPM. However, the morphology and distribution of the cold component are likely only marginally resolved. 
As shown in \se{thermal} below, the characteristic scale of the accretion shock that surrounds our filaments is $r_{\rm shock}\sim 50\kpc$, while the radius of the cold stream is $r_{\rm stream}\sim 0.25r_{\rm shock}\sim 12\kpc$. Thus, we have $\sim 80$ cells across the diameter of the cold stream, and many more across the diameter of the entire filament, sufficient to resolve the driving scale of turbulence and a partial inertial range, thus resolving most of the support due to turbulence and vorticity, if present.

\smallskip
This work is organised as follows: In \se{Methods}, we introduce the simulation, our method for selecting filament slices for analysis, and our division of the filament cross-section into three distinct radial zones. In \se{Radial_prop}, we analyse the radial structure of gas and dark matter in filaments, including their thermal structure. In \se{tcool}, we discuss the thermal stability of gaseous filaments by comparing their cooling and free-fall time scales. In \se{virial}, we address whether the gaseous filaments are in virial equilibrium within the potential well set by the dark matter filament. In \se{fil_dyn}, we decompose the forces that support gaseous filaments against their self-gravity and use a simple model to explain their origin. We discuss the implications of our results and compare them to previous work in \se{disc}. Finally, in \se{conclusions}, we conclude. Throughout, we assume a flat $\Lambda$CDM cosmology with $\Omega_{\rm m}= 1 - \Omega_{\Lambda} = 0.3089$, $\Omega_{\rm b} = 0.0486$, $h = 0.6774$, $\sigma_8 = 0.8159$, and $n_s = 0.9667$ \citep{Planck16}.

\section{Methods}
\label{sec:Methods}

\subsection{Simulation Method}
\label{sec:Sims}

We use the highest resolution version of \hypa{} introduced in \citet{mandelker2019shattering} and \citet{mandelker2021thermal}. We briefly describe key aspects of the simulation below and refer the reader to those papers for additional details. The simulation was performed with the quasi-Lagrangian moving-mesh code \texttt{AREPO} \citep{springel2010pur}. The goal of the simulation was to zoom-in not just on a single halo, as is commonly done, but rather on two protoclusters and the cosmic-web elements that connect them. To select our target region, we first considered the 200 most massive haloes in the $z\sim 2.3$ snapshot of the Illustris TNG100\footnote{http://www.tng-project.org} magnetohydrodynamic cosmological simulation \citep{springel2018first,nelson2018first, pillepich2018first}. These have virial masses\footnote{The virial radius, $\Rv$, is defined using the \citealp{Bryan98} spherical overdensity criterion, and the virial mass, $\Mv$, is the total mass within $\Rv$} $\Mv$ in the range $\sim (1-40)\times 10^{12}\msun$. We then selected from all these halo pairs with a comoving separation in the range $(2.5-4.0)h^{-1}\Mpc \sim (3.7-5.9)\Mpc$. There are 48 such halo pairs, each of which is either connected by a cosmic-web filament or else embedded in the same cosmic sheet. We randomly selected one pair consisting of two haloes with $\Mv\sim 5\times 10^{12}\msun$ each, separated by a proper distance of $D\sim 1.2\Mpc$ at $z\sim 2.3$. By $z=0$, the two haloes evolve into mid-size groups with $\Mv\sim (1.6-1.9)\times 10^{13}\msun$ separated by $\sim 2.7\Mpc$, such that their comoving distance has decreased by $\lsim 30\%$.

\smallskip
The zoom-in region is the union of a cylinder with length $D$ and radius $R_{\rm ref}=1.5\times R_{\rm v,max}\sim 240\kpc$ that extends between the two halo centers, and two spheres of radius $R_{\rm ref}$ centred on either halo, where $R_{\rm v,max}$ is the larger of the two virial radii at $z=2.3$. We trace all dark matter particles within this volume back to the initial conditions of the simulation at $z=127$, refine the corresponding Lagrangian region to higher resolution, and rerun the simulation to a final redshift, $z_{\rm fin}$. In this paper, we use the highest resolution version of this simulation, dubbed ZF4.0 in \citet{mandelker2021thermal}, which has a dark matter particle mass of $m_{\rm dm}=8.2\times 10^4\,\msun$ and a Plummer-equivalent gravitational softening of $\epsilon_{\rm dm}=250\pc$ comoving. Gas cells are refined so that their mass is within a factor of 2 of the target mass, $m_{\rm gas}=1.5\times 10^4\msun$. Gravitational softening for gas cells is twice the cell size, down to a minimal gravitational softening $\epsilon_{\rm gas}=0.5\epsilon_{\rm dm}=125\pc$. The simulation was run until $z_{\rm fin}\sim 2.9$, though our analysis here focuses on $z\sim 4$.

\smallskip
The simulations were performed with the same physics model used in the TNG simulations, described in detail in \citet{Weinberger2017Sim} and \citet{pillepich2018Sim}. This includes radiative cooling down to $T=10^4{\rm K}$ from Hydrogen and Helium \citep{Katz96}, metal-line cooling \citep{Wiersma09}, radiative heating from a spatially constant and redshift-dependent ionising ultraviolet background (UVB) \citep{FG09}, partial self-shielding of dense gas from the UVB \citep{Rahmati13}, and additional heating from the radiation field of active galactic nuclei (AGN) within $3\Rv$ of haloes containing actively accreting supermassive black holes \citep{Vogelsberger13}. Star-formation occurs stochastically in gas with densities greater than $n_{\rm thresh}=0.13\cmc$, which is placed on an artificial equation of state meant to mimic the unresolved multiphase ISM \citep{Springel03}. We include feedback from both supernova and AGN following the Illustris-TNG model.

\smallskip
To identify dark matter (sub)haloes in the simulation, we first apply a Friends-of-Friends (FoF) algorithm with a linking length $b = 0.2$ to dark matter particles. We then assign gas and stars to FoF groups based on their nearest-neighbour dark matter particle. Finally, we apply \texttt{SUBFIND} \citep{Springel01,Dolag09} to the total mass distribution in each FoF group. The most massive \texttt{SUBFIND} object in each FoF group is identified as the central halo, whose virial radius $\Rv$ is defined using the \citet{Bryan98} spherical overdensity criterion, and the virial mass $\Mv$ is the total mass of dark matter, gas, and stars within $\Rv$.

\subsection{Filament Selection}
\label{sec:Fil_selec}

\begin{figure*}
    \centering
    \includegraphics[trim={0.0cm 0.0cm 0.0cm 0.0cm}, clip, width =0.98 \textwidth]{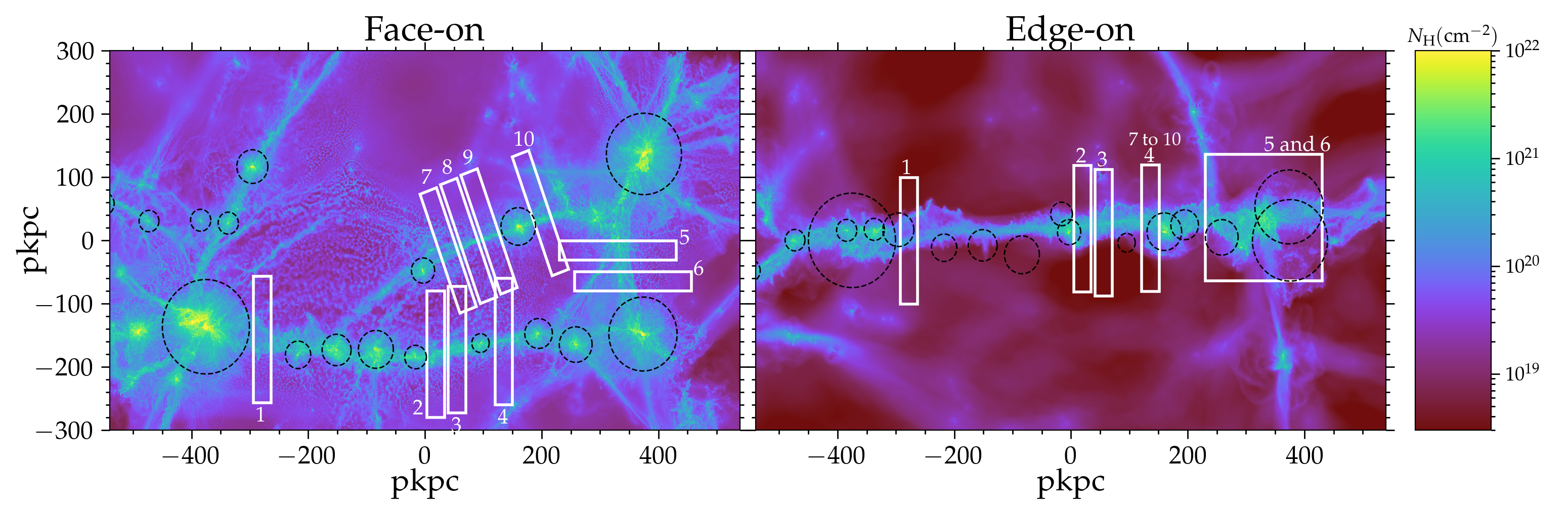}
    \caption{Hydrogen column density, $N_{\rm H}$,  of the large-scale cosmic sheet in our simulation at $z\sim 4$. Coordinates are given in proper physical distances. The left panel shows the ``face-on'' view of the sheet, while the right panel shows an ``edge-on'' view.  The integration depth for both panels is $\pm 60 \kpc$. Haloes with $\Mv\ge 10^{10}\,\msun$ within $\pm 60 \kpc$ of the sheet midplane are marked in both panels by black circles, whose radii correspond to the halo virial radii. The three largest haloes in the bottom-left, bottom-right, and top-right of the left-hand panel all have virial masses $\Mv\sim 10^{12}\,\msun$. White boxes mark the ten filament slices selected for analysis, numbered from $1$ to $10$, selected not to contain any haloes with $\Mv\ge 10^{10}\,\msun$ (see \se{Fil_selec}).}
    \label{fig:slices}
\end{figure*}

We focus our analysis on $z\sim 4$. At this time, our system consists of a well-defined cosmic sheet containing several prominent co-planar filaments, with end-points at massive haloes with $\Mv\sim 10^{12}\,\msun$, and along which nearly all haloes with $\Mv>10^9\,\msun$ are located. The sheet is formed by the collision of two smaller sheets at $z\gsim 5$, while most of the filaments merge by $z\sim 2.5$ \citep{mandelker2019shattering,mandelker2021thermal}. \Fig{slices} shows the Hydrogen column density, $N_{\rm H}$, of the sheet at $z\sim 4$ in face-on (left) and edge-on (right) projections\footnote{Projection maps of temperature, metallicity, and HI column density in the same region can be found in \citet{mandelker2021thermal}.}. In order to minimise contamination of the filament properties by haloes, we avoid analysing filament regions that contain haloes with $\Mv\ge 10^{10}\,\msun$, marked by black circles in \fig{slices}. This is $\sim 1\%$ of the mass of each of the three haloes at the nodes of the cosmic-web in this region, located at the bottom-left, bottom-right, and top-right of the left-hand panel in \fig{slices}. We selected ten such slices from three separate filaments, marked with white rectangles and numbered from $1-10$ in \fig{slices}. Each slice is $30\kpc$ thick along the filament axis, slightly larger than the typical characteristic scale radius of the filaments, $r_0$, as described in \se{thermal} below. However, we note that using a slice thickness of $(20-40)\kpc$ does not change our key results. We note that most of our slices contain 1-2 haloes with masses $10^9<\Mv/\msun<10^{10}$, some of which appear assoiated with local distortions in the density/temperature structure of the filament. However, these distortions average out when stacking the slices, and have no impact on our results. Likewise, all of our slices contain several haloes with $\Mv<10^8\msun$, though none of these appear to have any impact on the filament structure.

\smallskip
The orientation of the filament axis in each slice is determined by eye by approximating a straight line to the ridge line of maximal $N_{\rm H}$ in each slice, as seen in the face-on projection through the sheet (left-hand panel of \fig{slices}). Given the relatively small length of our slices compared to the total length of the filament, small changes to the axial directions in each slice do not change any of our results. We hereafter define the $z$-axis to refer to the local filament axis in each slice, while the $(x,y)$ plane represents the filament cross-section with the $y$-axis always aligned within the sheet. 

\smallskip
To find the filament centre in each slice, we project the gas density in each slice along its axial direction and initially place the filament centre at the point of maximum gas density. We then refine this using the ``shrinking-cylinders'' method. We begin by calculating the centre of mass of a cylinder of gas with radius $R_0=50\kpc$ and length $L=30\kpc$ about our initial centre, i.e., spanning the entire thickness of the slice, and update the centre to this position. We repeat this process five times, where the radius of the cylinder at each step is half that of the previous step, leading to a final radius of $50/2^5\sim 1.6\kpc$, while keeping $L$ unchanged. Changing the number of iterations from $4-6$ has no effect on our results. We experimented with alternative definitions of the filament centre, such as using shrinking-circles of gas density projected along the length of the slice rather than shrinking cylinders in 3D, using spheres rather than cylinders/circles, and using total (gas plus dark matter) mass rather than gas mass. All these definitions yield centres within $\sim (1-4)\kpc$ of each other for nearly all cases, and the differences have no bearing on our results.

\subsection{Three Radial Zones}
\label{sec:zones}

\begin{figure}
    \centering
    \includegraphics[trim={2.5cm 0.0cm 1.6cm 0.6cm}, clip, width =0.48 \textwidth]{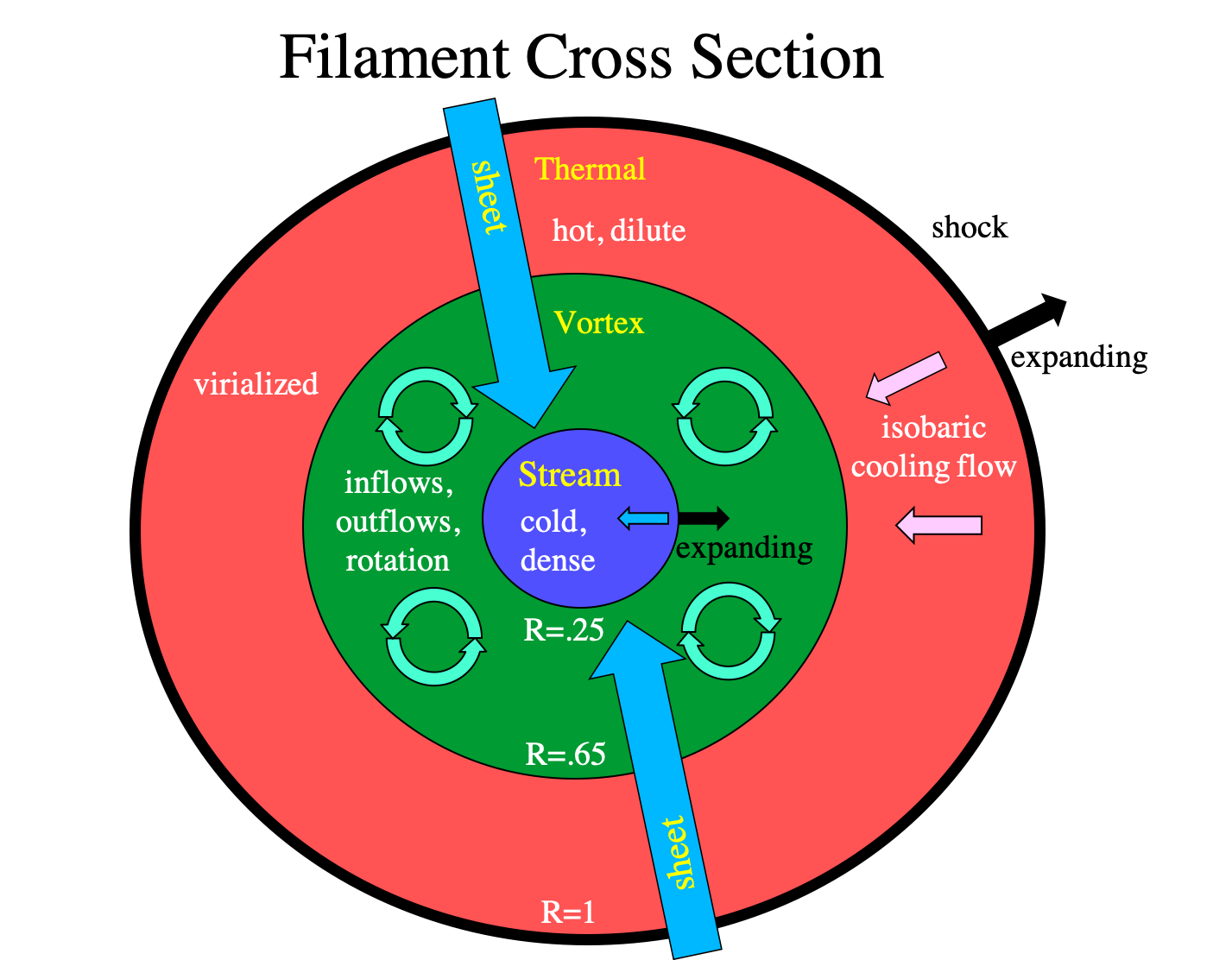}
    \caption{A schematic diagram showing the structure of the filament cross section, marking the key physical characteristics of the three zones described in \se{zones} and detailed throughout the paper. The stream is surrounded by a cylindrical accretion shock, at a normalised radius of ${\rm R}=1$, which expands outwards as the filament grows. The shock-heated gas in the outer Thermal \textbf{(T)} zone, at ${\rm R}\gsim 0.65$, is at the virial temperature of the underlying dark matter filament, with low density and inwardsradial velocity driven by an isobaric cooling flow. In this region, the thermal pressure gradients balance both gravity and the ram pressure from the inflowing gas. In the intermediate Vortex \textbf{(V)} zone, at $0.65\gsim {\rm R}\gsim 0.25$, the dynamics is dominated by a quadrupolar vortex structure which is induced by shearing motions between the sheet that feeds the filament and the post-shock filament gas. These vortices induce a complex combination of inflows, outflows, and rotation in this zone. Gravity here is balanced by a combination of thermal pressure gradients, centrifugal forces, and non-thermal random motions. The inner stream \textbf{(S)} zone, at ${\rm R}\lsim 0.25$, is a dense, cold, isothermal core which represents the cold streams that feed massive galaxies. While the stream boundary expands as the filament grows, the gas in this region is also characterised by both ithats and outflows due to non-radial inflow along the sheet. 
    }
    \label{fig:cartoon}
\end{figure}

As will be made clear and expanded upon throughout \ses{Radial_prop}-\sem{fil_dyn} below, we find that the filament cross section can be broadly divided into three radial zones, with a fourth zone outside the stream boundary. Since the details of each zone are motivated at different points in the paper, yet we refer to each throughout the paper and mark them on most figures, we introduce them here for the benefit of the reader. A schematic diagram showing this proposed structure is presented in \fig{cartoon}. 

\begin{enumerate}

    \item Outer thermal \textbf{(T)} zone, $0.65 r_{\rm shock}\lsim r \lsim r_{\rm shock}$: The post-shock region, where the gas is hot (at the virial temperature, see \se{virial}) and the dynamics are dominated by thermal and ram pressure.
    
    \item Intermediate vortex \textbf{(V)} zone, $0.25 r_{\rm shock}\lsim r\lsim 0.65 r_{\rm shock}$: The gas is cooling and the dynamics are dominated by a quadrupolar vortex structure and increasing centrifugal forces (see \se{fil_dyn}).
    
    \item Inner stream \textbf{(S)} zone, $0\lsim r \lsim 0.25 r_{\rm shock}$: A dense, isothermal core in the inner filament, representing the cold stream that penetrates the CGM of massive haloes.
    
\end{enumerate}

In addition, we define a fourth zone outside the filament boundary, the Pre-shock \textbf{(PS)} zone at $r\gsim r_{\rm shock}$. This is the region outside the accretion shock surrounding the filament, where cold and diffuse gas is free-streaming towards the filament. 
While $r_{\rm shock}$ is well-defined (\se{thermal}), the boundaries between the \textbf{T} and \textbf{V} zones and between the \textbf{V} and \textbf{S} zones are not as sharp. We crudely assign a width of $\sim 0.05r_{\rm shock}$ to each of them, $r/r_{\rm shock}\sim (0.6-0.65)$ and $(0.2-0.25)$ respectively, marked in the figures below. This roughly corresponds to the width of transition regions of various zone characteristics in stacked data, as detailed throughout the paper. However, we caution that the precise width of each zone likely depends on filament properties and is not universal.

\section{Filament Radial Structure}
\label{sec:Radial_prop}

In this and the following sections, we analyse the radial structure of various filament properties. These include the gas density, temperature, and thermal pressure (\se{thermal}), the baryon fraction (\se{fbar}), the dark matter density and radial velocity dispersion (\se{DM_prop}), the mass-per-unit-length of gas and the total mass (\se{Line_Mass}), the ratio of cooling time to free-fall time (\se{tcool}), the virial parameter (\se{virial}), and the internal kinematics and dynamics (\se{fil_dyn}). We compute all profiles as a function of the cylindrical radius, i.e., the distance from the local filament axis as defined in \se{Fil_selec}, with each radial bin representing a cylindrical shell extending the thickness of the slice, $L=30\kpc$. 

\subsection{Thermal Properties and Characteristic Radii}
\label{sec:thermal}

\begin{figure*}
    \centering
    \includegraphics[scale = 0.32]{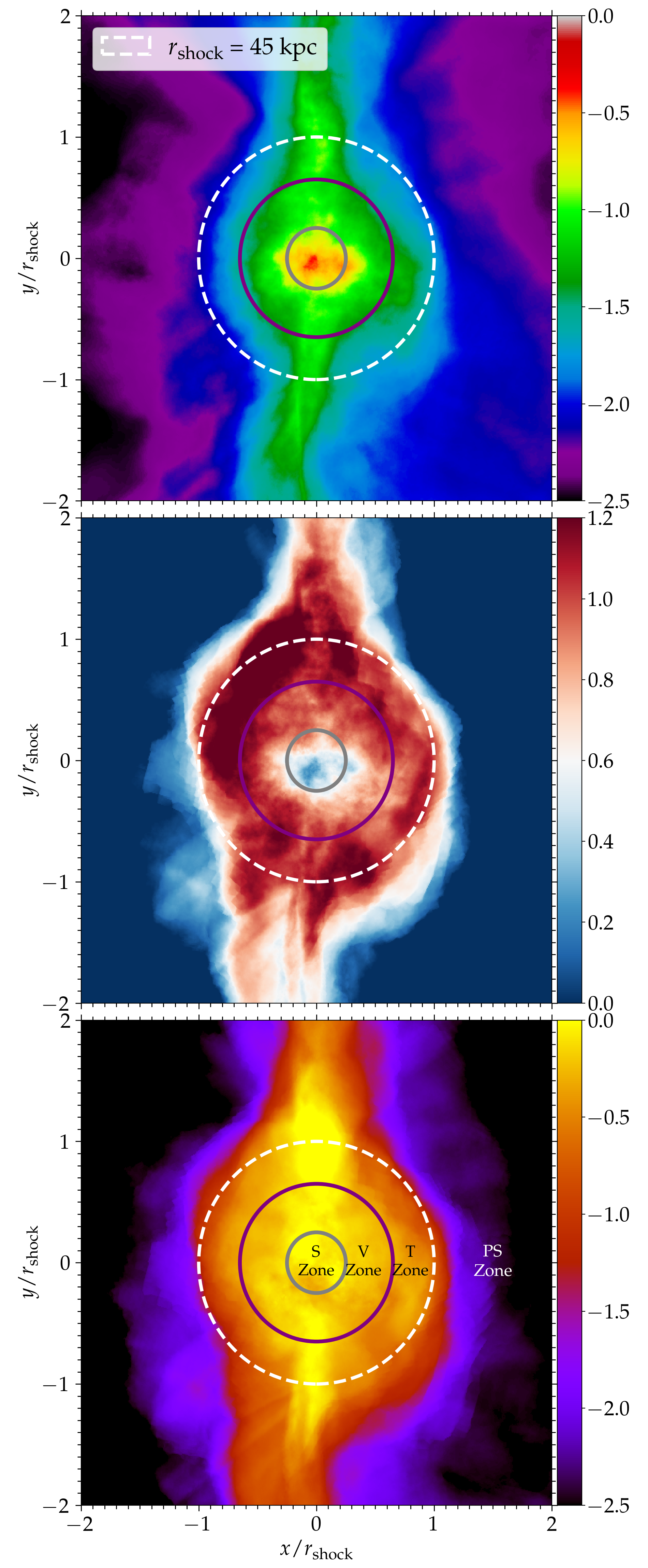}
    \includegraphics[scale = 0.32]{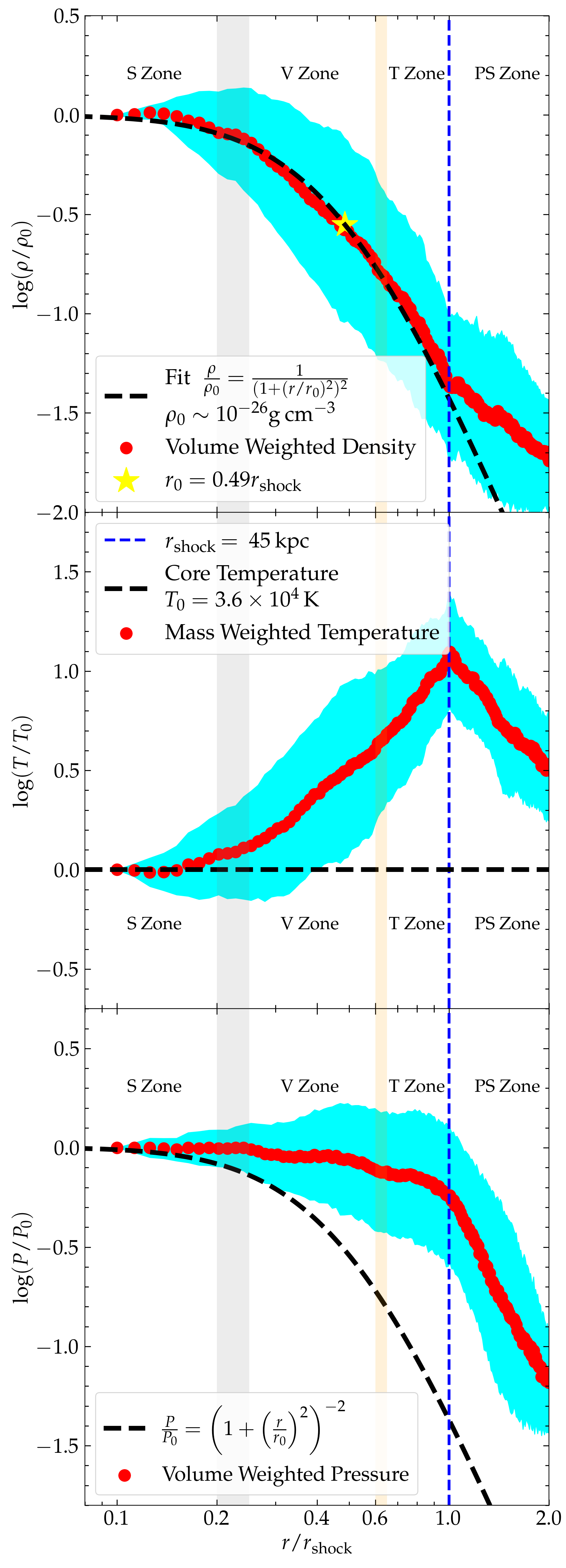}
    \caption{Thermal structure of filaments, stacked among the ten slices. We show projected maps integrated along the filament axis (left) and radial profiles (right) of gas density (top), temperature (middle), and thermal pressure (bottom) of the text for details regarding the averaging and stacking procedures. The temperature profiles reach a well-defined maximum that defines a shock radius, $r_{\rm shock}$, marked by white circle in the projection maps and vertical blue lines in the radial profiles. The temperature declines towards smaller radii, reaching an isothermal core in the \textbf{S} Zone, $r\lsim 0.25 \, r_{\rm shock}$, where the gas is dense and cold. In the \textbf{V} zone, $r/r_{\rm shock}\sim (0.25-0.65)$, the pressure is roughly constant while the temperature increases by a factor of $\sim 3$. In the \textbf{T} zone, $r/r_{\rm shock}\sim (0.65-1.0)$, the pressure drops by a factor of $\sim 2$, the temperature increases by a factor of $\sim 5$, and the density develops a slope of $\rho \propto r^{-3}$. Beyond the shock radius, in the \textbf{PS} zone, the density becomes dominated by the sheet and develops a shallower profile. The boundaries between the \textbf{S} and \textbf{V} zones and between the \textbf{V} and \textbf{T} zones are marked by concentric circles in the projection maps and by shaded regions of width $0.05 \, r_{\rm shock}$ in the radial profiles. At $r<r_{\rm shock}$, the density profile is well fit by an infinite self-gravitating isothermal cylinder in hydrostatic equilibrium, shown by the black dashed line, with a scale radius of $r_0\sim 0.5\,r_{\rm shock}$ marked by a yellow star in the density profile. However, this model is a very poor fit to the pressure profile, where strong gradients are only present in the \textbf{T} zone, while the \textbf{S} and \textbf{V} zones are nearly isobaric.} 
    \label{fig:thermal profiles}
\end{figure*}

In \fig{thermal profiles}, we address the distribution of gas density (top), temperature (centre), and thermal pressure (bottom), stacked among all ten slices. We present stacked projection maps on the left, and stacked radial profiles on the right. The projection maps represent weighted-averages along the filament axis (the $z$ direction) across the slice thickness of $L=30\kpc$. The average density and pressure are weighted by volume, while the temperature is weighted by mass. Prior to stacking, we align the $y$-axes such that they lie in the sheet containing the filaments (see \fig{slices}), while the $x$-axes remain perpendicular to the sheet. We then normalize each property by its value at the filament centre, and take the average in log-space among all ten slices. The cross-sections are roughly circular with high-density, low-temperature cores surrounded by a low-density, high-temperature medium, and roughly constant pressure throughout. The vertical high-density region extending from the filament in the $y$-direction breaking the circular symmetry represents the sheet. There are well-defined accretion shocks around both the filament and the sheet, visible in both the temperature and pressure maps. The filamentary accretion shocks identified here have typical Mach numbers of order $\mathcal{M}\gsim 10$ (Appendix \ref{sec:Mach}). While the accretion-shock around the sheet is discussed in \citet{mandelker2019shattering,mandelker2021thermal,Pasha.etal.22}, those works did not explicitly identify filamentary accretion shocks as seen here. Such filamentary accretion shocks have been both predicted theoretically \citep[e.g.,][]{Klar.Mucket.2012, birnboim2016stability} and seen in other cosmological simulations \citep{ramsoy2021rivers}. While the sheet clearly breaks the circular symmetry of the filament cross-section, for most of our analysis (except \fig{radial inflows} below) we ignore this complication and treat the filament as circular. A more accurate analysis distinguishing between the on-sheet and off-sheet properties and dynamics is deferred to future work. 

\smallskip
When generating the profiles in the right-hand column, we first use 150 linearly spaced bins out to a radius of $R=100\kpc$ for each individual slice, namely a bin size of $\Delta r = 100\kpc/150\approx0.67\kpc$. While this is larger than the typical cell size near the filament centre, it is slightly smaller than the typical cell size at large distances from the filament axis (see \fig{bin sizes and cell sizes} below). However, due to the unstructured nature of the grid, all of our bins are sufficiently populated for good statistics, with $\sim(4000-6000)$ cells per bin in the \textbf{S} and \textbf{V} zones, and $\sim(2000-4000)$ cells per bin in the \textbf{T} Zone. 
The density and pressure (temperature) profiles represent the volume-weighted (mass-weighted) average of all gas cells in each cylindrical shell. In each slice, the temperature profile reaches a well-defined maximum, which we define as the shock-radius, $r_{\rm shock}$. Once $r_{\rm shock}$ is identified, we recompute each profile using 150 linearly spaced bins from $r/r_{\rm shock}=0.1-2.0$ and normalise them by their central values. We then compute the mean (red dots) and standard deviation (cyan shaded regions) in log-space among all ten slices in each bin to produce the stacked profiles shown in \fig{thermal profiles}. Hereafter, all stacked profiles and projection maps presented used binonproportional to $r_{\rm shock}$ of each slice. The location of the shock seen on the temperature and pressure maps is indeed at $r\simeq r_{\rm shock}$, marked by a dashed-white circle in each projection. 

\smallskip
At small radii, the filaments are characterised by a core of constant density and temperature extending to roughly $\sim (0.2-0.25)\,r_{\rm shock}$, namely the \textbf{S} zone defined in \se{zones}\footnote{Note that our centering procedure described in \se{Fil_selec} does not guarantee that each slice will be centered on its densest or coldest point. This leads to slight offsets in the density maximum and temperature minimum with respect to the centres of the projection maps shown in \fig{thermal profiles}. Nonetheless, after stacking the radial profiles do show a well-defined central core.}. Outside of this radius, the temperature begins increasing towards a maximum at the shock radius, marked in each profile panel with a vertical dashed blue line. On average, the peak temperature at the shock is $\sim (10-20)$ times larger than the core temperature. This is consistent with the shock Mach number, which is measured to be $\mathcal{M}\sim (10-20)$ for all slices (see Appendix \ref{sec:Mach}). Within the \textbf{V} zone, $r\sim (0.25-0.65)r_{\rm shock}$, the pressure remains roughly constant, decreasing by only $\sim 10\%$ from its central value. The temperature increases by a factor of $\sim 3$ from its central value, often considered the threshold for `cool' gas in studies of multiphase gas mixing \citep[e.g.,][]{Scannapieco.Bruggen.2015,Gronke.Oh.2018,mandelker2020KHI4}, though the density declines by a larger factor in this region. The cold and dense filament core in the \textbf{S} zone thus seems to be in pressure equilibrium with the hot and diffuse mixed gas in the \textbf{V} zone. In the \textbf{T} zone, $r\sim (0.65-1.0)r_{\rm shock}$, the pressure decreases by another factor of $\lsim 2$, the temperature rapidly increases by a factor of $\sim 5$, and the density reaches a slope of roughly $\rho\propto r^{-3}$ at $r\lsim r_{\rm shock}$. As will be discussed further below, thermal pressure gradients are indeed not the dominant force in the \textbf{S} and \textbf{V} zones, becoming important only in the \textbf{T} zone. Outside the shock, in the \textbf{PS} zone, the pressure and temperature rapidly decline, while the slope of the density profile becomes noticeably shallower. In this region, the density is dominated by the underlying sheet, rather than by the filament itself \citep[see also][]{ramsoy2021rivers}. The boundaries between the four radial zones are marked by vertical lines/shaded regions in the radial profiles and by concentric circles in the projections. 

\begin{figure}
    \centering
    \includegraphics[trim={0.0cm 0.0cm 0.0cm 0.0cm}, clip, width =0.48 \textwidth]{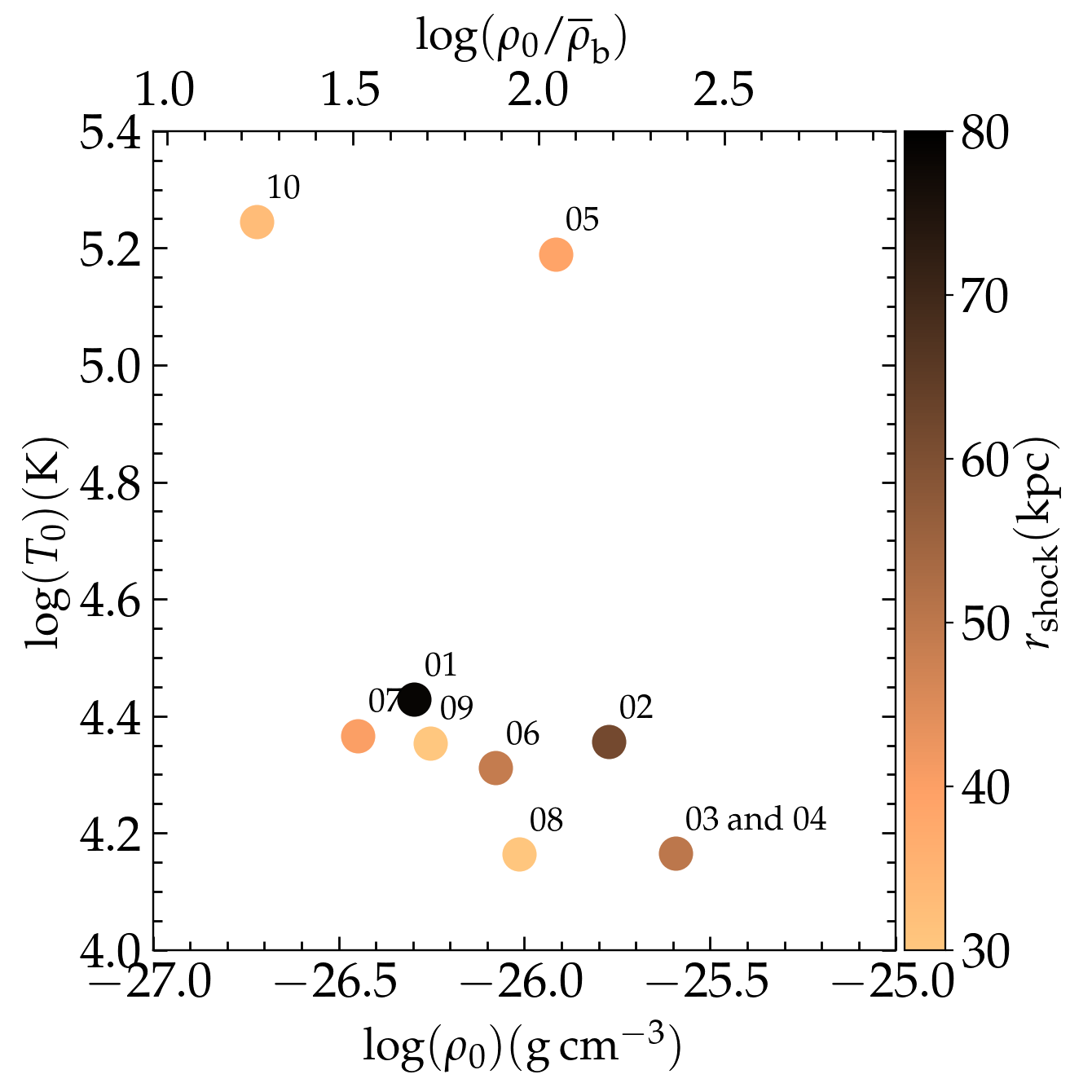}
    \caption{Distribution of central densities, $\rho_0$ on the $x$-axis, central temperatures, $T_0$ on the $y$-axis, and shock radii, $r_{\rm shock}$ shown in colour, for our ten filament slices. The central density is noramlised by the average baryonic density at $z=3.93$, $\overline{\rho}_{\rm b}=\Omega_{\rm b}\rho_{\rm crit}(1+z)^3=1.00\times 10^{-28}{\rm g \: cm^{-3}}$ (labelled at the top). The corresponding slice numbers are denoted. Note that slices 03 and 04 in the lower right corner overlap with each other.
    }
    \label{fig:rho_0 and T_0 scatterings}
\end{figure}

\smallskip
We show the values of central density, $\rho_0$, central gas temperature, $T_0$, and shock radius, $r_{\rm shock}$, for our ten slices in \fig{rho_0 and T_0 scatterings}. Most of the slices have central densities $\rho_0 \sim (10^{-26.5}-10^{-25.5})\cmc$, corresponding to a Hydrogen number density of $n_{\rm H}\sim 0.01\cmc$ and to a baryonic overdensity of $\lsim 100$. Their central temperatures are $T_0\sim 2\times 10^4\K$, placing the filament core in approximate thermal equilibrium with the UVB \citep[e.g.,][]{mandelker2020KHI4}. The shock radii are typically $r_{\rm shock}\sim (35-50)\kpc$, with an average value of $\sim 45\kpc$ among the ten slices. The typical core size is thus $r_{\rm core}\sim 0.25\,r_{\rm shock}\sim 12 \kpc$. These values of $\rho_0$, $T_0$, and $r_{\rm core}$ are in good agreement with those predicted for cold streams feeding massive high-$z$ galaxies from the cosmic-web \citep{dekel2009cold,mandelker2018cold,mandelker2020LABs}.It thus seems that the cold streams predicted to penetrate the CGM of massive haloes and feed their central galaxies can be associated with the isothermal cores of cosmic-web filaments, motivating our definition of the \textbf{S} zone.

\smallskip
Slices 05 and 10 seem to have much hotter cores, with $T_0\gsim 10^5\K$. Both of these slices lie close to the intersection between two filaments (see \fig{slices} left), leading to a highly disturbed morphology where the cold region of the filament is off-centre. However, even in these cases, the minimum temperature in the filament is $\lsim 3\times 10^4\K$. Similarly, slices 01 and 02 have unusually extended shocks, with $r_{\rm shock}\sim (70-80)\kpc$. Slice 01 is adjacent to a massive, $\sim 10^{12}\msun$ halo, which likely influences both the extent and the Mach number of the shock, the latter of which is $\gsim 50\%$ larger than the rest of the sample (see Appendix \ref{sec:Mach}). The large shock radius in slice 02 seems to be due to curvature in the filament in this region, making it difficult to track the filament spine. However, these slice-to-slice variations do not influence our stacked results.

\smallskip
To further characterise the filament radial structure, we fit the stacked density profile shown in the upper-right hand panel of \fig{thermal profiles} to the density profile of an infinitely long, self-gravitating, isothermal cylinder in hydrostatic equilibrium \citep{ostriker64}, 
\begin{equation}
\label{eq:isothermal_density}
    \rho_{\rm isothermal}(r)=\rho_0\left[1+\left(\dfrac{r}{r_0}\right)^2\right]^{-2},
\end{equation}
with 
\begin{equation}
\label{eq:r0}
    r_0=\sqrt{\frac{2k_{\rm B}T_0}{\mu m_{\rm p}\pi G \rho_0}},
\end{equation}
where $\rho_0$ is the central density, $G$ is the gravitational constant, $k_{\rm B}$ is the Boltzmann constant, $m_{\rm p}$ is the proton mass, $\mu$ is the mean molecular weight, and $T_0$ is the isothermal temperature. While the filament is clearly not isothermal within $r_{\rm shock}$, several previous works have attempted to model intergalactic filaments as such \citep[e.g.,][]{Harford2011isothermal,mandelker2018cold,ramsoy2021rivers}. It is, therefore, interesting to see how well this model can describe the actual density profile. This is shown by the dashed black line in the upper-right-hand panel of \fig{thermal profiles}. As we can see, it actually fits the density profile quite well, except at large radii, $r\gsim r_{\rm shock}$, where the density becomes dominated by the sheet rather than the filament. Similar results were found in \citet{ramsoy2021rivers}, who attempted to fit the density profile with a combination of an isothermal cylinder and an isothermal sheet, though we make no such attempt here.

\smallskip
When fitting our stacked density profile to \equ{isothermal_density}, the only free parameter is $r_0/r_{\rm shock}$, since the density has been normalized to its central value so effectively $\rho_0=1$\footnote{Fitting for $\rho_0$ as well produces a value extremely close to $1$.}. From the fit, we derive a characteristic scale radius of $r_0\sim 0.5\,r_{\rm shock}$, marked by a yellow star on the density profile. This is roughly twice as large as the radius of the cold, dense, isothermal core that represents the cold stream. It is slightly smaller than the outer radius of the \textbf{V} zone, though there is no clear relation between these two radii. 

\smallskip
We can use the isothermal fit to the density profile to find the corresponding temperature and pressure profiles. The isothermal temperature is given by 
\begin{equation}
\label{eq:isothermal_temperature}
    T_{\rm isothermal}=\frac{r_0^2\mu m_{\rm p}\pi \rho_0 G}{2k_{\rm B}},
\end{equation}
{\no}and the isothermal pressure profile is given by 
\begin{equation}
\label{eq:isothermal_pressure}
    P_{\rm isothermal}(r)=\frac{\pi G\rho_0^2r_0^2}{2}\left[1+\left(\frac{r}{r_0}\right)^2\right]^{-2}.
\end{equation}
Plugging the average value of $\rho_0\sim 10^{-26}\cmc$ and the derived value of $r_0\sim 0.5r_{\rm shock}\sim 22\kpc$ into \equ{isothermal_temperature} returns a value of $T_0\sim 3.6\times 10^4 \, \K$, which is roughly the average value of $T_{\rm isothermal}$ measured among our ten slices (see \fig{rho_0 and T_0 scatterings}). The density profile thus resembles that of an isothermal filament at the core temperature. We discuss this further in \se{comparison}. The model pressure profile, unsurprisingly, drastically underpredicts the pressure outside the core region (\textbf{S} zone), where the temperature increases and deviates strongly from isothermality. 
This suggests that thermal pressure gradients are not the primary force supporting the filaments against gravity, at least not in the \textbf{S} and \textbf{V} zones where the pressure is nearly constant both on and off the sheet. From the projection map we see that in the \textbf{T} zone, the off-sheet pressure declines more rapidly, leading to stronger gradients in this region. This will be discussed further in \se{sim_dynamics}. 

\begin{figure}
    \centering
    \includegraphics[trim={0.0cm 0.0cm 0.0cm 0.0cm}, clip, width =0.48 \textwidth]{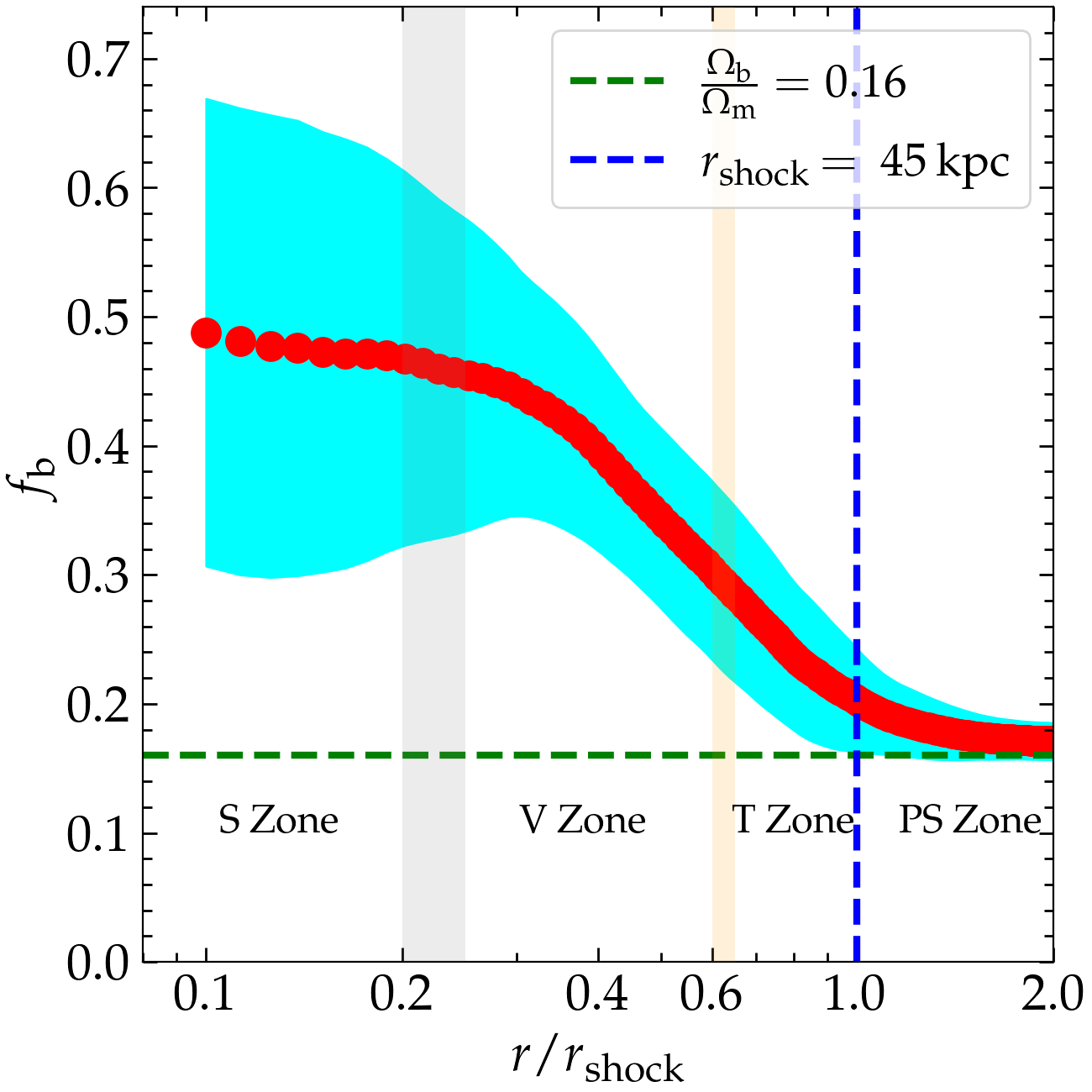}
    \caption{Baryon fraction as a function of radius. 
    As in \fig{thermal profiles}, red dots indicate the mean and the cyan-shaded region indicates the standard deviation among our ten slices, while the vertical lines/shaded regions mark the boundaries between the four filament zones. In the \textbf{PS} Zone, $r\gsim r_{\rm shock}$, $f_{\rm b}$ converges to the universal baryon fraction of $\sim 0.16$, marked by the horizontal dashed green line. The gas is more centrally concentrated than dark matter due to efficient cooling, so $f_{\rm b}$ rises through the \textbf{T} and \textbf{V} zones, saturating at $f_{\rm b}\sim 0.5$ in the \textbf{S} zone, where the gas and dark matter masses are comparable. 
    }
    \label{fig:fbar}
\end{figure}

\begin{figure*}
    \centering
    \includegraphics[scale = 0.33]{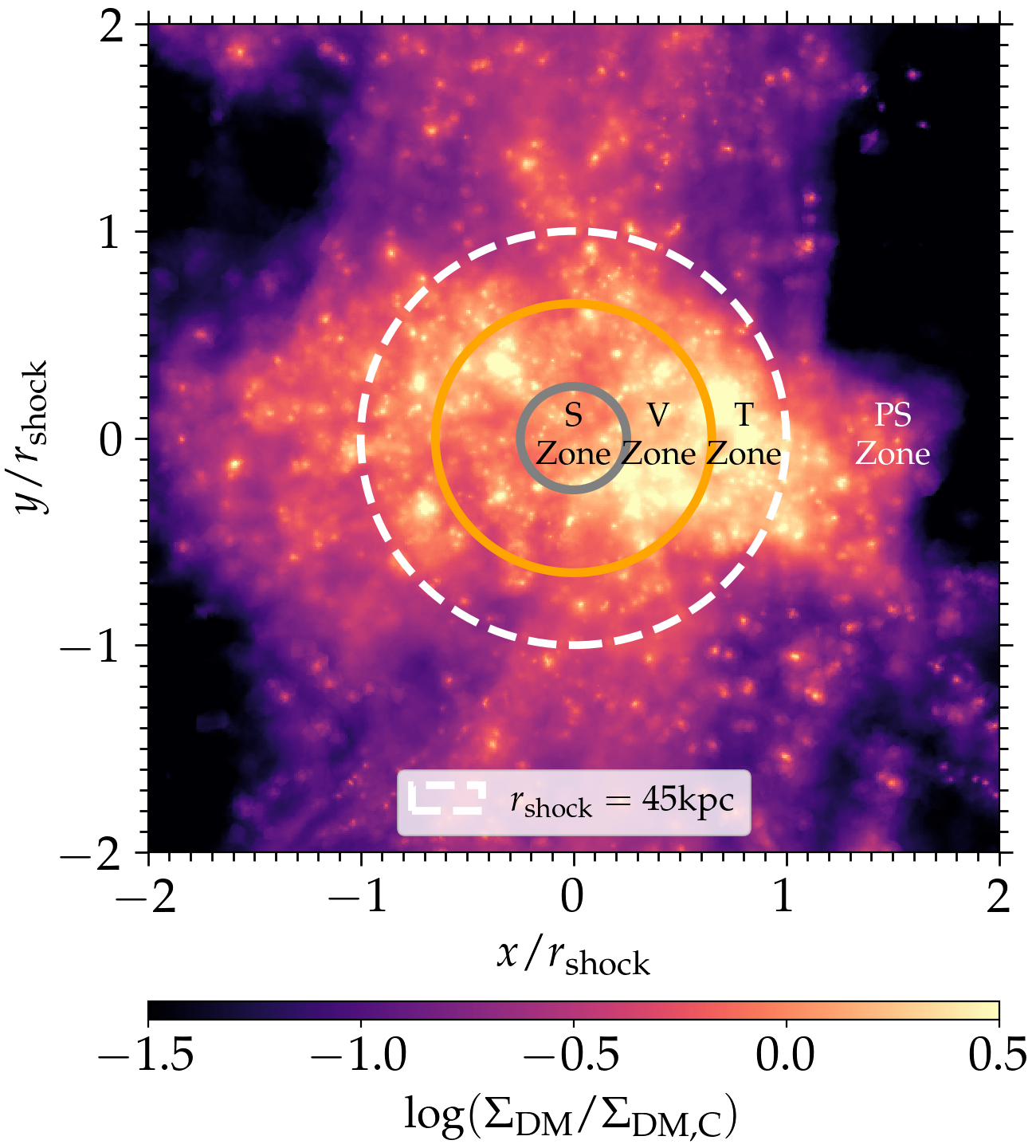}
    \includegraphics[scale = 0.37]{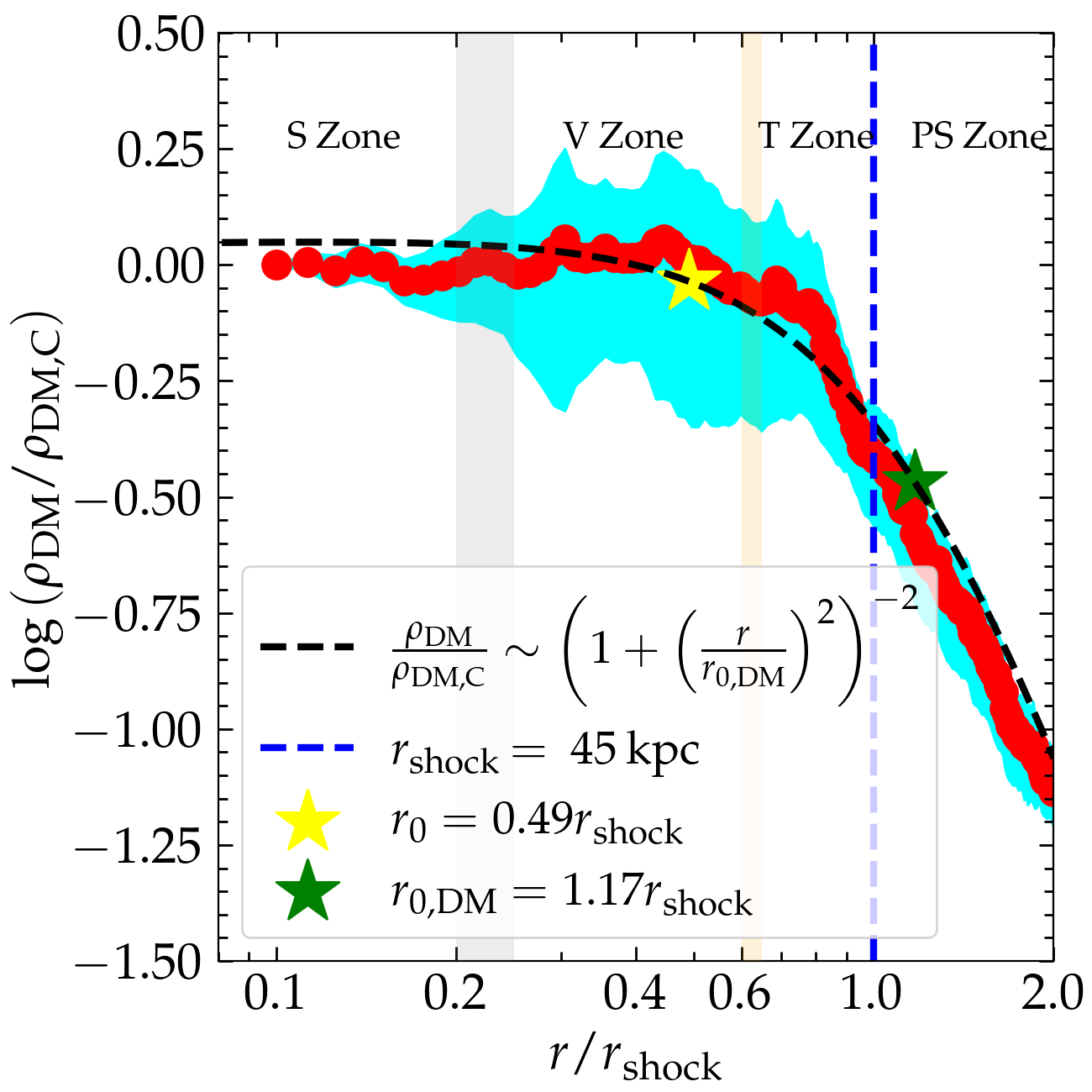}
    \includegraphics[scale = 0.37]{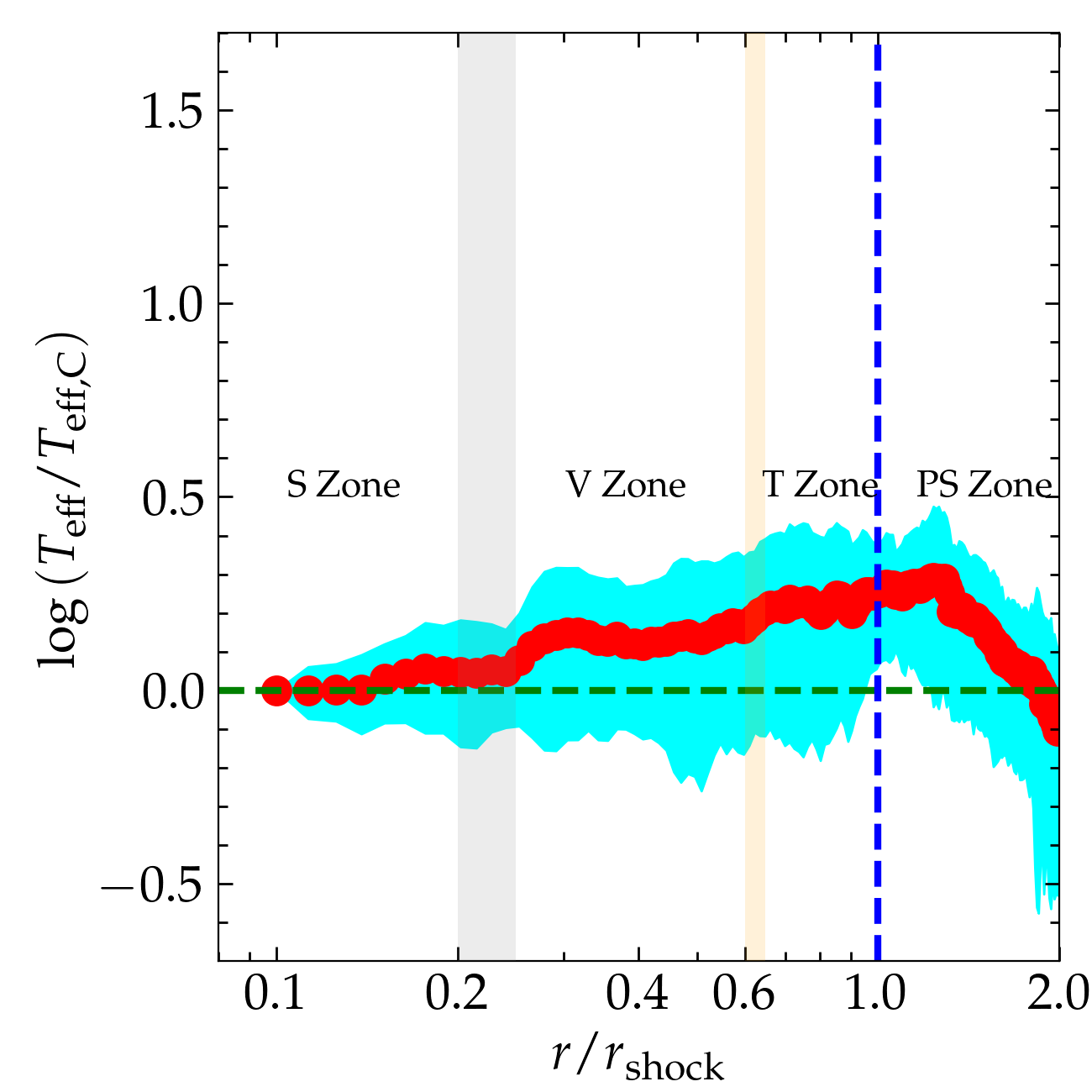}
    \caption{\textbf{\textit{Left panel: }}Projected map of the dark matter density averaged along the filament axis and stacked among our ten slices. The stacking procedure and slice orientation are the same as in the projected maps of gas properties shown in \fig{thermal profiles}. The shock radius, $r_{\rm shock}$, identified based on the gas temperature profiles, is marked by a white circle. As in \fig{thermal profiles}, we also mark the boundaries between the four radial zones with concentric circles. \textbf{\textit{Middle and right panels: }} Structure of the dark matter filaments, stacked among our ten slices. We show the radial profiles of dark matter density (middle) and effective temperature (right, computed using the dark matter radial velocity dispersion following \equnp{DM_temp}). These were normalised, stacked, and averaged in the same way as the gas profiles shown in \fig{thermal profiles}, with red points (cyan shaded region) representing the mean (standard deviation) in log space. The $y$-axes in these two panels span the same range as the upper and middle right-hand panels of \fig{thermal profiles}. The black dashed line in the density profile shows a fit to the same parametric isothermal model used for the gas (\equnp{isothermal_density}), which matches the data well. The yellow and green stars mark the scale radii for the gas and dark matter density profiles, respectively. The latter is slightly larger than the shock radius, $r_{\rm shock}$, marked with a vertical blue line. The vertical shaded regions mark the boundaries between the \textbf{S} and \textbf{V} zones and the \textbf{V} and \textbf{T} zones. The effective dark matter temperature increases by a factor of $\sim 2$ from the filament core to a peak near $r_{\rm 0, DM}$, and is thus more isothermal than the gas, whose temperature increases by a factor of $\sim 20$ over the same radial range. 
    }
    \label{fig:DM prop}
\end{figure*}

\subsection{Baryon Fraction}
\label{sec:fbar}

In \fig{fbar} we show the baryon fraction as a function of radius, stacked among our ten slices as described in \se{thermal}. The baryon fraction in each radial bin, $f_{\rm b}(r)$, is given by the ratio of the baryonic mass (gas and stars) to the total mass (gas, stars, and dark matter) enclosed within the radius~$r$. In practice, the stellar mass is negligible, contributing $\sim 0.3 \%$ of the total baryon mass, and $f_{\rm b}$ can be thought of as the gas fraction. In the \textbf{PS} zone, $r\gsim r_{\rm shock}$, $f_{\rm b}(r)$ converges to the universal baryon fraction, $f_{\rm b}\sim 0.16$, with very little scatter among the different slices. Within the filament core, the \textbf{S} zone, $f_{\rm b}$, saturates at roughly $0.5$ on average, with a large scatter of $\sim (0.3-0.7)$. The gas is thus more centrally concentrated than dark matter due to efficient cooling, resulting in a comparable contribution of gas and dark matter in the core. This is similar to galaxies in dark matter haloes, for example, in the Milky Way the baryon-to-total mass ratio within the solar circle is $\sim 0.5$.

\subsection{Dark Matter Properties}
\label{sec:DM_prop}

Following the discussion in \se{thermal} and \se{fbar}, we now examine the radial structure of the dark matter in our filament slices. In the left panel of \fig{DM prop}, we show a projected map of the dark matter density, oriented and stacked among all ten slices in an identical way to the projection map of gas density shown in \fig{thermal profiles}. Similar to gas, the dark matter filament has a roughly circular cross-section if ignoring the contribution of the overdense pancake along the $y$ direction. However, unlike the gas density which is very centrally concentrated, the dark matter density appears approximately constant within $r_{\rm shock}$, marked with a white circle. The dark matter distribution is also much clumpier than the gas, due to the presence of low mass haloes with $\Mv<10^{10}\msun$ (recall that higher mass haloes have already been removed when selecting the slices). Removing haloes with lower and lower masses prior to projecting the density results in smoother and smoother maps, until the clumpiness disappears once haloes with $\Mv\gsim 10^8 \msun$ are removed. This clumpiness thus does not seem to be due to internal fragmentation within the filament, but rather to the cosmological halo mass function.

\smallskip
In the middle panel of \fig{DM prop}, we present the radial profile of dark matter density, normalised by its central value and stacked among our ten filament slices, using the same procedure described in \se{thermal} for the gas profiles. We fit the dark matter density profile to the same parametric isothermal model used for the gas (\equnp{isothermal_density}). The fit is shown in the middle panel of \fig{DM prop} with a black-dashed line and is a good representation of the data. From \fig{fbar} we know that the central dark matter density is comparable to the central gas density, namely an overdensity of $\sim 30$ with respect to the mean matter density\footnote{Corresponding to an overdensity of $\sim 200$ with respect to the mean baryonic density, as shown in \fig{rho_0 and T_0 scatterings}.} at $z\sim 4$. From the fit, we extract the scale radius of the DM density $r_{0,{\rm DM}}\simeq 1.17\, r_{\rm shock}$, which is marked in \fig{DM prop} along with the scale radius of the gas density profile, $r_0\lsim 0.5\,r_{\rm shock}$. The radius of the DM density core is $r_{\rm core,\,DM}\sim 0.8r_{\rm shock}$, approximately 4 times larger than the gas density core. This is consistent with the baryon fraction profile (\fig{fbar}), which showed that the gas was much more centrally concentrated than dark matter. This is unlike the case studied in \citet{ramsoy2021rivers}, where gas and dark matter were found to have similar concentrations and density profiles. We address this in \se{comparison}. 

\smallskip
In the right-hand panel of \fig{DM prop}, we present the radial profile of dark matter effective temperature, normalised by its central value and stacked among our ten filament slices as described above. This is derived from the radial velocity dispersion of dark matter, which we compute in the standard way as $\sigma_r^2 = \langle v_r^2\rangle-\langle v_r\rangle^2$, where $v_r$ is the radial velocity with respect to the filament axis and $\langle \dots\rangle$ represents a mass-weighted average\footnote{Note that since in our simulation every dark matter particle has the same mass this is simply the arithmetic variance.} over all dark matter particles in a given cylindrical shell at radius $r$. The effective dark matter temperature\footnote{By assuming a mean particle mass of $\mu m_{\rm p}$, \equ{DM_temp} defines the temperature the gas would have if it were in equipartition with the dark matter, similar to the way the halo virial temperature is typically defined.} is then defined via 
\be 
\label{eq:DM_temp}
k_{\rm B}T_{\rm eff}(r)=\frac{3}{2}\mu m_{\rm p}\sigma_r^2(r).
\ee
{\no}From the profile, we see that $T_{\rm eff}$ increases by a factor of $\sim 2$ from the filament core to a peak just outside $r_{\rm shock}$, near $r_{\rm 0,DM}$. This is much less than the factor $\sim 20$ increase in gas temperature, showing that the isothermal approximation is more valid for dark matter than for gas. The average central value of $T_{\rm eff}$ in the filament core is $T_{\rm eff, C}\sim 9.6\times 10^5 \Kdegree$. Using the average central dark matter density, our fit to $r_{0, {\rm DM}}$, and \equ{isothermal_temperature}, we calculate the effective ``isothermal'' temperature of the dark matter, $T_{\rm iso,DM}=1.2\times 10^6\, \Kdegree$. This is very similar to the measured central dark matter temperature, supporting the validity of the isothermal model for the dark matter. We note that dark matter haloes are also often modelled as isothermal spheres, though as singular isothermal spheres with a central density cusp, $\rho\propto r^{-2}$, rather than a central core as we find for filaments. 

\subsection{Mass per unit length (line-mass)}
\label{sec:Line_Mass}

\begin{figure}
    \centering
    \includegraphics[trim={0.0cm 0.0cm 0.0cm 0.0cm}, clip, width =0.48 \textwidth]{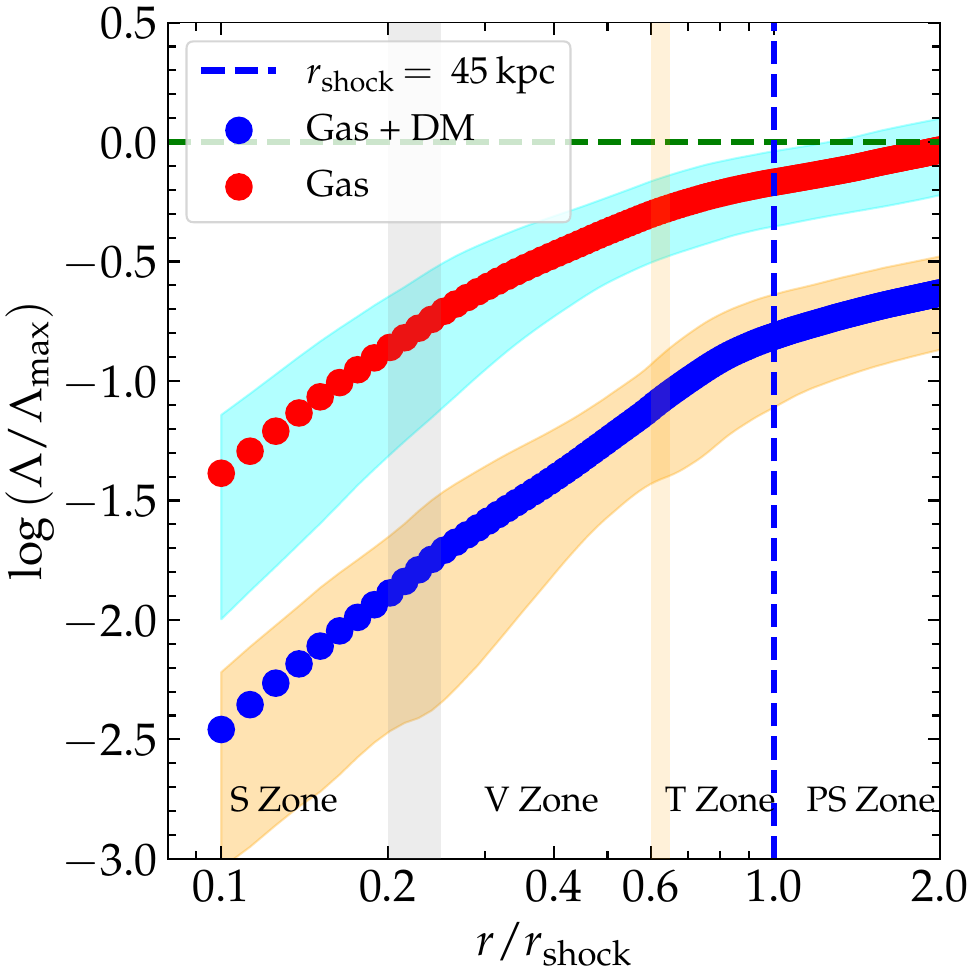}
    \caption{Profiles of the filament line-mass, $\Lambda$, accounting for only gas (red points) and total mass (mostly gas and dark matter, blue points), averaged among our ten filament slices. Each profile has been normalised by the corresponding, temperature-dependent, maximal line-mass for hydrostatic equilibrium, $\Lambda_{\rm max}$ (\equnp{Lmax}). Cyan and orange shaded regions represent $1-\sigma$ scatter of $\Lambda_{\rm gas}$ and $\Lambda_{\rm tot}$ among our ten filament slices, respectively. Note that since all slices have the same length (\se{Fil_selec}), this result is also the same as the total line-mass of a single filament consisting of all our ten slices connected end-to-end. While the total line-mass is clearly larger than the line-mass of gas only, the higher temperature of the dark matter results in a higher value of $\Lambda_{\rm max}$, leading to a lower normalised line-mass. While the normalised line-mass for gas approaches unity near the filament edge, the total line-mass is well below the critical value.}
    \label{fig:line mass}
\end{figure}

In \fig{line mass} we show radial profiles of the mass per-unit-length (hereafter line-mass), $\Lambda(<r)$, normalised and stacked among our ten filament slices. For a given slice, $\Lambda(<r)$ is simply defined as the total mass (of gas and/or dark matter, see below) interior to radius $r$, divided by the thickness of the slice, $L=30\kpc$. 
Motivated by our fits of the gas and dark matter density to isothermal profiles (\se{thermal}, \se{DM_prop}), prior to stacking we normalise the line-mass in each slice by the maximal possible value of a self-gravitating isothermal cylinder in hydrostatic equilibrium \citep{ostriker64},
\be 
\label{eq:Lmax}
\Lambda_{\rm max}=\frac{2c_s^2}{G},
\ee
where $c_s^2=k_{\rm B}T/(\mu m_{\rm p})$ is the isothermal sound speed squared, and $T$ is the effective isothermal temperature given by the fit of the model using the average central density, the characteristic scale radius, $r_0$, and \equ{isothermal_temperature}.  
This is similar to the central temperature in the filament core. Filaments with $\Lambda\sim \Lambda_{\rm max}$ are unstable to gravitational fragmentation and radial collapse \citep{mandelker2018cold,Aung.etal.19}, while much lower mass filaments are likely gravitationally stable.

\smallskip
In \fig{line mass}, we show separately the normalised line-mass profiles of gas alone (red) and total mass (blue, primarily gas and dark matter). Note that the total line-mass is obviously larger than the line-mass of gas alone, $\Lambda_{\rm tot}(r_{\rm shock})\sim 9\times 10^8\ \msun \kpc^{-1}$ while $\Lambda_{\rm gas}(r_{\rm shock})\sim 1.6\times 10^8\ \msun \kpc^{-1}\sim f_{\rm b}\Lambda_{\rm tot}(r_{\rm shock})$. However,
the gas has a much lower effective 
temperature than the dark matter, $\sim 3.6\times 10^4\,{\rm K}$ and $\sim 1.2\times 10^6\,{\rm K}$ respectively, while the effective 
temperature derived from the density profile of total mass 
is $\sim 8.6\times 10^5\,{\rm K}$. This leads to a much smaller value of $\Lambda_{\rm max}$ for the gas alone compared to the total mass — $2.3\times 10^8\, {\rm M}_\odot \, \kpc^{-1}$ 
and $5.5\times 10^9\, {\rm M}_\odot \, \kpc^{-1}$ respectively. 
As a result, the normalised line-mass of gas alone is larger than the normalised total line-mass. 

\smallskip
For the gas, $\Lambda/\Lambda_{\rm max}\sim 0.2$ near the outer boundary of the \textbf{S} zone at $r\sim r_{\rm core}\sim 0.25 \, r_{\rm shock}$, $\Lambda/\Lambda_{\rm max}\sim 0.7$ near $r_{\rm shock}$, and it approaches unity at $2\,r_{\rm shock}$. This is consistent with the results of \citet{mandelker2018cold} (see their equation 40), who predicted that cold streams feeding massive high-$z$ galaxies should have $\Lambda\sim \Lambda_{\rm max}$, which could lead to gravitational fragmentation and star-formation in filaments outside of galaxies. However, we stress that this estimate ignores non-thermal support such as turbulence, which is important at small radii (see \se{sim_dynamics} below), and the non-isothermality of the gas which is important at large radii (see \fig{thermal profiles}). The net effect on the gravitational fragmentation of star formation in filaments is thus unclear and is left for future work. When considering the total filament line-mass we have $\Lambda/\Lambda_{\rm max}\sim 0.02$ and $0.15$ within $r_{\rm core}$ and $r_{\rm shock}$ respectively, suggesting that gravitational fragmentation of the dark matter in the filament is unlikely, owing to its much larger effective temperature.

\section{Thermal Equilibrium}\label{sec:thermal equi}
\label{sec:tcool}

In analogy with the gaseous haloes around massive galaxies, groups, and clusters, we estimate the thermal stability of the shock-heated gas surrounding the isothermal core by examining the ratio of the gas cooling time, $t_{\rm cool}$, to the free-fall time, $t_{\rm ff}$. 

\smallskip
The gas cooling time, $t_{\rm cool}$, is given by
\begin{equation}
\label{eq:tcool}
    t_{\rm cool}=\frac{u\rho}{\mathcal{L} n_{\rm H}^2},
\end{equation}
where $u$ is the gas internal energy per unit mass, $\rho$ is the gas density, $n_{\rm H}$ is the hydrogen number density, and $\mathcal{L}=\mathcal{C}-\mathcal{H}$ is the net cooling rate, i.e., cooling minus heating, per unit density squared. All these properties are stored for each cell in the simulation, allowing us to evaluate $t_{\rm cool}$ for each cell. We then compute the average cooling time as a function of radius, $t_{\rm cool}(r)$, by taking the mass-weighted average of \equ{tcool} among all cells in each radial bin with temperatures $T>T_0=3.6\times 10^4\K$, the temperature of the cold isothermal core. The gas with $T\lsim T_0$ is near thermal equilibrium with UVB and often undergoes net heating, while virtually all gas with $T>T_0$ is undergoing net cooling. 

\smallskip
For an infinite cylinder, the free-fall time is given by
\begin{equation}
\label{eq:tff}
    t_{\rm ff}(r)=\sqrt{\frac{1}{4G\rho_{\rm mean}(<r)}},
\end{equation}
where $\rho_{\rm mean}(<r)=\Lambda(<r)/(\pi r^2)$ is the mean density interior to $r$. 

\smallskip
For isobaric cooling, $u\rho\propto P \sim {\rm const}$, so $t_{\rm cool}\propto \rho^{-2}$, while $t_{\rm ff}\propto \rho^{-1/2}$. Therefore, the ratio $t_{\rm cool}/t_{\rm ff}$ is smaller at smaller radii closer to the filament axis. When $t_{\rm cool}/t_{\rm ff}\lsim 1$, 
the system cannot maintain the shock-heated component, and the bulk of the gas cools and falls to the centre \citep[e.g.,][]{Rees77,White78,Birnboim2003shocks,birnboim2016stability,Fielding2017CGM,Stern2020maximum,Stern2021DLAs}. The outer radius where $t_{\rm cool}/t_{\rm ff}= 1$ and the volume-filling medium transitions from hot to cool corresponds to the `sonic radius' \citep{Bertschinger.1989,stern.etal.2020,stern.etal.2023}. Even if $t_{\rm cool}/t_{\rm ff}>1$, the hot gas may be unstable to local thermal instabilities where perturbations create dense clouds that cool and condense out of the hot medium and ``rain down'' to the centre. In studies of the hot CGM around massive galaxies or the ICM in galaxy clusters, this process is often referred to as precipitation and occurs when $t_{\rm cool}/t_{\rm ff}\lsim 10$ \citep[e.g.,][]{mccourt2012thermal,sharma2012thermal,Gaspari2012multiphase,Voit2015a,Voit2015b,Voit2015c}. The analogous threshold for cylindrical geometry may be different due to the different radial dependences of stabilising buoyancy forces. A detailed study of this is beyond the scope of this paper, and instead we adopt the threshold of $t_{\rm cool}/t_{\rm ff}\sim 10$ as representative of the threshold for condensation to occur.

\begin{figure}
    \centering
    \includegraphics[trim={0.0cm 0.0cm 0.0cm 0.0cm}, clip, width =0.48 \textwidth]{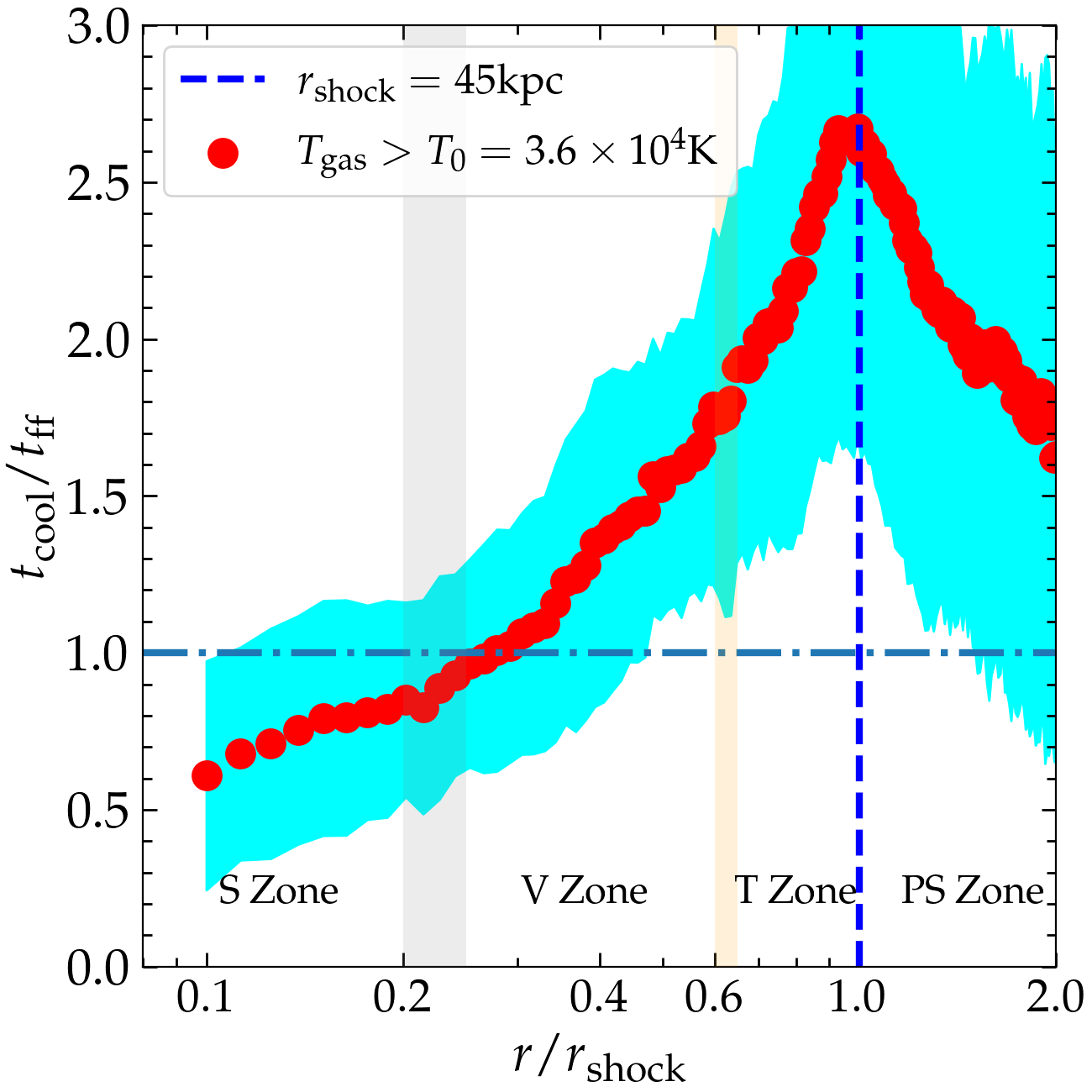}
    \caption{Stacked profile of ratio between cooling and free fall times for gas with $T>T_0=3.6\times 10^4 \K$, the temperature of the cold stream. Near $r_{\rm shock}$, $t_{\rm cool}\lsim 3t_{\rm ff}$ implying that the shocked medium is likely unstable to condensation and precipitation. In the \textbf{S} Zone, $t_{\rm cool}\lsim t_{\rm ff}$ consistent with the formation of a cold isothermal core at these radii.
    }
    \label{fig:tcool_tff}
\end{figure}

\smallskip
\Fig{tcool_tff} shows the ratio $t_{\rm cool}(r)/t_{\rm ff}(r)$ as a function of $r/r_{\rm shock}$, stacked among the ten slices by taking the average and standard deviation. 
At $r \sim r_{\rm shock}$, $t_{\rm cool}/t_{\rm ff}\sim 3$, which implies that while the shock may be stable to monolithic cooling it is likely unstable to precipitation and condensation. Indeed, we see several cases of cold clouds encompassed by hot gas in the \textbf{T} and \textbf{V} zones in individual slices. A more detailed analysis of these clouds is beyond the scope of the current study, and is left for future work. The cooling radius, where $t_{\rm cool}/t_{\rm ff}\sim 1$, occurs near the outer edge of the \textbf{S} zone, consistent with this being the boundary of the dense isothermal core. These results suggest that the cold stream is built by cooling of the post-shock filament gas, and that this cooling occurs roughly isobarically (see \fig{thermal profiles}).

\section{Virial Equilibrium}
\label{sec:virial}

In this section, we wish to ascertain to what extent the gas is in virial equilibrium within the gravitational potential of the filament. While dark matter haloes are expected to be bound and virialized in three dimensions, intergalactic filaments will at most be bound and virialized in the two dimensions perpendicular to the filament axes, while remaining unbound along their axis. Previous studies have speculated that intergalactic filaments may result from cylindrical collapse ending in virial equilibrium per-unit-length \citep{FG84,mandelker2018cold}. In this section, we test this hypothesis in our simulations. 

\subsection{The Virial Theorem per-unit-length}

An infinite, self-gravitating, collisionless and isolated cylinder in virial equilibrium (per-unit-length) obeys \citep{mandelker2018cold} 
\be 
\label{eq:virial_ideal}
2\mathcal{K}=G\Lambda^2,
\ee
{\no}where $\mathcal{K}$ is the kinetic energy per-unit-length and $\Lambda$ is the line-mass. If the motions are confined to the plane perpendicular to the filament axis, this becomes the two-dimensional version of the virial theorem. In a confined, gaseous system such as ours, we must account for additional terms related to thermal pressure, surface pressure, and magnetic pressure. 

\smallskip
As we show in Appendix \ref{sec:simplification} (\fig{plasma_beta}), magnetic support is negligible in the filament gas. Therefore, we use the virial theorem for an unmagnetized gaseous system. For such a system in time-independent equilibrium, the virial theorem states \citep[][Chapter 6.1]{krumholz2015notes}
\be
\label{eq:vir_thm}
    2(\mathcal{T}-\mathcal{T}_s)+\mathcal{W}=0,
\end{equation}
where $\mathcal{T}$ is the total thermal plus kinetic energy in the system, $\mathcal{T}_s$ is the confining pressure on the surface, including both thermal and ram pressure, and $\mathcal{W}$ is the total gravitational energy of the system. In integral form, these terms can be expressed as 
\be 
\label{eq:vir_tau}
\mathcal{T} = \int_V \left(\frac{1}{2}\rho v^2+\frac{3}{2}P\right){\rm d}V,
\ee
\be 
\label{eq:vir_taus}
\mathcal{T}_s = \oint_S \left({\vec{r}} \cdot \overleftrightarrow{\pi}\right) \cdot {\rm d}{\vec{S}},
\ee 
\be 
\label{eq:vir_grav}
\mathcal{W} = -\int_V \left(\rho {\vec{r}}\cdot{\vec{\nabla}}\Phi \right) {\rm d}V.
\ee
{\no}In \equs{vir_tau}-\equm{vir_grav}, $\rho$ is the gas density, ${\vec{v}}$ is its velocity, $P$ is its thermal pressure, $\Phi$ is the gravitational potential\footnote{Note that \equ{vir_grav} depends only on the gravitational acceleration, $\vec{\nabla}\Phi$, and therefore does not depend explicitly on the zero point of the potential, which cannot be defined at $r=\infty$ for an infinite cylinder as we are assuming.}, and $\overleftrightarrow{\pi}$ is the fluid pressure tensor given by 
\be
\label{eq:Pi_def}
    \pi_{ij}=\rho v_iv_j + \delta_{ij}P,
\end{equation}
{\no}where $\delta_{ij}$ is the Kronecker delta. 
We define the virial parameter 
\begin{equation}
\label{eq:vir para}
    \alpha_{\rm vir}=\frac{2(\mathcal{T}-\mathcal{T}_s)}{|\mathcal{W}|}.
\end{equation}
{\no}Note that this differs from the standard virial parameter for unmagnetized systems only in the inclusion of the surface pressure term, which is important for non-isolated systems\footnote{See \cite{shaw2006statistics} for the importance of including an analogous surface pressure term for collisionless particles when estimating the virial equilibrium of dark matter haloes.}. When $\alpha_{\rm vir}>1$, the system expands due to kinetic plus thermal energy, while for $\alpha_{\rm vir}<1$ the system collapses due to the combined effects of surface pressure and gravity. 

\smallskip
When evaluating \equs{vir_tau}-\equm{vir_grav} for our filament slices, we recall that we are only interested in virial equilibrium per-unit-length, in the two dimensions perpendicular to the filament axis. We thus make an approximation of cylindrical symmetry and treat the filament as an infinite cylinder. We are here neglecting the non-axisymmetric nature of the sheet, an approximation which will be validated by the results below. Therefore, we approximate the gravitational acceleration as 
\be 
\label{eq:grad_phi}
\vec{\nabla} \Phi \simeq \frac{2G\Lambda_{\rm tot}(<r)}{r}\hat{r},
\ee
{\no}with $\Lambda_{\rm tot}(<r)$ being the total line-mass interior to radius $r$. Similarly, the gas density and line-mass are related by 
\be 
\label{eq:lambda_gas}
2\pi r \rho\simeq \frac{{\rm d} \Lambda_{\rm gas}}{{\rm d} r}.
\ee 
{\no}We, therefore, approximate the gravitational potential energy as 
\be 
\label{eq:W(r)}
    \mathcal{W}(r)\simeq -2GL\int_0^r\Lambda_{\rm tot}\frac{{\rm d} \Lambda_{\rm gas}}{{\rm d}r'}~{\rm d}r',
\end{equation}
{\no}where $L=30\kpc$ is the filament slice thickness. We note that this factor $L$ is only included for consistency with the formalism presented in \citet{krumholz2015notes}, and in practice, we normalise all the energy terms in \equs{vir_tau}-\equm{vir_grav} by $L$ to obtain energy per-unit-length before computing $\alpha_{\rm vir}$. We numerically evaluate the integral in \equ{W(r)} by performing the following double-sum:
\be
\label{eq:double_sum}
    \int_0^{\Lambda_{\rm gas}(r)}{\rm d}\Lambda_{\rm gas}\Lambda_{\rm tot}=\sum_{r'<r}\left[\sum_{<r'}(m_{\rm gas}+m_{\rm dm}+m_{\rm star})\sum_{r'}m_{\rm gas}\right],
\end{equation}
where the sum $\sum_{<r'}$ in the square brackets is taken over all particles (dark matter and stellar) and gas cells interior to radius $r'$. The sum $\sum_{r'}$ is taken over all gas cells within the cylindrical shell of thickness $\Delta r'$ at $r'$. 

\smallskip
We numerically evaluate the volumetric kinetic and thermal terms (\equnp{vir_tau}) as 
\be
\label{eq:tau_sum}
    \mathcal{T}(r)=\sum_{<r}\frac{1}{2}m_{\rm gas}v^2+\sum_{<r}\frac{3}{2}k_{\rm B} T\equiv \mathcal{T}_{\rm dyn}(r)+\mathcal{T}_{\rm therm}(r),
\ee
where $v=\sqrt{v_x^2+v_y^2+v_z^2}$ is the total velocity. Note that we include $v_z^2$ in $\mathcal{T}$ despite our assumption of filaments as infinite cylinders when computing $\mathcal{W}$. This is because even when evaluating the energy per-unit-length, such that there is no net velocity or acceleration along the $z$ direction, we still expect the velocity dispersion to be three-dimensional locally, such that $v_z^2$ remains an important contribution to the kinetic energy density of the system. We comment in \se{ 5.3} on the effect of removing $v_z^2$ from the kinetic term. 

\smallskip
The surface term, \equ{vir_taus}, comprises three surfaces for each radius $r$, one in the cylindrical shell, and two others at either end of the cylindrical slice. However, the latter two surfaces are irrelevant for evaluating the virial equilibrium per-unit-length. Therefore, we do not consider them here. We comment in \se{ 5.3} on the effect of including them in the calculation. 
The surface term is thus given by 
\be 
\label{eq:taus_r}
\mathcal{T}_{s}(r) = \int_{-L/2}^{L/2} \int_{0}^{2\pi} \left(\rho r v_r^2 + \rho z v_rv_z + Pr \right) r \text{d}\phi \text{d}z. 
\ee 

\smallskip
The second term, namely the integral of $\rho z v_rv_z$, is not compatible with our assumption of an infinite cylinder, nor with the evaluation of $\mathcal{T}_{s}(r)$ per-unit-length. For an infinite cylinder in equilibrium per-unit-length, $v_{\rm z}$ would be constant (or zero). Likewise, \equs{grad_phi}-\equm{W(r)} are only valid for $\rho$ which does not depend on $z$, and in such cases $v_{\rm r}$ should also be independent of $z$. This term thus reduces to the integral of $z{\rm d}z$ over a symmetric interval and is, therefore, $0$. In light of this, we neglect this term moving forward. In practice, it turns out that this term is negligible anyway, at most a few percent of the integral of $\rho r v_r^2$ at all radii $r\gsim 0.15\,r_{\rm shock}$, because $\rho v_r v_z$ is indeed roughly independent of $z$, consistent with our assumed symmetry. 

\smallskip
In practice, when numerically evaluating $\mathcal{T}_{s}$ we perform the following sum over gas cells, 
\be
\label{eq:taus_sum}
    \mathcal{T}_s=\frac{\sum_r\rho r v_r^2V}{\Delta r} + \frac{\sum_r PrV}{\Delta r}\equiv \mathcal{T}_{{\rm dyn},s} + \mathcal{T}_{{\rm therm},s},
\ee
where $V$ is the volume of each gas cell and $\Delta r$ is the width of the bin. Therefore, we have approximated ${\rm d}S \simeq V/\Delta r$. We have verified that this approximation is roughly independent of our choice of $\Delta r$, as long as this is within a factor of $\sim (2-3)$ of the typical cell size in the simulation.

\subsection{Rapidly Flowing Gas on and off the Sheet}
\label{sec:inflows}

\begin{figure*}
    \centering
    \includegraphics[trim={0.0cm 0.0cm 0.0cm 0.0cm}, clip, width =0.98 \textwidth]{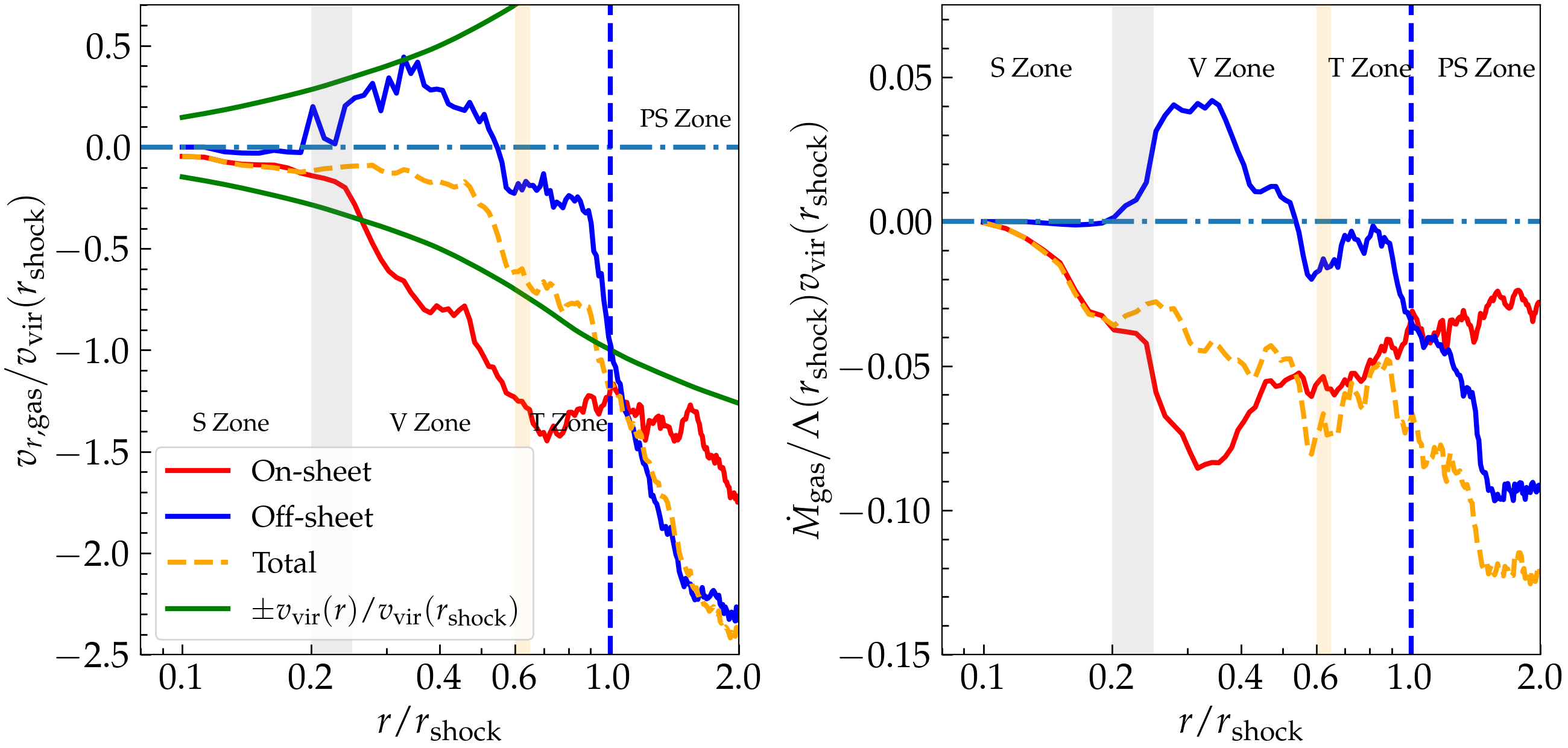}
    \caption{Profiles of gas radial velocities (left) and gas mass flow rate (right). Negative values refer to inflow towards the filament axis while positive values refer to outflowing gas. We distinguish between gas outside the sheet (blue lines) and within the sheet (red lines), where the sheet region is crudely defined as $|x|<10\kpc\sim 0.22\,r_{\rm shock}$ (see \fig{thermal profiles}). The dashed-orange line in each panel shows the results for all gases regardless of their position relative to the sheet. The value of $v_r$ in each bin is the mass-weighted average radial velocity, and the value of $\dot{M}$ in each bin is $\sum mv_r/\Delta r$, where $\Delta r$ is the width of the bin. 
    Prior to stacking, the $v_r$ profiles for each slice are normalised by the corresponding virial velocity at $r_{\rm shock}$, $v_{\rm vir}(r_{\rm shock})=(G\Lambda_{\rm shock})^{1/2}$ (\equ{v_vir}), while the profiles of $\dot{M}$ were normalised by $\Lambda_(r_{\rm shock})v_{\rm vir}(r_{\rm shock})$. 
    In the left-hand panel, the green lines show the profiles of $\pm v_{\rm vir}(r)/v_{\rm vir}(r_{\rm shock})$. 
    The sheet is inflowing at all radii, with $|v_r|>v_{\rm vir}(r)$ outside the \textbf{S} zone and with $|v_r|<v_{\rm vir}(r)$ inside the \textbf{S} zone. On the other hand, the off-sheet gas is inflowing in the \textbf{T} zone and outflowing in the \textbf{V} zone, but always with $|v_r|<v_{\rm vir}(r)$.
    }
    \label{fig:radial inflows}
\end{figure*}

Finally, before we calculate the corresponding profiles, we wish to remove gas that is rapidly accreting onto or outflowing from the filaments. This gas is not expected to be in equilibrium within the filament, but rather to be unbound or on its first infall, and therefore will bias our results \citep[see][for a similar discussion in the context of DM haloes]{Lochhaas.etal.21}. We do this by applying a cut in the radial velocity of gas cells included in the energy terms described above. At each radius $r$, we calculate the corresponding ``virial'' velocity (from \equnp{virial_ideal}) 
\be
\label{eq:v_vir}
    v_{\rm vir}(r)=\sqrt{G\Lambda_{\rm tot}(r)},
\ee
and only consider gas cells where $|v_r|<v_{\rm vir}(r)$. We apply this threshold to all terms appearing above, except for $\Lambda_{\rm tot}$ which sets the gravitational potential. In practice, this removes almost exclusively gas which is rapidly inflowing along the sheet onto the filament. To illustrate this, we present in the left-hand panel of \fig{radial inflows} radial profiles of the mass-weighted average radial velocity, stacked among our ten filament slices as in previous figures. In the right-hand panel, we show the radial mass flux, $\dot{M}$, stacked among the ten slices. In both panels, we show separately the profiles for material within and outside the sheet. We crudely define the sheet region for all the slices as anything within $|x|<10\kpc\sim 0.22\,r_{\rm shock}$ in the frame of the projection maps in \figs{thermal profiles} and \figss{DM prop}. In the left-hand panel, we also show the profile of $\pm v_{\rm vir}$. It is evident that the radial velocities towards the filament are quite different within and outside the sheet, and that our velocity threshold of $|v_r|<v_{\rm vir}(r)$ for the virial parameter predominantly removes rapidly inflowing gas (with $v_r<0$) along the sheet in the \textbf{T} and \textbf{V} zones. The mean inflow velocity along the sheet exceeds $v_{\rm vir}$ at $r\gsim 0.3\,r_{\rm shock}$, while outside the sheet it only exceeds $v_{\rm vir}$ at $r>1.2r_{\rm shock}$. 

\smallskip
Besides their use for our calculation of the virial parameter, it is worth discussing the structure of the inflow velocity and mass accretion profiles in detail. In the \textbf{PS} zone at $r\lsim 2r_{\rm shock}$, and throughout the \textbf{T} zone, the off-sheet gas is decelerating as a result of the strong thermal pressure gradients generated by the shock (see \se{sim_dynamics}). At $r>2r_{\rm shock}$, the off-sheet gas is either accelerating or maintaining constant velocity. However, note that the gas does not completely stall at $r_{\rm shock}$, as might have been expected for a strong accretion shock, because the cooling time is only slightly longer than the free-fall time at $r\lsim r_{\rm shock}$ (\fig{tcool_tff})\footnote{Note that while there is some slice-to-slice variation in the width of the radial shell around $r_{\rm shock}$ where the off-sheet gas decelerates, this is never less than $\gsim 0.3r_{\rm shock}$, and is $\sim 1r_{\rm shock}$ on average, as seen in \fig{radial inflows}.}. This generates a cooling flow and allows the gas to continue flowing towards the centre, albeit at a reduced velocity. This is supported by the fact the mass-flux of off-sheet material remains constant throughout the \textbf{T} zone, suggesting that the material does not ``pile-up'' behind the shock. On the other hand, the on-sheet gas maintains a roughly constant inflow velocity throughout the \textbf{T} zone, as it interacts only weakly with the shock. This is similar to how the inflow velocity of cold streams feeding massive galaxies at high-$z$ through their shock-heated haloes are seen in simulations to maintain roughly constant inflow velocities throughout the hot CGM \citep{dekel2009cold,Goerdt.etal.15}. 

\smallskip
Within the \textbf{V} zone, at $0.2\lsim r/r_{\rm shock}\lsim 0.6$, the off-sheet gas is outflowing with an average outflow velocity smaller than $v_{\rm vir}$. This seems to be caused by a strong quadrupolar vortex structure that develops in this region, as discussed in \se{toy_model}. These vortices cause the gas outside the sheet to swirl around and can lead to outwards radial motions. In this same region, the gas within the sheet decelerates due to a combination of shear against the shock-heated gas and interaction with the vortices, both of which drain momentum from the inflowing gas \citep{mandelker.etal.2016,padnos.etal.2018,mandelker.etal.2019}, and also due to a non-zero impact parameter of the sheet with respect to the filament, which decreases the radial component of the velocity closer to the filament centre. In general, mass accretion into the \textbf{S} zone is dominated by flow outside the sheet in the \textbf{PS} zone, and by flow along the sheet in the \textbf{T} and \textbf{V} zones.

\smallskip
Within the \textbf{S} zone, the off-sheet component is not well-defined due to our crude definition of the sheet as everything within $|x|<10\kpc\sim 0.22\,r_{\rm shock}$. The sheet material continues inflowing towards the filament axis, slowly decelerating from $\sim 0.1 \, v_{\rm vir}(r_{\rm shock})$ in the outer \textbf{S} zone towards smaller radii.

\begin{figure}
    \centering
    \includegraphics[trim={0.0cm 0.0cm 0.0cm 0.0cm}, clip, width =0.48 \textwidth]{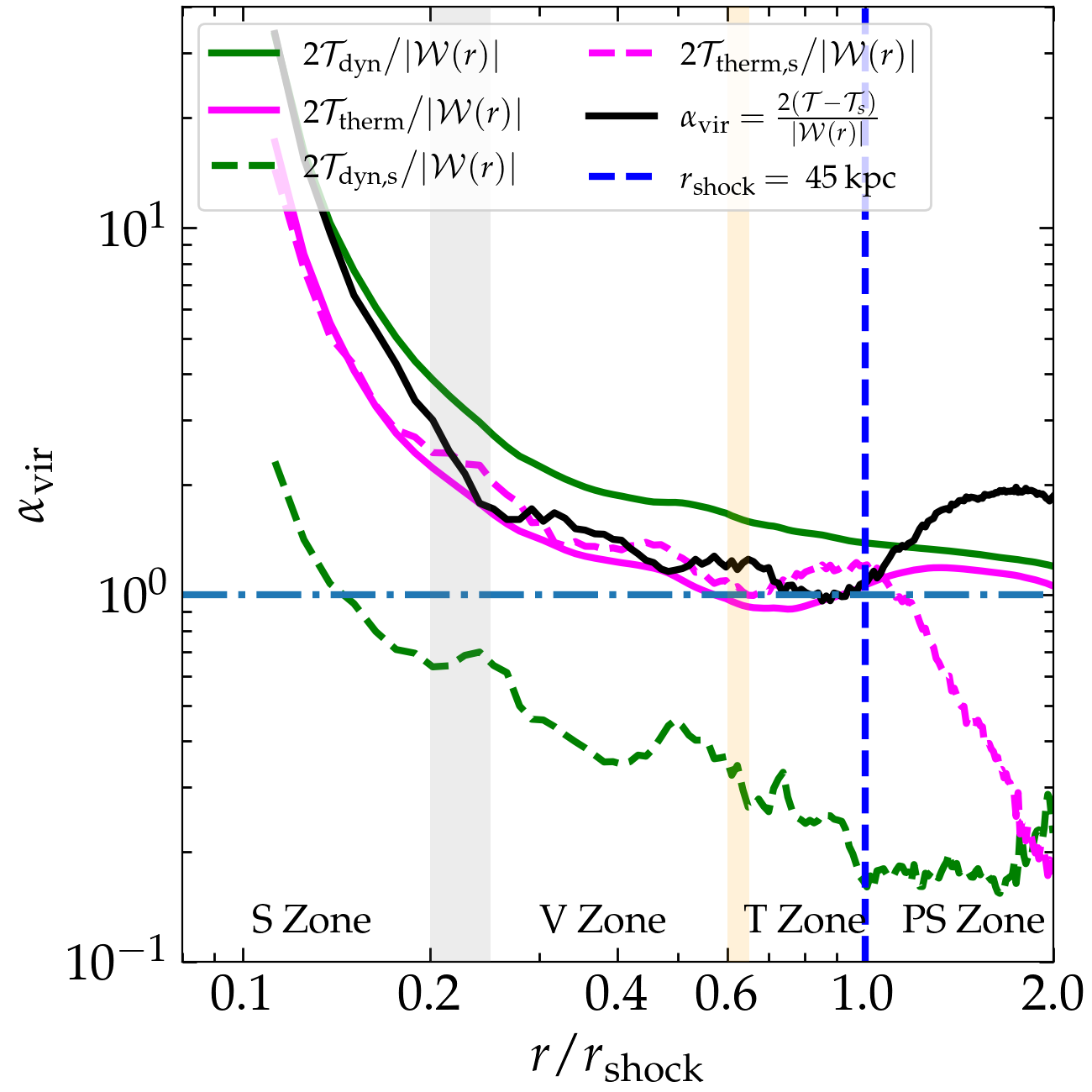}
    \caption{Radial profile of the virial parameter, $\alpha_{\rm vir}$ (black line), along with the ratios of the various energy terms in the numerator of \equ{vir para} to the gravitational energy, $|\mathcal{W}(r)|$. Specifically, these are the volumetric kinetic and thermal terms, $\mathcal{T}_{\rm dyn}$ (green solid line) and $\mathcal{T}_{\rm therm}$ (solid magenta line), and the surface kinetic and thermal terms, $\mathcal{T}_{{\rm dyn},s}$ (green dashed line) and $\mathcal{T}_{{\rm therm},s}$ (solid magenta line). We see that $\alpha_{\rm vir}$, $\mathcal{T}_{\rm therm}/|\mathcal{W}|$, and $\mathcal{T}_{{\rm therm},s}/|\mathcal{W}|$ are all of order unity in the \textbf{T} and outer \textbf{V} zones, where the gas temperature is hot and comparable to the effective temperature of the dark matter. Therefore, the filaments are in approximate virial equilibrium per unit-length with $r_{\rm vir}\sim r_{\rm shock}$.
    }
    \label{fig:virial parameter}
\end{figure}

\subsection{Virial Equilibrium and the Virial Radius}\label{sec: 5.3}

Using \equs{vir para}, \equm{W(r)}-\equm{tau_sum}, and \equm{taus_sum}, we evaluate $\alpha_{\rm vir}$ for each slice as a function of radius for gas with $|v_r|<v_{\rm vir}(r)$. We discuss the sensitivity of our results to this threshold below. We then stack all the ten slices by weighting $\alpha_{\rm vir}(r)$ by $|\mathcal{W}(r)|$. This is equivalent to treating all ten slices as one long cylinder when computing the virial parameter per-unit-length, in accordance with our modelling of the filaments as infinite cylinders. 

\smallskip
We present the stacked profile of $\alpha_{\rm vir}(r)$ in \fig{virial parameter} in black, along with the ratios of the four energies in the numerator of $\alpha_{\rm vir}$ (volumetric and surface kinetic and thermal energies) to the gravitational energy. Due to the inclusion of the surface terms, the virial parameter should be well-defined at each radius rather than just at the outer boundary of the system. Examining the profile of $\alpha_{\rm vir}(r)$ is useful to locate the radius where $\alpha_{\rm vir}\sim 1$, which may be considered the ``virial radius'' of the system. The profile monotonically decreases from $\alpha_{\rm vir}\sim 10$ in the \textbf{S} zone 
to $\alpha_{\rm vir}\sim 1$ in the \textbf{T} and outer \textbf{V} zones, $0.5\lsim r/r_{\rm shock}\lsim 1$. It then increases towards larger radii as the surface thermal term sharply declines outside the shock. We may thus associate a virial radius of $r_{\rm vir}\sim r_{\rm shock}$ with our filament sample. 

\smallskip
We further note that the volumetric and surface thermal terms are both $\sim 1$ in the \textbf{T} zone, where the gas temperature is roughly equal to the effective temperature of dark matter defined by its velocity dispersion (\se{DM_prop}). This implies that dark matter is also expected to be in approximate virial equilibrium (per-unit-length) within $r_{\rm shock}$. This is consistent with previous estimates of the virial radii of filaments feeding massive haloes at high-$z$ \citep[][equation 14]{mandelker2018cold}. For a filament feeding a $10^{12}\,\msun$ halo at $z\sim 4$, these authors predict $r_{\rm vir}\sim 55\kpc$, slightly larger than our average $r_{\rm shock}\sim 45\kpc$ and within the range of $r_{\rm shock}$ values among our ten slices (\fig{rho_0 and T_0 scatterings}). Note that this is different from the case of galaxy clusters, where the shock radius is found to be larger than the virial radius \citep{lau2015gasprofile,zinger2018quenching,aung2021shock}.

\smallskip
Removing $v_z^2$ from $\mathcal{T}$ and $\mathcal{T}_s$ reduces $\alpha_{\rm vir}$ by $\sim 30\%$, such that $\alpha_{\rm vir}\sim 0.6$ near $r_{\rm shock}$. This is consistent with the approximate configuration between the three dimensions of kinetic energy. Our inclusion of $v_z^2$ in the calculation is thus justified because if small-scale motion along this dimension were restricted, the same kinetic energy would likely be split between the remaining two dimensions. On the other hand, including the axial surface terms in $\mathcal{T}_s$ results in $\alpha_{\rm vir}<0$ at all radii. The exact numerical value of $\alpha_{\rm vir}$, in this case, is of little consequence, but suffice to say that the system is clearly not virialized or even bound along its axis, as expected. Finally, regarding our velocity cut, excluding the gas with $|v_r|>2v_{\rm vir}$ does not affect our results, showing that the exclusion is dominated by a very rapidly flowing gas. Changing our threshold in $v_r$ to only exclude inflowing gas with $v_r<-v_{\rm vir}$ while including rapidly outflowing gas retains $\alpha_{\rm vir}\sim 1$ in the 
\textbf{T} and outer \textbf{V} zones, 
showing that our estimation of a virial radius at $r_{\rm vir}\sim r_{\rm shock}$ is robust. At smaller radii, where the system is still far from equilibrium based on \fig{virial parameter}, $\alpha_{\rm vir}$ becomes negative due to a much larger surface kinetic term. Including a rapidly flowing gas also results in negative $\alpha_{\rm vir}$ everywhere, consistent with our interpretation that the gas currently inflowing towards the filament along the sheet is not yet in equilibrium.

\smallskip
In summary, the gas within $r_{\rm shock}$, and by extension the dark matter as well, is in virial equilibrium within the potential well set by the dark matter filament, allowing us to define a virial radius $r_{\rm vir}\sim r_{\rm shock}$. Within the \textbf{T} zone, the gas is approximately at the virial temperature, and additional kinetic motions are largely offset by surface pressure. However, in the \textbf{V} and \textbf{S} zones, these kinetic motions grow stronger with respect to the gravitational potential, and the surface pressure terms cannot keep up. This drives the system out of equilibrium in these regions, consistent with the outflow velocities seen in \fig{radial inflows}. 


\section{Dynamical Equilibrium}
\label{sec:fil_dyn}

\smallskip
In this section, we study the dynamical stability of gaseous filaments. In particular, we seek to address what the dominant force is that supports the filament against gravitational collapse towards its axis. We begin in \se{summation} by describing the mathematical formalism of our force decomposition and present results from our simulation in \se{sim_dynamics}. In \se{toy_model}, we interpret the simulation results using an analytic toy model for the internal filament dynamics. In Appendix \se{sim_numerics}, we discuss the robustness of the results presented in \se{sim_dynamics} to our numerical method.

\subsection{Force Decomposition}
\label{sec:summation}

\subsubsection{Basic Framework}

\smallskip
As we show in Appendix \ref{sec:simplification} (\fig{plasma_beta}), magnetic fields are not dynamically important in the filaments, with typical plasma $\beta$ values of $\beta=P_{\rm thermal}/P_{\rm magnetic}>10^5$. Therefore, the equation of motion for gas in our simulation can be approximated by the Euler equation: 
\be 
\label{eq:euler}
\pdv{\vec{v}}{t}+\left(\vec{v}\cdot \vec{\nabla} \right) \vec{v} = -\frac{1}{\rho} \vec{\nabla} P - \vec{\nabla} \Phi.
\ee
\no We follow \citet{lau2013weighing} in using \equ{euler} to determine the relative importance of the different terms in maintaining dynamical equilibrium and supporting the system against gravitational collapse. Specifically, we use a modified version of the ``summation method'' detailed in that paper. While that work focused on gas in massive galaxy clusters assuming spherical symmetry, we generalise their method to study gas dynamics in intergalactic filaments assuming cylindrical symmetry. We describe the method below, both for general considerations and for formulas specific to cylindrical geometry. 

\subsubsection{Primary Decomposition}

Gauss's law relates the total mass in any arbitrary volume to the gravitational potential at the boundary of that volume, 
\be
\label{eq:gauss}
    M_{\rm grav}\equiv \frac{1}{4\pi G}\oint_{S}\vec{\nabla}\Phi\cdot {\rm d}\vec{S} = M_{\rm tot},
\ee
{\no}where $M_{\rm tot}$ is the total mass enclosed by an imaginary closed surface $S$ over which the surface integral is taken. Combining \equ{euler} and \equ{gauss}, we have the following. 
\begin{equation}
\label{eq:summation}
    M_{\rm grav}=-\frac{1}{4\pi G}\oint_{S}\left(\pdv{\vec{v}}{t}+\left(\vec{v}\cdot\vec{\nabla}\right)\vec{v}+\frac{1}{\rho}\vec{\nabla} P\right)\cdot{\rm d}\vec{S}.
\end{equation}
{\no}Each of the terms in the integral on the right-hand-side of \equ{summation} represents an acceleration. When combined, these terms must balance gravitational acceleration. From the left-hand side of \equ{summation}, we learn that these terms can be considered as ``mass terms'', contributing to the total mass interior of the surface $S$. Thus, we define 
\begin{equation}
\label{eq:Mtot}
    M_{\rm grav}=M_{\rm accel}+M_{\rm therm}+M_{\rm inertial},
\end{equation}
{\no}where 
\begin{equation}
\label{eq:acceleration term}
    M_{\rm accel} \equiv -\frac{1}{4\pi G}\oint \pdv{\vec{v}}{t}\cdot {\rm d}\vec{S},
\end{equation}
{\no}is the \textit{acceleration term} representing the temporal change of the velocity field normal to the bounding surface, \footnote{Note that in our Eulerian representation, this represents a temporal change of the velocity field inside the filament, rather than the local acceleration experienced by a test particle as it moves through the filament.} 
\begin{equation}
\label{eq:thermal term}
    M_{\rm therm}\equiv -\frac{1}{4\pi G}\oint \frac{1}{\rho}\vec{\nabla} P\cdot {\rm d}\vec{S},
\end{equation}
{\no}is the \textit{thermal term} representing the support of the filament against gravity by thermal pressure gradients, and 
\begin{equation}\label{eq:inertial term}
    M_{\rm inertial}\equiv -\frac{1}{4\pi G}\oint \left(\vec{v}\cdot\vec{\nabla} \right)\vec{v}\cdot{\rm d}\vec{S},
\end{equation}
{\no}is the \textit{inertial term}. As we demonstrate below and expand upon in \se{toy_model}, this term can be decomposed into several fictitious forces resulting from the fluid motion, each of which can either help support the filament against gravity or else work together with gravity to induce 
collapse. 

\smallskip
In order to estimate the overall dynamical equilibrium of the filament and the relative contribution of each term to the support of the filament against gravity, we analyse \equs{acceleration term}-\equm{inertial term} as a function of the radius of each filament slice and compare their sum (\equnp{Mtot}) to the true mass enclosed within the radius $r$ in the simulation. For each radius $r$, we assume a Gaussian surface that is a cylinder of radius $r$ and length $L=30\kpc$, the thickness of each filament slice. We comment on the impact of $L$ on our results in \se{sim_dynamics}. Unlike the discussion of virial equilibrium presented in \se{virial}, here we are interested in the dynamical state in three dimensions rather than per-unit-length. Furthermore, Gauss's theorem, in this case, requires a closed surface around the three-dimensional volume of interest. This results in three surface integrals for each radius, one on the cylindrical shell itself, hereafter $S_0$ with ${\rm d}\vec{S}_0={\rm d}S_0{\hat{r}}$, and two others on either end of the cylinder, hereafter $S_{\pm}$, with ${\rm d}\vec{S}_{\pm}={\rm d}S_{\pm}{\pm\hat{z}}$. Then, we have the following:
\begin{equation}
\label{eq:S1}
    \int_{S_0}\text{d}S=r\int_{0}^{2\pi}\text{d}\phi\int_{-L/2}^{L/2}\text{d}z, 
\end{equation} 
and
\begin{equation}
\label{eq:S2}
    \int_{S_{\pm}}\text{d}S=\left(\int_0^r r'\text{d}r'\int_{0}^{2\pi}\text{d}\phi\right)\bigg|_{z=\pm L/2}.
\end{equation}
Now let 
\be 
\label{eq:integral_decomp}
\int_{\mathcal R}\text{d}S\equiv \int_{S_0}\text{d}S, \quad \int_{\mathcal A}\text{d}S\equiv \int_{S_+}\text{d}S-\int_{S_-}\text{d}S,
\ee 
{\no}where ${\mathcal R}$ stands for radial and ${\mathcal A}$ stands for axial. Each of the integrals in \equs{acceleration term}-\equm{inertial term} is decomposed into two parts, $\oint_{\rm cylinder}\text{d}S=\int_{\mathcal R}\text{d}S+\int_{\mathcal A}\text{d}S.$

\smallskip
Using this notation, we can explicitly write out all the terms in \equs{acceleration term}-\equm{inertial term} as follows: 
\be
\label{eq:accel}
    M_\text{accel,r}(r)=-\frac{1}{4\pi G}\int_{\mathcal R}\pdv{v_r}{t}~{\rm d}S,
\ee 
\be
\label{eq:accel_2}
    M_\text{accel,a}(r)=-\frac{1}{4\pi G}\int_{\mathcal A}\pdv{v_z}{t}~{\rm d}S,
\ee 
\be
\label{eq:Mtherm_r}
    M_\text{therm,r}(r)=-\frac{1}{4\pi G}\int_{\mathcal R}\frac{1}{\rho}\pdv{P}{r}~{\rm d}S,
\ee 
\be
\label{eq:Mtherm_a}
    M_\text{therm,a}(r)=-\frac{1}{4\pi G}\int_{\mathcal A}\frac{1}{\rho}\pdv{P}{z}~{\rm d}S,
\ee 
\be 
\label{eq:radial inertial term}
    M_{\rm inertial, r}(r)=-\frac{1}{4\pi G}\int_{\mathcal R}\left(v_r\pdv{v_r}{r}+\frac{v_\phi}{r}\pdv{v_r}{\phi}+v_z\pdv{v_r}{z}-\frac{v_\phi^2}{r}\right)~{\rm d}S,
\ee 
\be 
\label{eq:axial inertial term}
    M_{\rm inertial, a}(r)=-\frac{1}{4\pi G}\int_{\mathcal A}\left(v_r\pdv{v_z}{r'}+\frac{v_\phi}{r'}\pdv{v_z}{\phi}+v_z\pdv{v_z}{z}\right)~{\rm d}S.
\ee 
{\no}Note that all the integrals in \equs{accel}-\equm{axial inertial term} are functions of radius, $r$, which we have highlighted on the left-hand-side of each of these equations. \Equs{accel}, \equm{Mtherm_r}, and \equm{radial inertial term} represent radial accelerations toward or away from the filament axis, while \equs{accel_2}, \equm{Mtherm_a} and \equm{axial inertial term} represent axial accelerations along the filament axis.

\subsubsection{Further decomposing the Inertial Terms}

The last term in the integrand of \equ{radial inertial term} represents the centrifugal acceleration and is hereafter referred to as the \textit{rotation term},
\be
\label{eq:rotation}
    M_{\rm rot}(r)=\frac{1}{4\pi G}\int_{\mathcal R}\frac{v_\phi^2}{r}~{\rm d}S.
\ee 
{\no}When evaluating \equ{rotation}, we further decompose the azimuthal velocity at radius $r$ into \textit{mean} and \textit{residual} components. The mean rotation is given by 
\be
\label{eq:mean_rot}
    v_{\phi,{\rm mean}}(r)=\frac{1}{2\pi r L}\int_{-L/2}^{L/2}\int_{0}^{2\pi} v_\phi r\text{d}\phi \text{d}z.
\ee 
{\no}Note that \equ{mean_rot} represents the volume-weighted average azimuthal velocity at radius $r$, rather than the mass- or density-weighted average. This is necessary for consistency with \equ{rotation}. The residual azimuthal velocity in each cell is then given by 
\begin{equation}
\label{eq:rand_rot}
    v_{\phi, \rm res}(r,\phi,z)=v_\phi(r,\phi,z)-v_{\phi,\rm mean}(r).
\end{equation}
With these definitions, it is straightforward to show that
\be 
\begin{array}{c}
\label{eq:decomposition of the rotation term}
M_{\rm rot}(r)=\dfrac{1}{4\pi G}\bigints_{\mathcal R}\left(\dfrac{v_{\phi,\rm mean}^2}{r} + \dfrac{v_{\phi,\rm res}^2}{r}\right){\rm d}S \\
\equiv M_{\rm rot, mean}(r)+M_{\rm rot, res}(r),
\end{array}
\ee
{\no}where we have further decomposed the rotation term into a mean and a residual component, resulting from the mean and residual rotation velocity, respectively.

\smallskip
The first three terms in the integrand of \equ{radial inertial term} can be interpreted in two ways. The first, which we will demonstrate in detail in \se{toy_model}, is the additional fictitious forces resulting from the fluid motion. The second contributes to ram pressure, turbulent pressure, and shear forces acting on the fluid. The ram (turbulent) pressure in the radial direction is the flux of radial momentum associated with mean (residual) radial motions, while radial shear forces represent flux of radial momentum caused by non-radial motions. 
These three terms in \equ{radial inertial term} represent the radial momentum advected into a volume element by radial, azimuthal, or axial motions, respectively. If one defines mean and residual radial/axial motions analogously to \equs{mean_rot}-\equm{rand_rot}, such that the mean is only a function of $r$, then it is straightforward to show that these terms can be written as follows: 
\be
\label{eq:Mram}
    M_{{\rm ram}, r}(r)=-\frac{1}{4\pi G}\int_{\mathcal R}v_{r,{\rm mean}}\pdv{v_{r,{\rm mean}}}{r}~{\rm d}S,
\ee 
\be
\label{eq:Mturb_r}
    M_{{\rm turb}, rr}(r)=-\frac{1}{4\pi G}\int_{\mathcal R}v_{r,{\rm res}}\pdv{v_{r,{\rm res}}}{r}~{\rm d}S,
\ee
\be
\label{eq:Mturb_p}
    M_{{\rm shear}, r\phi}(r)=-\frac{1}{4\pi G}\int_{\mathcal R} \left(\frac{v_{\phi, {\rm mean}}}{r}+\frac{v_{\phi, {\rm res}}}{r}\right)\pdv{v_{r,{\rm res}}}{\phi}~{\rm d}S ,
\ee
\be
\label{eq:Mturb_z}
    M_{{\rm shear}, rz}(r)=-\frac{1}{4\pi G}\int_{\mathcal R} \left(v_{z, {\rm mean}}+v_{z, {\rm res}}\right)\pdv{v_{r,{\rm res}}}{z}~{\rm d}S.
\ee
{\no}We expect the first terms in \equs{Mturb_p} and \equm{Mturb_z} are negligible compared to the second, because $v_{\phi, {\rm mean}}(r)$ and $v_{z, {\rm mean}}(r)$ can be taken outside the integral and the average of the derivatives of $v_{r,{\rm res}}$ should be small. Analogous divisions\footnote{We note that this distinction between ram and turbulent pressure is not unique. If one defines the mean motion with respect to some local smoothing kernel around each fluid element (e.g., as was done in the study of CGM dynamics by \citealp{Lochhaas.etal.22}), then the mean motion is no longer only a function of $r$ and all three terms contribute to both the ram and turbulent pressure.} into the ram, turbulent and shear forces in the axial direction can be applied to the three terms in \equ{axial inertial term}. 

\smallskip
Finally, we also divide the gravitational term (\equnp{Mtot}) into its radial and axial components:
\be
\label{eq:Mgrav_r}
    M_{\rm grav, r}(r) = \frac{1}{4\pi G}\int_{\mathcal{R}} \frac{\partial\Phi}{\partial r} {\rm d}S, 
\ee
and 
\be
\label{eq:Mgrav_a}
    M_{\rm grav, a}(r) = \frac{1}{4\pi G}\int_{\mathcal{A}} \frac{\partial\Phi}{\partial z} {\rm d}S.
\ee
These represent the radial and axial gravitational fields in the filament, respectively. Note that since the gravitational field is $\vec{g}=-\vec{\nabla}\Phi$, $M_{\rm grav, r}$ and $M_{\rm grav, a}$ are defined to be positive when the radial and axial gravitational fields are directed inwardstowards the filament centre. However, the thermal and inertial terms in \equs{Mtherm_r}-\equm{axial inertial term} and \equs{decomposition of the rotation term}-\equm{Mturb_z} are defined as positive when the respective forces are directed \textit{outwards}, away from the filament centre. 
The equation of force equilibrium within the filament in the radial directions is 
\be
\label{eq:summation_radial}
    M_{\rm grav, r} = M_{\rm accel, r} + M_{\rm inertial, r} + M_{\rm therm, r}, 
\ee
with an analogous equation for the axial direction.

\subsubsection{Numerical Calculation}
\label{sec:dyn_num}

Using our simulations, we numerically evaluate \equs{accel}-\equm{Mgrav_a} as a function of radius, $r$, for each filament slice. Full details of how this is done, including convergence and robustness tests, are provided in the Appendix \se{sim_numerics}. Here, we briefly review the key points of our fiducial method. For each slice, we begin by depositing all fluid properties, including the gravitational potential, onto a uniform cylindrical grid extending from $(0.05-1.1)r_{\rm shock}$ with a typical cell-size of $\Delta\sim 300\pc$, comparable to the size of the smallest gas cells in the simulated filaments. We assign to each grid-cell the fluid properties of its nearest-neighbour gas cell in the simulation. Despite not being inherently conservative, this method conserves mass, momentum, and energy to better than $\lsim 2\%$ both globally and locally. All numerical derivatives and integrals are evaluated on these grids. Before computing any partial derivatives, we smooth the relevant quantity with a one-dimensional Gaussian in the direction along which we are evaluating the derivative, with a width of $\sigma=0.5\kpc$. Evaluating both sides of \equ{gauss}, we find that our numerical estimate of $M_{\rm grav}$ 
deviates from $M_{\rm tot}$ by at most $\lsim 5\%$ at each radius.  

\smallskip
After computing the radial profiles of each mass term for each filament slice, we normalise these by $M_{\rm tot}(<r)$ and stack the ten filament slices by taking the average of each normalised mass term in each radial bin. While it is the ratio of each mass term to $M_{\rm grav}(r)$ that tells us the contribution of the respective force to the filament support at that radius, we normalise the mass terms by $M_{\rm tot}(<r)$, which is a smooth, monotonically increasing function within $\lsim 5\%$ of $M_{\rm grav}(r)$, which exhibits small fluctuations. Hereafter we use the notation $M_{...}/M_{\rm tot}$, $M_{...}/M_{\rm grav}$, and $F_{...}/F_{\rm grav}$ interchangeably and think of these as ratios of forces/accelerations rather than ratios of masses.

\smallskip
Finally, we note that we did not directly compute the two acceleration terms in \equs{accel}-\equm{accel_2}, representing a temporal change in the velocity field in the filament. Direct evaluation of these terms requires the use of multiple snapshots. As detailed in Appendix \se{sim_numerics}, this is difficult due both to the time between adjacent snapshots being comparable to both the cooling time and the eddy crossing time of filament gas, and to complications in identifying the same volume elements in each adjacent timestep while the entire filament moves through the sheet. Therefore, we use \equ{summation_radial} and the corresponding equation for the axial terms to infer the acceleration terms from the difference between the gravitational, thermal and inertial terms.

\subsection{Results}
\label{sec:sim_dynamics}

In what follows, we present the results of our analysis of the dynamical state of filaments. We begin in \se{dynamical stabilities} by analysing where the velocity field in the filaments is in steady state, with $M_{\rm accel}\propto \partial \vec{v}/\partial t\simeq 0$, and where it is changing over time. We then analyse the various components of the thermal and inertial terms in \se{detailed force decomposition}, to better characterise the force balance and dynamical state in the filaments. Throughout, we emphasise key trends and physical processes in the three radial zones. 

\subsubsection{Overall Force Balance}
\label{sec:dynamical stabilities}

\begin{figure}
    \centering
    \includegraphics[trim={0.0cm 0.0cm 0.0cm 0.0cm}, clip, width =0.48 \textwidth]{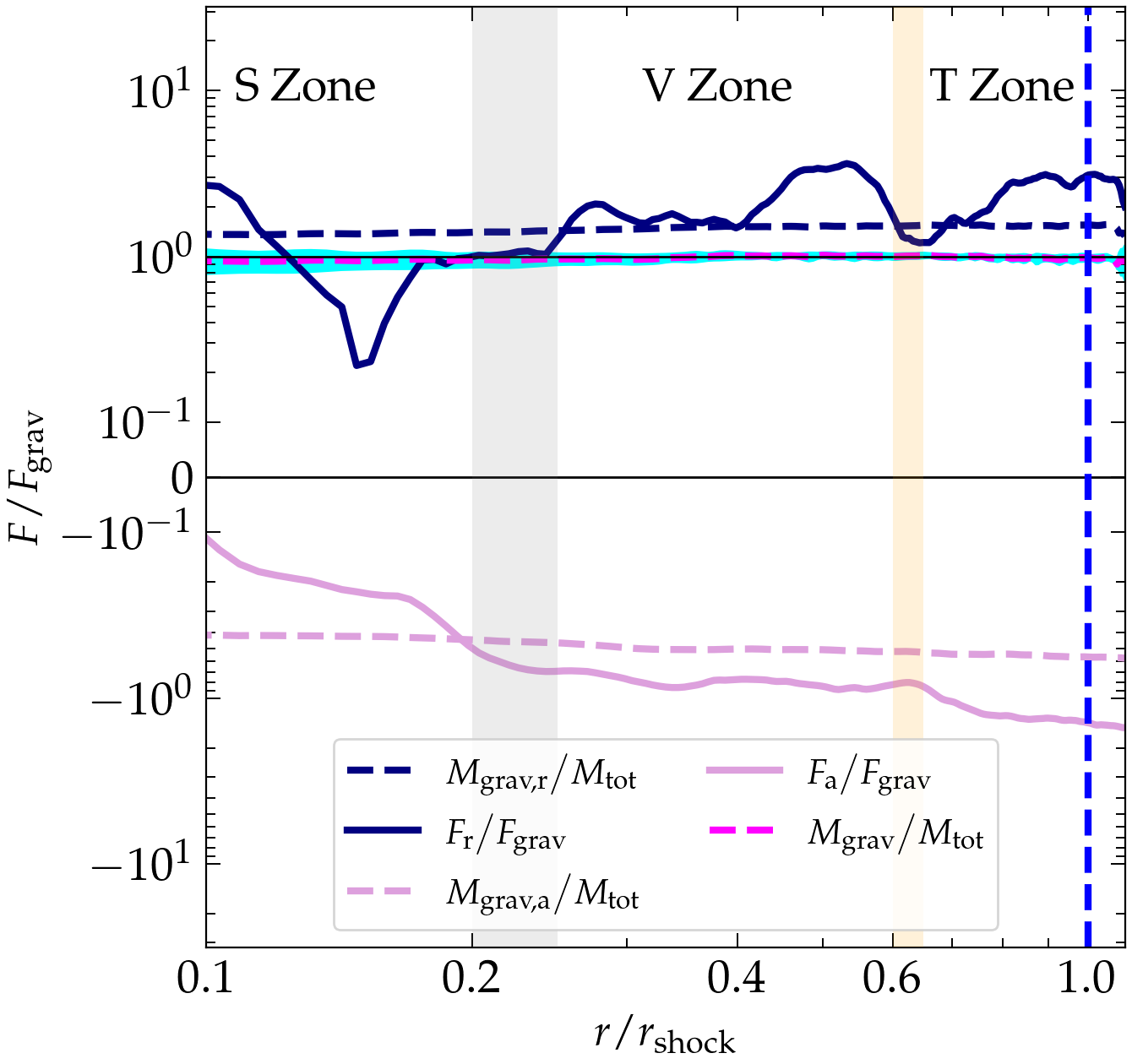}
    \caption{Force equilibrium in the filaments. We compare the sum of thermal and inertial forces (solid lines) with the gravitational force (dashed lines), in both the radial and axial directions, as a function of $r/r_{\rm shock}$. The magenta dashed line and cyan shaded region show $M_{\rm grav}(r)/M_{\rm tot}(<r)$, 
    which is $\simeq 1$ everywhere with a $\lsim 5\%$ error at $r/r_{\rm shock}\lsim 0.3$ and $\gsim 0.8$ due to our numerical method. The navy and pink dashed lines show the ratios $M_{\rm grav,r}(r)/M_{\rm tot}(<r)$ and $M_{\rm grav,a}(r)/M_{\rm tot}(<r)$, respectively. The former is positive, indicating a radial gravitational field directed inwards, while the latter is negative, indicating an axial gravitational field directed outwards since the filament is unbound along its axis. Solid navy and pink lines show $(M_{\rm therm,r}+M_{\rm inertial,r})/M_{\rm tot}(<r)$ and $(M_{\rm therm,a}+M_{\rm inertial,a})/M_{\rm tot}(<r)$, respectively (\equsnp{Mtherm_r}-\equmnp{axial inertial term}). Radii, where the navy or the pink solid lines are equal to the corresponding dashed lines, indicate regions in a dynamical steady state without temporal acceleration in the corresponding direction (\equs{accel} and/or \equm{accel_2} are equal to 0). If, however, the solid lines are greater (smaller) than the corresponding dashed lines, this indicates a negative (positive) temporal acceleration at this radius. Over a fairly large radial range in the \textbf{V} and \textbf{T} zones, $r/r_{\rm shock}\sim (0.25-0.40)$ and $(0.60-0.80)$, there is a dynamical steady-state in the radial direction. $\partial v_r/\partial t > 0$ at $r/r_{\rm shock}\sim (0.40-0.60)$ and $(0.80-1.0)$, while $\partial v_r/\partial t < 0$ in the \textbf{S} zone at $r/r_{\rm shock}\lsim 0.25$. In the axial direction, there is no dynamical steady state at $r/r_{\rm shock}\gsim 0.2$.}
    \label{fig:summation method}
\end{figure}

In \fig{summation method}, we assess to what extent thermal and inertial forces balance gravity within the filament, as a function of $r/r_{\rm shock}$. The dashed dark-blue (hereafter navy) and pink lines show the stacked profiles of $M_{\rm grav, r}/M_{\rm tot}$ and $M_{\rm grav, a}/M_{\rm tot}$, 
respectively, representing the normalized radial and axial gravitational fields. Notice that the radial term is positive, while the axial term is negative. As discussed following \equs{Mgrav_r}-\equm{Mgrav_a}, this implies that the radial gravitational field is directed inwards towards the filament axis, as expected, while the axial gravitational field is directed outwards. The latter is a manifestation of the fact that filaments are unbound along their axis (see \se{virial}), and are being stretched apart by large-scale tidal forces. Note that the influence of the background Hubble expansion on this result is negligible. 
For reference, the dashed magenta line and the cyan-shaded region show the mean and 1-$\sigma$ standard deviation among our ten slices of the radial profile of $M_{\rm grav}/M_{\rm tot} = (M_{\rm grav, r}+M_{\rm grav, a})/M_{\rm tot}$. This ratio is $\simeq 1$ everywhere, as Gauss's Theorem tells us, and must be (\equnp{gauss}). Although our numerical method introduces an error $\sim 5\%$ in $r/r_{\rm shock}\lsim 0.3$ and $\gsim 0.8$ (see also \fig{Grad Phi together}), this is negligible compared to the variations in $M_{\rm grav, r}$ and $M_{\rm grav, a}$. 

\smallskip
The solid navy and pink lines show the stacked profiles of $(M_{\rm therm, r}+M_{\rm inertial,r})/M_{\rm tot}$ and $(M_{\rm therm, a}+M_{\rm inertial,a})/M_{\rm tot}$, respectively (\equsnp{Mtherm_r}-\equmnp{axial inertial term}). In a steady state where thermal and inertial forces balance gravity and the filament velocity field is constant in time, these must equal $M_{\rm grav, r}/M_{\rm tot}$ (\equnp{summation_radial}) and $M_{\rm grav, a}/M_{\rm tot}$, respectively. 
If they are larger (smaller) than the corresponding terms $M_{\rm grav,r}$ and $M_{\rm grav,a}$ at some radius $r$, then the velocity field at that radius accelerates outwards (inwards) in the corresponding direction.

\smallskip
In the radial direction, the filament exhibits a dynamical steady-state with no temporal acceleration throughout the inner halves of both the \textbf{V} zone and the \textbf{T} zone, $0.25\lsim r/r_{\rm shock}\lsim 0.4$ and $0.6\lsim r/r_{\rm shock}\lsim 0.8$. However, each zone contains regions with net temporal acceleration where the velocity field is not in a steady state.

\smallskip
In the \textbf{T} zone, at $r\gsim 0.8r_{\rm shock}$, there is an outwards radial acceleration which appears to be due to the outwards expansion of the shock. While a full analysis of filament properties as a function of time is beyond the scope of this paper, we repeated the analysis presented in \se{thermal} to identify the shock radius in our ten filament slices at $z\sim 4.05$ and $z\sim 3.82$, roughly $55\Myr$ before and after our fiducial snapshot at $z\sim 3.93$. The typical $r_{\rm shock}$ increases from $\sim 36\kpc$ to $\sim 51\kpc$ during this interval, 
corresponding to a shock velocity of $v_{\rm shock}\sim 130\kms$, while the sound speed in the post-shock gas is $c_{\rm s}\sim 85\kms$. The outer \textbf{T} zone near $r_{\rm shock}$ has not reached equilibrium after being hit by the shock, which explains the positive acceleration in this region.

\smallskip
In the \textbf{V} zone, there is an outwards radial acceleration at $0.4\lsim r/r_{\rm shock}\lsim 0.6$. This is the region where the radial velocity of the sheet material begins to decrease and the off-sheet material begins to move outwards (\fig{radial inflows}). As we show in \se{toy_model} below, this is also the location of peak vorticity in the filament, where a quadrupolar vortex structure induced by the non-radial flow of the sheet towards the filament centre, dominates the dynamics. 

\begin{figure*}
    \centering
    \includegraphics[trim={0.0cm 0.0cm 0.0cm 0.0cm}, clip, width =0.98 \textwidth]{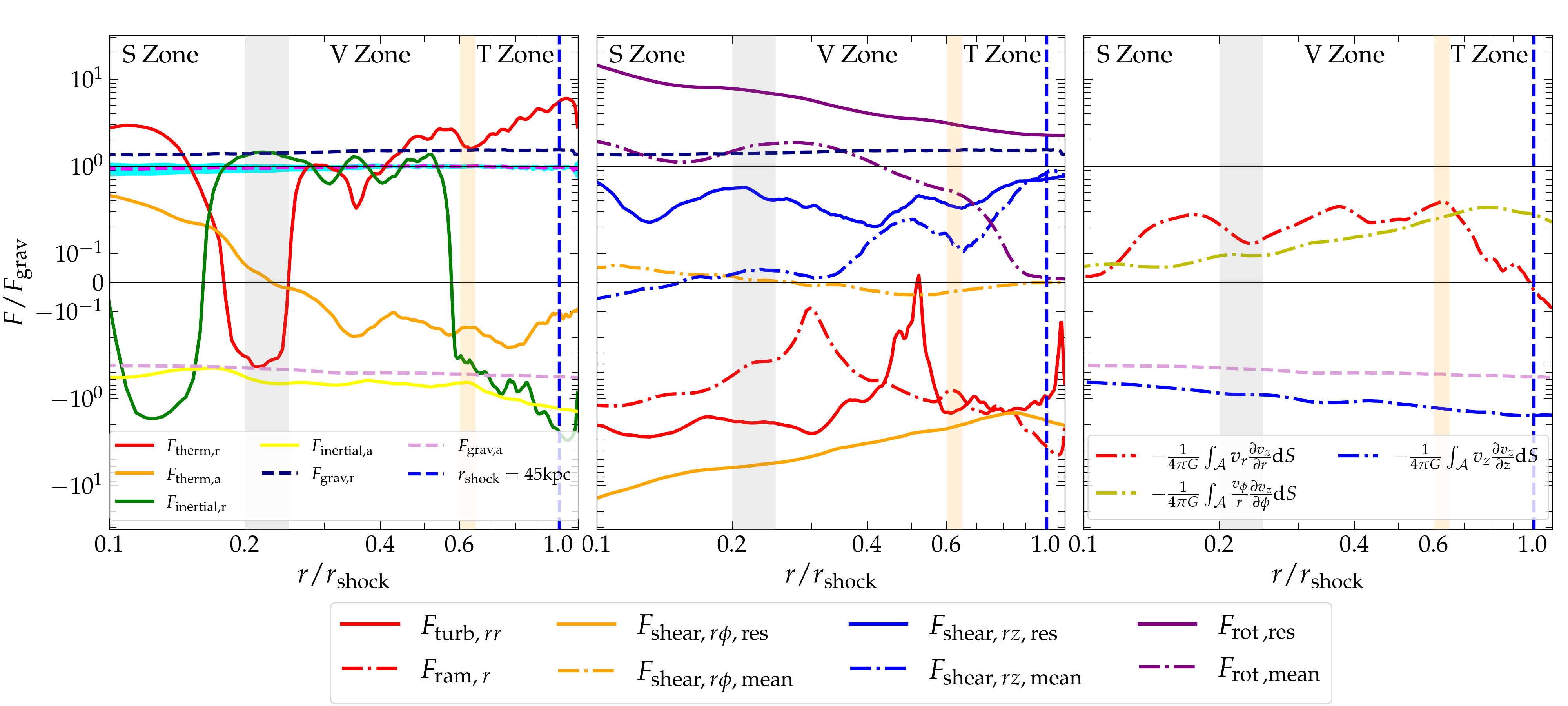}
    \caption{Detailed force balance in filaments. \textit{\textbf{Left panel:}} We compare the radial thermal and inertial forces (solid red and green lines) with the radial gravitational force (navy dashed line, as in \fig{summation method}), and the axial thermal and inertial forces (solid orange and yellow lines) with the axial gravitational force (pink dashed line). \textit{\textbf{Centre panel:}} We show the four constituent forces comprising the radial inertial force (\equnp{radial inertial term}), namely the centrifugal force (purple lines), the ram/turbulent pressure forces (red lines) and the shear forces (orange and blue lines for azimuthal and axial shear, respectively). For each force, we further distinguish between their mean (dot-dashed lines) and residual (solid lines) components. 
    The legend for this panel is in the box located below all of the panels. 
    \textit{\textbf{Right panel:}} We show the three constituent forces that comprise the axial inertial force (\equnp{axial inertial term}). These are the ram/turbulent pressure forces (blue lines) and the two shear forces (red and yellow lines for radial and azimuthal shear, respectively). 
    \textbf{\uline{Based on all three panels, we conclude:}} In the \textbf{T} Zone, the radial thermal force is directed outwards and is stronger than the gravitational force inwards, while the inertial forces are directed inwards and are dominated by the ram pressure, which mainly offsets the excess thermal pressure. In the \textbf{V} Zone, the radial thermal force is directed outwards and decreases from $\sim 1.5|F_{\rm grav,r}|$ to $\sim 0.6|F_{\rm grav,r}|$ as $r$ decreases. Radial inertial forces are directed outwards with a roughly constant amplitude of $\sim 0.6|F_{\rm grav,r}|$, and are dominated by centrifugal forces. The mean rotation has a roughly constant amplitude of $\sim |F_{\rm grav,r}|$. The residual rotation is much larger, but is largely offset by ram/turbulent pressure and azimuthal shear. In the \textbf{S} Zone, the radial thermal and inertial forces become negative at $r\gsim 0.15\, r_{\rm shock}$ and $r\lsim 0.15\,r_{\rm shock}$, respectively, resulting in a net inwards acceleration in this region. 
    The axial gravitational force is directed outwards, representing the tidal stretching of the filament along its axis. It is roughly balanced by axial ram pressure forces at all radii, while the axial thermal pressure force is relatively small, directed outwards in the \textbf{S} zone and inwards the \textbf{T} and \textbf{V} zones. 
    }
    \label{fig:summation method another}
\end{figure*}

\smallskip
In the \textbf{S} zone, at $r\lsim 0.25r_{\rm shock}$, there is an inwards acceleration. This corresponds to the outer boundary of the isothermal core (\fig{thermal profiles}), where the thermal pressure decreases locally due to strong cooling (\fig{tcool_tff}). However, we do not interpret this as evidence that the cold-stream is collapsing with no support against gravity. Rather, when examining the radial velocity profiles (as in \fig{radial inflows}) at $z\sim 4.05$ and $z\sim 3.82$, roughly $55\Myr$ before and after our fiducial snapshot at $z\sim 3.93$, we see that the mild inwards velocity along the sheet in the \textbf{S} zone increases slightly with amplitude towards lower redshift, from $\lsim 0.05v_{\rm vir}(r_{\rm shock})$ to $\gsim 0.1 v_{\rm vir}(r_{\rm shock})$. This may be due to additional gas accumulation near the filament axis, due to both inflows along the sheet and the off-sheet cooling flow. This results in a slightly larger inwards gravitational acceleration in the central regions, which leads to slightly more rapid inflow along the sheet. However, since the radial velocities in this region are always small anyway, we defer a more detailed study of the temporal evolution of the central streams to future work.

\smallskip
Finally, we note that in the axial direction, there is no steady-state at $r\gsim 0.2\,r_{\rm shock}$, as the filament is continually stretched by tidal forces and the material accelerates towards the massive haloes on either end. Note that the acceleration term is positive in the \textbf{S} zone and negative outside it. This suggests that the velocity of the cold streams towards the haloes increases with time, whereas that of the hot filament gas decreases with time. We discuss this further below. 
\subsubsection{Detailed Force Decomposition}
\label{sec:detailed force decomposition}

In \fig{summation method another}, we show the individual profiles of all the forces that went into the curves shown in \fig{summation method}, to assess their individual contributions to the dynamical state of the filament. On the left, the solid lines show $F_{\rm therm, r}/F_{\rm grav}$ (red), $F_{\rm inertial, r}/F_{\rm grav}$ (green), $F_{\rm therm, a}/F_{\rm grav}$ (orange), and $F_{\rm inertial, a}/F_{\rm grav}$ (yellow). The dashed lines are as in \fig{summation method}, showing $F_{\rm grav,r}/F_{\rm grav}$ (navy), $F_{\rm grav,a}/F_{\rm grav}$ (pink), and $M_{\rm grav}/M_{\rm tot}$ (magenta). First, we note that the four non-gravitational forces obtain both positive and negative values. As discussed in the following \equs{Mgrav_r}-\equm{Mgrav_a}, a positive value implies an acceleration outwards that counteracts the collapse, while a negative value implies an acceleration inwards. In the centre and right panels, we show the profiles of all the individual components that comprise the radial and axial inertial terms, respectively (\equsnp{radial inertial term}-\equmnp{axial inertial term}). We begin with the radial direction, discussing the left and centre panels simultaneously, and focusing on the different radial zones. 

\smallskip
In the \textbf{T} zone, at $r\gsim 0.6r_{\rm shock}$, $F_{\rm therm,r}$ is directed outwards and is stronger than the gravitational force inwards, monotonically increasing from $\sim |F_{\rm grav,r}|$ at $r\sim 0.6\,r_{\rm shock}$ to $\sim 4|F_{\rm grav,r}|$ at $r\sim r_{\rm shock}$. This excess in the force of the external thermal pressure is partially balanced by a strong inertial force in the inside, $F_{\rm inertial,r}$, which increases in amplitude from $\sim (0.3-2)|F_{\rm grav,r}|$ in the same radial range. The inwards inertial force in this region is dominated by the $-v_r \partial v_r/\partial r$ term (centre panel, red lines), which increases in amplitude from $\sim (1.5-3)|F_{\rm grav,r}|$. As discussed in \se{summation}, this term contributes to the radial components of both the ram pressure force (\equnp{Mram}, dot-dashed red line) and the turbulent pressure force (\equnp{Mturb_r}, solid red line). The ram pressure force can be visualised by examining the radial velocity profiles in \fig{radial inflows}. While the velocity in the sheet is roughly constant in the \textbf{T} zone, the inwards velocity outside the sheet decreases from $\sim v_{\rm vir}$ at $r_{\rm shock}$ to $\sim 0$ at $0.6\, r_{\rm shock}$, resulting in $-v_r \partial v_r/\partial r<0$. 
In the outer \textbf{T} zone, $0.8\lsim r/r_{\rm shock}$ where there is a positive radial acceleration term (\fig{summation method}), the ram pressure force dominates over the turbulent pressure force, which actually becomes negligibly small at $r\gsim r_{\rm shock}$. 
In the inner \textbf{T} zone, $0.6\lsim r/r_{\rm shock}\lsim 0.8$ where the radial velocity field is in steady-state (\fig{summation method}), the ram pressure and turbulent pressure forces are comparable. The two shear forces, $r\phi$ and $rz$ (\equsnp{Mturb_p}-\equmnp{Mturb_z}, orange and blue lines in the centre panel, respectively), are dominated by residual motions (solid lines), while the shear generated by the mean flow (dot-dashed lines) is negligible. Together with the centrifugal forces (purple dot-dashed and solid lines for mean and residual rotation, respectively), these four terms combine to produce a net outwards force that is smaller than both the thermal and ram pressure forces throughout this region. In summary, strong thermal pressure gradients in the \textbf{T} zone support the filament against both gravity and ram pressure from the inflowing gas. At $r\lsim 0.8\, r_{\rm shock}$, the gas is roughly in a steady state, while at $r\gsim 0.8\, r_{\rm shock}$ the velocity field accelerates outwards in response to the shock. 

\smallskip
The \textbf{V} zone, $0.25 \lsim r/r_{\rm shock} \lsim 0.6$, is the only zone where both $F_{\rm therm,r}$ and $F_{\rm inertial,r}$ are directed outward. In the outer \textbf{V} zone, $0.4 \lsim r/r_{\rm shock} \lsim 0.6$ where there is a positive radial acceleration (\fig{summation method}), the thermal pressure is dominant and still stronger than the radial gravitational force and is comparable in amplitude to its value in the inner \textbf{T} zone. In the inner \textbf{V} zone, $0.25 \lsim r/r_{\rm shock} \lsim 0.4$ where the radial velocity field is in steady-state (\fig{summation method}), the thermal pressure term decreases and becomes comparable to the inertial term, which maintains a value of $\lsim 0.6|F_{\rm grav,r}|$ throughout the entire \textbf{V} zone. The change in sign of the inertial term, from negative in the \textbf{T} zone to positive in the outer \textbf{V} zone, is mainly due to a decrease in the force of ram pressure inwards and an increase in the centrifugal force outward. The mean rotation term increases from $\sim 0$ at $r\sim r_{\rm shock}$ to a constant $\sim |F_{\rm grav,r}|$ throughout the \textbf{S} and inner \textbf{V} zones, at $r\lsim 0.4 r_{\rm shock}$. The residual rotation term is $\sim 1.5|F_{\rm grav,r}|$ at $r\sim r_{\rm shock}$, and increases monotonically toward smaller $r$. Throughout the \textbf{V} and \textbf{S} zones, at $r < 0.6 r_{\rm shock}$, this is mainly offset by the inwards $r\phi$ shear force, $-(v_{\phi}/r)\partial v_r/\partial \phi$, with a smaller contribution from the ram/turbulent pressure. In \se{toy_model}, we will show that the behaviour of this trio of forces is generic, resulting from the quadrupolar vortex structure that dominates the filament dynamics throughout the \textbf{V} zone. 
To summarise the situation in the outer \textbf{V} zone, at $0.4\,r_{\rm shock}\lsim r\lsim 0.6\,r_{\rm shock}$, strong thermal pressure forces combined with increasing centrifugal forces and decreasing ram pressure forces lead to a net outwards acceleration of the velocity field. The lack of a steady state here may be linked to an evolving vorticity structure fuelled by ongoing accretion onto the inner filament (see \se{toy_model} below).

\smallskip
In the inner \textbf{V} zone, at $0.25 r_{\rm shock}\lsim r \lsim 0.4 r_{\rm shock}$ where the filament velocity field is in steady state with no acceleration (\fig{summation method}), the thermal force continues to be directed outwards but is now weaker than the gravitational force inwards, $\sim (0.3-0.6)|F_{\rm grav,r}|$. The inertial term continues to be directed outward, maintaining a constant amplitude of $\sim 0.6|F_{\rm grav,r}|$. The components of the inertial term behave similarly here to the outer \textbf{V} zone. The outwards inertial force is dominated by the centrifugal force, with the mean rotation term roughly balancing gravity with an amplitude of $\sim |F_{\rm grav,r}|$. The residual rotation term continues to increase towards smaller $r$, with amplitudes in the range $\sim (4-7)|F_{\rm grav,r}|$, though this is again compensated for by shear and turbulent pressure (see \se{toy_model} below). This trio yields a slight net inwards acceleration that lowers the total inertial term compared to the mean rotation. To summarise the situation in the inner \textbf{V} zone, at $0.25\,r_{\rm shock}\lsim r\lsim 0.4 r{\rm shock}$, the filament remains in force equilibrium with roughly equal contributions from thermal pressure forces and centrifugal forces in the support against gravity. 

\smallskip
In the \textbf{S} zone, at $r\lsim 0.25r_{\rm shock}$, there is a net inwards acceleration term (\fig{summation method}). This is mainly driven by the thermal pressure force becoming negative at $r\sim (0.15-0.25)r_{\rm shock}$, near the outer boundary of the isothermal core, where the cooling is maximal. The inertial term becomes negative at $r\lsim 0.15r_{\rm shock}$ due to a slight drop in centrifugal and residual shear forces (dot-dashed purple and solid blue lines). This suggests that the dynamics of the isothermal cores of high-$z$ filaments, representing the cold streams feeding massive high-$z$ galaxies (\se{thermal}), are far from a relaxed state of equilibrium, but are dominated by strong cooling flows and complex motions and accelerations of infalling and strongly rotating gas.

\smallskip
We broadly summarise the balance of the radial force throughout the filament, highlighting the key physical characteristics of each of the three radial zones. 
In the \textbf{T} (thermal) zone, thermal pressure forces dominate, roughly balancing gravity and ram pressure forces. In the \textbf{V} (vortex) zone, the thermal pressure forces become comparable to the centrifugal forces induced by mean rotation, and these combine to roughly balance gravity. At the same time, the dominant forces are actually residual rotation, shear, and turbulent pressure, which are induced by the vortex structure of the filament. Individually, each of these is stronger than gravity in absolute value, though they roughly cancel out, leaving a relatively small combined force, which is weaker than the mean centrifugal term. 
In the \textbf{S} (stream) zone, a drop in both centrifugal and thermal pressure forces due to the development of a strong cooling flow results in a net inwards acceleration of the velocity field. 

\smallskip
In the axial direction, the situation is much simpler. The axial inertial force is comparable in amplitude and opposite in sign to the axial gravitational force, which both exhibit a very weak radial dependence. $F_{\rm inertial,a}$ is dominated by axial ram pressure forces encapsulated in the term $-v_z\partial v_z/\partial z$, while the other two terms, representing turbulent forces, are very small. The axial thermal pressure force is comparatively small in amplitude, directed outwards in the \textbf{S} zone and inwards at larger radii. This is consistent with a picture where the cold stream gas penetrates the hot CGM and free-falls towards the massive central galaxies on either end, while the hot filament gas does not penetrate the virial shock and instead builds up in the CGM and IGM around the haloes, providing additional pressure confinement that slightly weakens the tidal stretching of the filament in the axial direction. We will discuss this further in \se{cold_streams}. 

\subsection{A Toy Model for the Filament Velocity Field}
\label{sec:toy_model}

In \se{sim_dynamics}, we saw that, in addition to thermal pressure gradients, the filament inertia plays a major role in its overall equilibrium. In particular, the residual rotation provides a very strong outwards force that significantly exceeds the inwards gravitational force at small radii, and is largely balanced by the shear and ram/turbulent forces, $(v_\phi/r)\partial v_r/\partial \phi$ and $v_r\partial v_r/\partial r$. In this section, we wish to understand the origin of these terms and in particular how they represent fictitious forces acting on the gas along its streamlines, rather than our description of them as ram/turbulent/shear forces in \se{sim_dynamics}. We begin in \se{kepler} by discussing the very simple case of an elliptical orbit in a Keplerian potential as an instructive example. In \se{filament's vorticity}, we discuss the vorticity structure of the filaments in our simulation, highlighting a characteristic quadrupolar structure. Finally, in \se{vortex}, we discuss how such dynamics can give rise to the inertia forces we find. 

\subsubsection{The Keplerian Orbit}
\label{sec:kepler}

Consider an elliptical Keplerian orbit of a test particle around a central mass $M$. The equation of motion in the radial direction, with respect to the central mass, is
\be
\label{eq:radial motion}
    r\dot{\phi}^2-\ddot{r}=\frac{GM}{r^2},
\ee 
{\no}where $r$ is the distance from the central mass and $\phi$ is the azimuthal angle. The azimuthal and radial velocities are $v_{\phi}=r\dot{\phi}$, and $v_r=\dot{r}$. The first term on the left-hand side of \equ{radial motion} is thus $v_\phi^2/r$ while the second is $d v_r/dt$. The latter term can be expressed in two possible ways, depending on whether the orbit is parametrised with respect to the radius, $r$, or the azimuthal angle\footnote{An ellipse is a one-dimensional curve with one independent parameter.}, $\phi$. 

If we parametrize the orbit in terms of $r$ we have 
\begin{equation}
\label{eq:r derivative}
    \dv{}{t}=\dot{r}\dv{}{r}=v_r\dv{}{r}.
\end{equation}

{\no}Inserting this into \equ{radial motion} yields 
\begin{equation}
\label{eq:r parametrization}
    \frac{v_\phi^2}{r}-v_r\dv{v_r}{r}=\frac{GM}{r^2},
\end{equation}
{\no}showing that in the frame of the particle, gravity is indeed balanced by the two relevant components of the inertial term in \equ{radial inertial term}, which act as fictitious forces along the orbit. We can explicitly evaluate \equ{r parametrization} for the Keplerian orbit by recalling that both the specific angular momentum, $l=v_{\phi}r$, and the specific energy, $\mathcal{E}=-GM/(2a)$ with $a$ the semi-major axis, are conserved. Therefore, 
\begin{equation}
\label{eq:vphi in r}
    v_{\phi}(r)=\frac{l}{r},
\end{equation}
{\no}and 
\begin{equation}
\label{eq:vr in r}
    v_r(r)=\sqrt{\frac{2GM}{r}-\frac{l^2}{r^2}-\frac{GM}{a}}.
\end{equation}
{\no}One can easily verify that \equs{vphi in r}-\equm{vr in r} obey \equ{r parametrization}. 

\smallskip
If we parametrize the orbit in terms of $\phi$, \equs{r derivative}-\equm{vr in r} become
\begin{equation}
    \dv[]{}{t}=\dot{\phi}\dv{}{\phi}=\frac{v_\phi}{r}\dv{}{\phi},
\end{equation}
\begin{equation}
\label{eq:phi parametrization}
    \frac{v_\phi^2}{r}-\frac{v_{\phi}}{r}\dv{v_r}{\phi}=\frac{GM}{r^2},
\end{equation}
\begin{equation}
\label{eq:vphi in phi}
   v_r(\phi)=\frac{GM}{l}e\sin\phi,
\end{equation}
\begin{equation}
\label{eq:vr in phi}
    v_{\phi}(\phi)=\frac{GM}{l}(1+e\cos\phi),
\end{equation}
{\no}where $e$ is the eccentricity of the orbit, which is a conserved quantity. Again, one can easily verify that \equs{vphi in phi}-\equm{vr in phi} obey \equ{phi parametrization} and that gravity is balanced by the two relevant components of the inertia term in \equ{radial inertial term}, which act as fictitious forces along the orbit.

\smallskip
This simple example shows that in a steady flow around a central mass without thermal pressure, the three major terms appearing in \equ{radial inertial term} act as fictitious forces that balance gravity. 


\subsubsection{The Velocity Structure of Filaments}
\label{sec:filament's vorticity}

\begin{figure}
    \centering
    \includegraphics[trim={0.0cm 0.0cm 0.0cm 0.0cm}, clip, width =0.48 \textwidth]{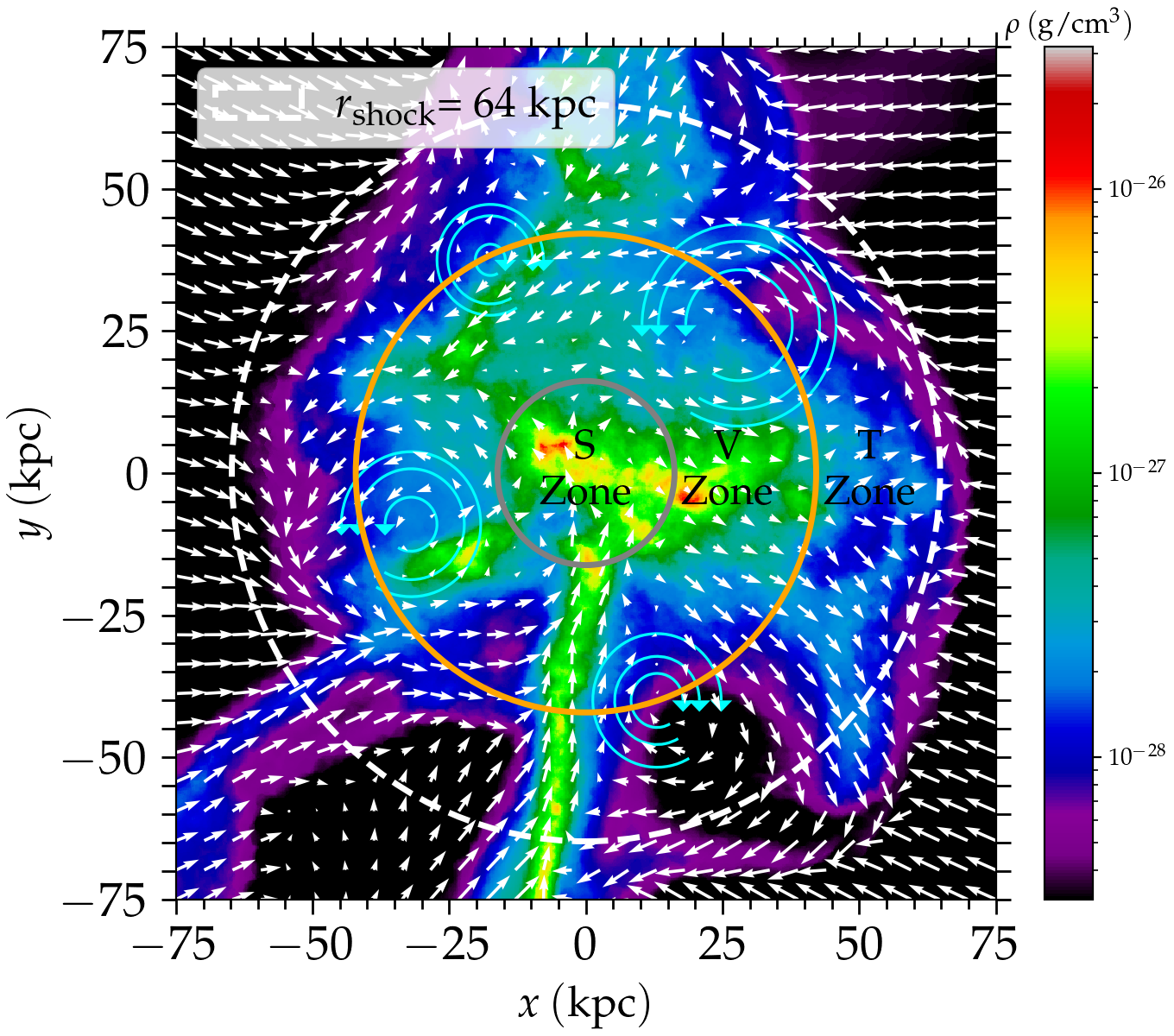}
    \caption{The projected velocity field of slice 2, a representative example among our 10 slices. The colour map encodes the average gas density projected along the line-of-sight, and arrows represent the projected 2D velocity field perpendicular to the line-of-sight. The longest arrows correspond to $290 \kms$, with lengths scaling as $l=30 \sinh^{-1}\left(v/30\kms\right)$, 
    such that a $30 \kms$ arrow is $\sim 3.36$ times shorter than the longest one. Four major vortices forming a quadrupolar pattern are present near the outer boundary of the \textbf{V} zone ($r\sim 0.6\,r_{\rm shock}$, orange circle). These are highlighted with schematic eddies to guide the eye. 
    }
    \label{fig:streamplot}
\end{figure}

\begin{figure*}
    \centering
    \includegraphics[scale = 0.54]{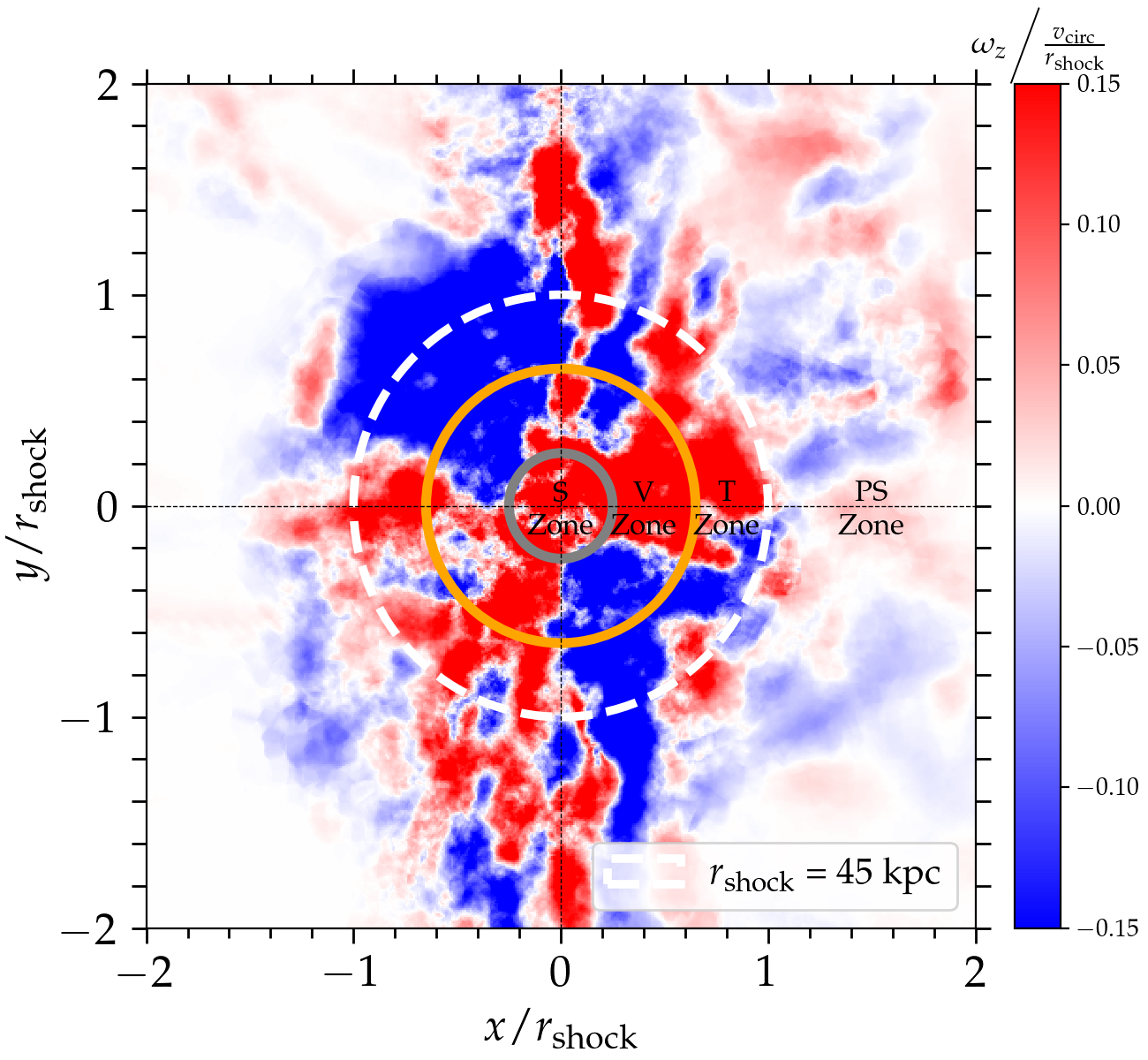}
    \includegraphics[scale = 0.52]{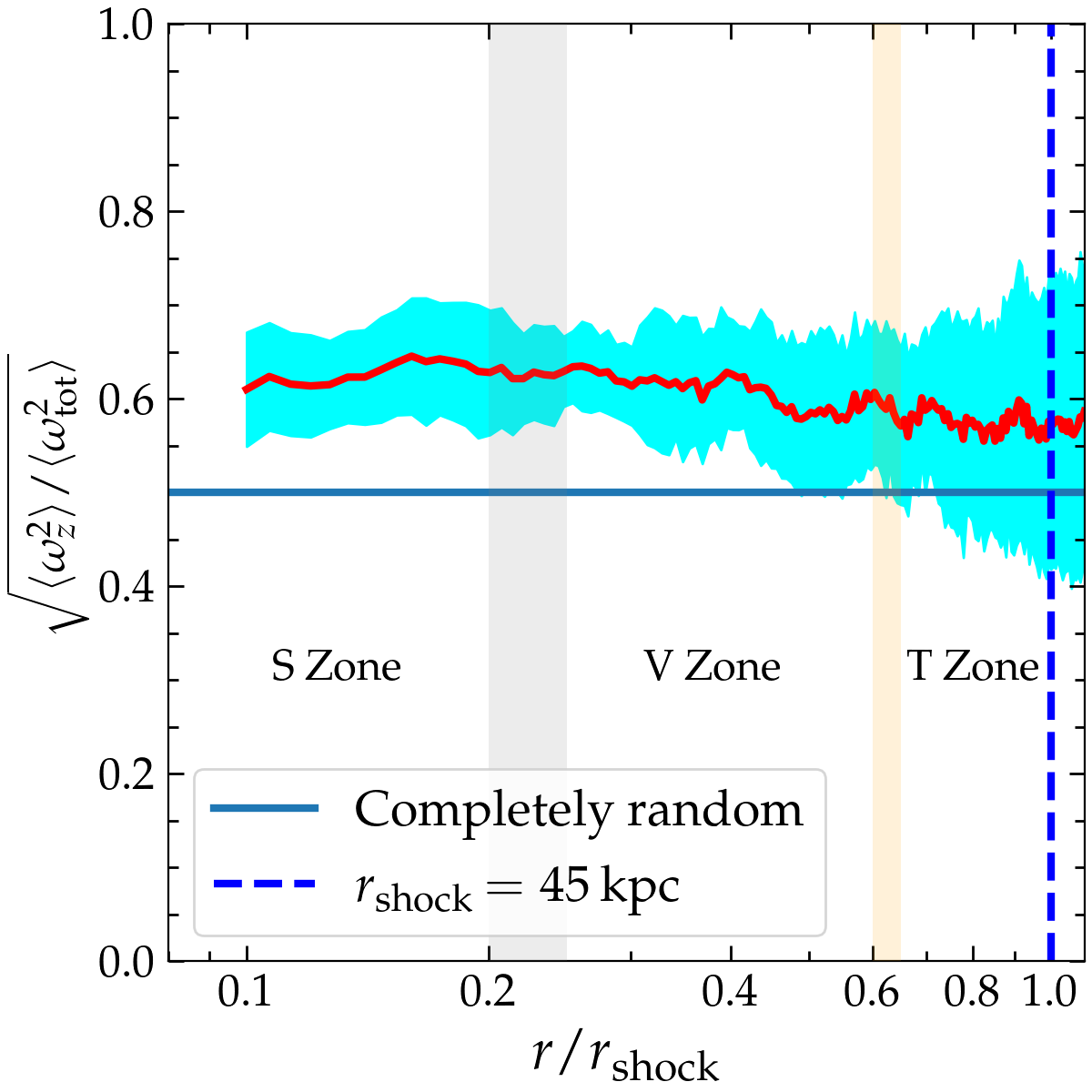}
    \caption{Vorticity along the filament axis. \textbf{\textit{Left:}} Projection map of the vorticity along the filament axis, mass-weighted along the line of sight, normalized by $v_{\rm circ}/r_{\rm shock}$, and stacked among the 10 filament slices. The same quadrupolar structure visible in the single slice shown in \fig{streamplot} is evident in the stacked map. 
    The upper-right and lower-left (first and third) quadrants have positive vorticity, while the upper-left and lower-right (second and fourth) quadrants have negative vorticity. \textbf{\textit{Right:}} The stacked profile of the contribution to the filament vorticity oriented along its axis. We calculate the mass-weighted average profiles of $\omega_z^2$ and $\omega_{\rm tot}^2$ for each slice and take the square root of their ratio. We then compute the average and standard deviation of this ratio among our ten slices. Filaments prefer vorticity to align with their axis, with a typical ratio of $\gsim 0.6$, compared to $0.5$, which is characteristic of a random distribution. }
    \label{fig:vorticity}
\end{figure*}

\smallskip
In order to model the inertia forces seen in \fig{summation method another} using an analysis similar to that presented in \se{kepler}, we must first characterise the gas velocity field within the filaments. In \fig{streamplot}, we show the velocity field in slice 2 (see \fig{slices}), projected along the filament cross-section and overlaid on a map of the average density along the filament's axis, in the same orientation as the stacked projections shown in \fig{thermal profiles}. The arrows represent the 2D gas velocity field perpendicular to the filament's axis, $v_x{\hat{x}}+v_y{\hat{y}}$, mass-weighted-averaged along the axis. Several features of the velocity field are apparent. Outside the shock radius (dashed-white circle), gas flows towards the filament from both inside and outside the sheet (aligned parallel to the $y$-axis). The cylindrical shock is evident in the flow pattern of off-sheet material whose velocity suddenly changes orientation and develops vorticity, as expected from a curved shock. The sheet gas, on the other hand, maintains its approximate trajectory due to its higher density, which destabilises the shock there \citep[see][for an analogous discussion of cold streams in hot haloes]{Dekel.Birnboim.06}. Inside $r_{\rm shock}$, the velocity is broadly characterised by four vortices centred near the transition between the \textbf{T} and \textbf{V} zones at $r\sim 0.6\,r_{\rm shock}$, in a quadrupolar pattern with the upper-right and lower-left quadrants spinning counter-clockwise (positive vorticity), and the upper-left and lower-right quadrants spinning clockwise (negative vorticity). Such a quadrupolar vorticity structure has been seen in previous studies of intergalactic filaments using lower-resolution cosmological simulations \citep[e.g.,][]{Pichon.Bernardeau.99,Codis.etal.12, Codis.etal.15b,Laigle.etal.15,xia2021intergalactic,ramsoy2021rivers}, and is thought to arise from asymmetric inflow into filaments from sheets with non-zero impact parameters. While this is the only source of vorticity for dark matter, an additional source of vorticity for the gas is shear due to the inflow along the sheet interacting with the hot gas in the \textbf{T} and \textbf{V} zones, combined with the vorticity creation due to the curved shocks. 
We defer a more detailed study of the origin of this vorticity structure in our simulations to future work.

\smallskip
The quadrupolar vortex structure is a generic feature of filament dynamics in our simulation, not unique to slice 2. In \fig{vorticity} (left), we show a stacked projection map of $\omega_z$, the vorticity component parallel to the filament axis. In each slice, we compute the mass-weighted average value of $\omega_z$ along the line-of-sight, and weigh this by $v_{\rm circ}(r_{\rm shock})/r_{\rm shock}$ prior to stacking, where $v_{\rm circ}(r)=\sqrt{2G\Lambda_{\rm tot}(r)}=\sqrt{2}v_{\rm vir}$ is the circular velocity of the cylindrical filament. Despite slice-to-slice variation in the detailed velocity field and in the location and strength of the four vortices, the stacked projection nicely recovers the quadrupolar structure, with the vorticity being primarily positive (negative) in the upper-right and lower-left (upper-left and lower-right) quadrants\footnote{Note that while all of our previous analysis was symmetric under the transformation $z\rightarrow -z$, not so $\omega_z$. Although the vorticity in each slice located in the same filament has the same orientation, we take care to orient the three different filaments (see \fig{slices}) such that $\omega_z$ is mostly positive (negative) in the first and third (second and fourth) quadrants prior to stacking.}. 

\smallskip
In the right-hand panel of \fig{vorticity}, we address the fraction of vorticity aligned with the filament axis. For each slice, we compute the mass-weighted average of both $\omega_{z}^2$ and $\omega_{\rm tot}^2=(\omega_x^2+\omega_y^2+\omega_z^2)$ at each radius $r$, and take the square-root of their ratio. We then present the mean and standard deviation of this ratio among the ten slices. If the vorticity were randomly oriented, this ratio would be $\sim 0.5$, which is the average of ${\rm cos}(\theta)$ uniformly distributed between $0$ and $1$. Instead, the typical ratio is $\gsim 0.6$ within the shock and increases slightly toward smaller radii, displaying a preference for the vorticity to align with the filament axis. This is broadly consistent with the results of \citet{Laigle.etal.15} who found a preference for the filament vorticity to align with its axis, with an excess probability of $\sim 20\%$ having an angle less than $60^{\circ}$, corresponding to a cosine greater than $0.5$.

\subsubsection{Ideal Vortex Model}
\label{sec:vortex}

\begin{figure}
    \centering
    \includegraphics[trim={0.0cm 0.0cm 0.0cm 0.0cm}, clip, width =0.48 \textwidth]{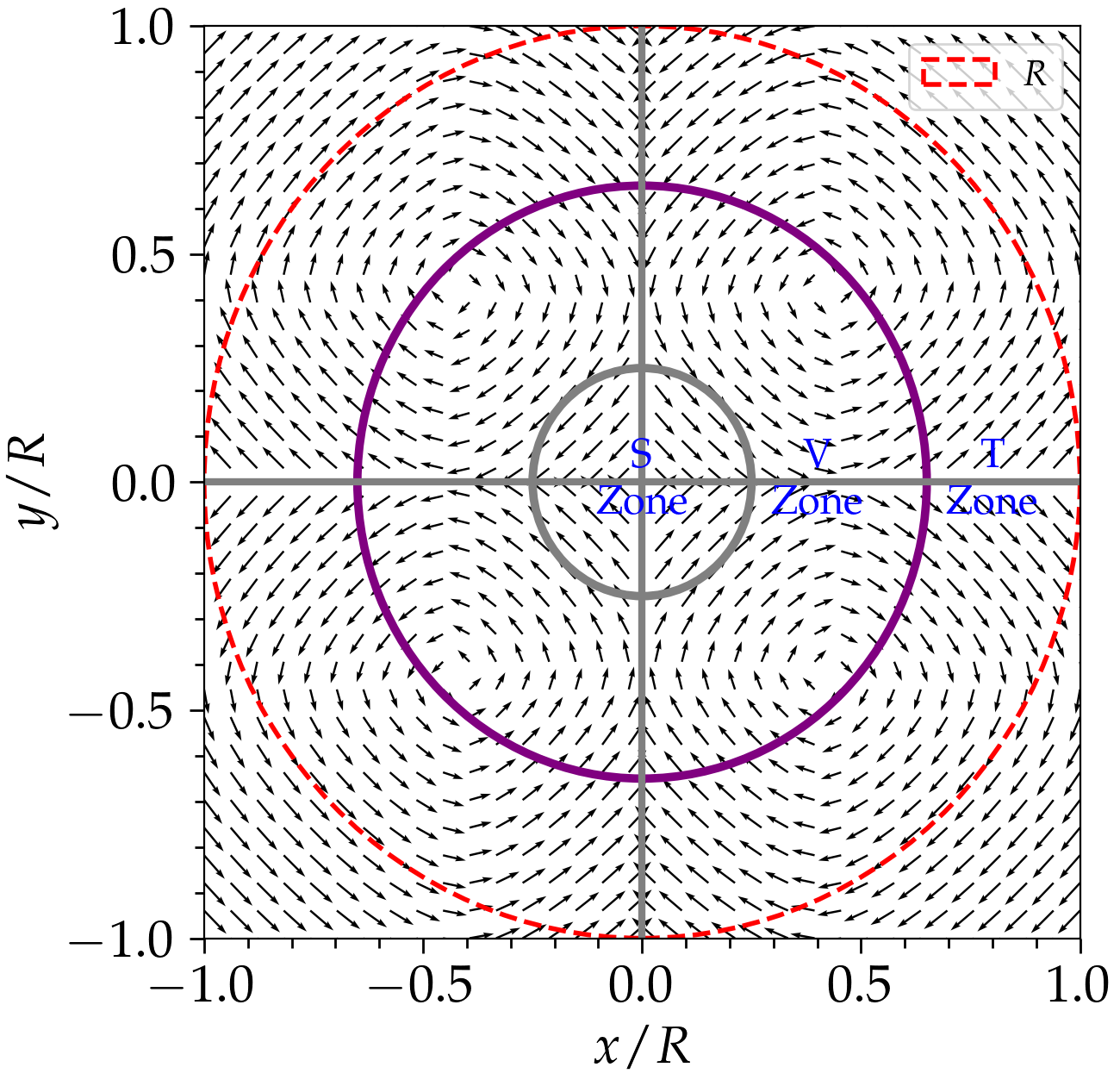}
    \caption{Our ideal vortex model is used to interpret the main features of the radial inertial forces from \fig{summation method another}. The red dashed circle represents the ``shock'' radius, $R$, and we have added circles at $0.6R$ and $0.25R$, schematically representing the borders between the three zones in our simulated filaments. Each quadrant has a rotational vortex centred at $(x,y)=(\pm a,\pm a)\simeq(\pm 0.42\, R, \pm 0.42\, R)$. These form a quadrupolar structure, with positive (negative) vorticity in the first and third (second and fourth) quadrants. Within each quadrant, the rotation velocity scales as $r'\,^{0.5}$, where $r'$ is the distance to the vortex centre.
    }
    \label{fig:vortex velocity}
\end{figure}

\smallskip

While the gas dynamics in filaments is inherently three-dimensional and complex, the quadrupolar vortex structure seen in \figs{streamplot}-\figss{vorticity} appears to be the most robust feature common to all filament slices. Therefore, we examine to what extent such a velocity field alone can reproduce the robust features of the radial inertial forces seen in \fig{summation method another} (centre panel). To this end, we construct a simple two-dimensional toy model for the filament velocity field, characterised by four ideal vortices in a quadrupolar configuration, such that the vorticity is positive (negative) in the first and third (second and fourth) quadrants, following \figs{streamplot}-\figss{vorticity}. This model is shown in \fig{vortex velocity}. In the frame of each vortex (hereafter the primed frames), its velocity contribution is given by 
\be
\label{eq:ideal_vort}
    \vec{v}=v_{\phi'}\hat{\phi}'=\pm \beta r'^{\alpha}\hat{\phi'},
\ee 
{\no}where the plus sign applies to the first and third quadrants and the minus sign applies to the second and fourth quadrants. $\alpha$ sets the slope of the rotation curve with respect to the vortex centres while $\beta$ is a normalisation constant. $r'$ is the distance to the centre of each vortex, which is assumed to be at $(x,y)=(\pm a,\pm a)$. 

\smallskip
The vorticity corresponding to such a velocity field, in the vortex frame, is:
\be
    \vec{\omega}=\nabla\times\vec{v}=\frac{1}{r'}\pdv{(r'v_{\phi'})}{r'}\hat{z}'=\pm \beta (\alpha+1) r'^{\alpha-1}\hat{z}'.
\ee 
{\no}For any $\alpha\not=-1$, the vorticity is non-zero\footnote{$\alpha=-1$ corresponds to a constant specific angular momentum profile.} and finite at all $r'>0$. Transforming the velocity field to the filament frame (hereafter the unprimed frame) yields
\be
    \vec{v}=\mp \beta r'^\alpha \sin(\phi')\hat{x}\pm\beta r'^\alpha \cos(\phi')\hat{y},
\ee 
{\no}where in every plus/minus combination, the upper symbol applies to the first and third quadrants, and the lower symbol applies to the second and fourth quadrants. The transformation of $r'$ and $\phi'$ into the unprimed coordinates $x$ and $y$ depends on the quadrant. For example, in the first quadrant $r'=[(x-a)^2 + (y-a)^2]^{1/2}$ and $\phi'=\arctan[(y-a)/(x-a)]$. We can then write the full velocity field in terms of the polar coordinates in the filament frame, $r$ and $\phi$ using $x=r\cos(\phi)$ and $y=r\sin(\phi)$. We use this model within the boundaries $|x|,|y|\le R$, where $R$ can be thought of as analogous to the shock radius of the filament, $r_{\rm shock}$. Hereafter, we place the vortices at a radius of $r=0.6\,R$, motivated by \fig{streamplot}, so $a/R=0.6/ \sqrt{2}\simeq 0.42$. The trends at $r<a$ and $r>a$ are independent of this choice. 

\smallskip
A visual inspection of \fig{streamplot} indicates that, the rotation velocity increases away from the vortex centres. Therefore, we take $\alpha>0$. We further require that the ram/turbulent pressure term, $-v_r\partial v_r/\partial r$ be negative, based on the middle panel of \fig{summation method another}, which turns out to require $\alpha<1$. We use $\alpha=0.5$ as our fiducial value, but note that we obtain similar trends with any value of $\alpha\sim(0.1-0.9)$. The value of the normalisation constant $\beta$ is arbitrary, and hereafter we take $\beta=1$.

\smallskip
Given parameters $a$, $\beta$, and $\alpha$, we compute the velocity field and the resulting fictitious forces using the same method we used in \se{summation}. We set up a cylindrical grid with linear spacing in the radial and azimuthal directions and evaluate the velocity in each bin. Note that since our assumed velocity field is explicitly two-dimensional, all axial derivatives vanish, i.e., $\partial/\partial z=0$. Furthermore, in our highly idealised model, both the mean azimuthal and radial velocities are zero, $v_{\phi,{\rm mean}}=v_{r,{\rm mean}}=0$. 
This leaves only the three components in the radial inertial term, which depend only on the residual velocities, namely the turbulent pressure (\equnp{Mturb_r}), residual azimuthal shear (\equnp{Mturb_p}), and residual centrifugal force (\equnp{decomposition of the rotation term}) (red, orange, and purple solid lines in the centre panel of \fig{summation method another}). We evaluate these following our methodology described in Appendix \se{sim_numerics}. 
Finally, we normalise all forces by $(r/R)^{\gamma}$, with $\gamma=1.48$. This accounts for the radial dependence of the gravitational force, $F_{\rm g}$, the denominator in all the profiles shown in \fig{summation method another}, which is given by the total mass interior to radius $r$ (blue line in \fig{line mass}).

\begin{figure}
    \centering
    \includegraphics[trim={0.0cm 0.0cm 0.0cm 0.0cm}, clip, width =0.48 \textwidth]{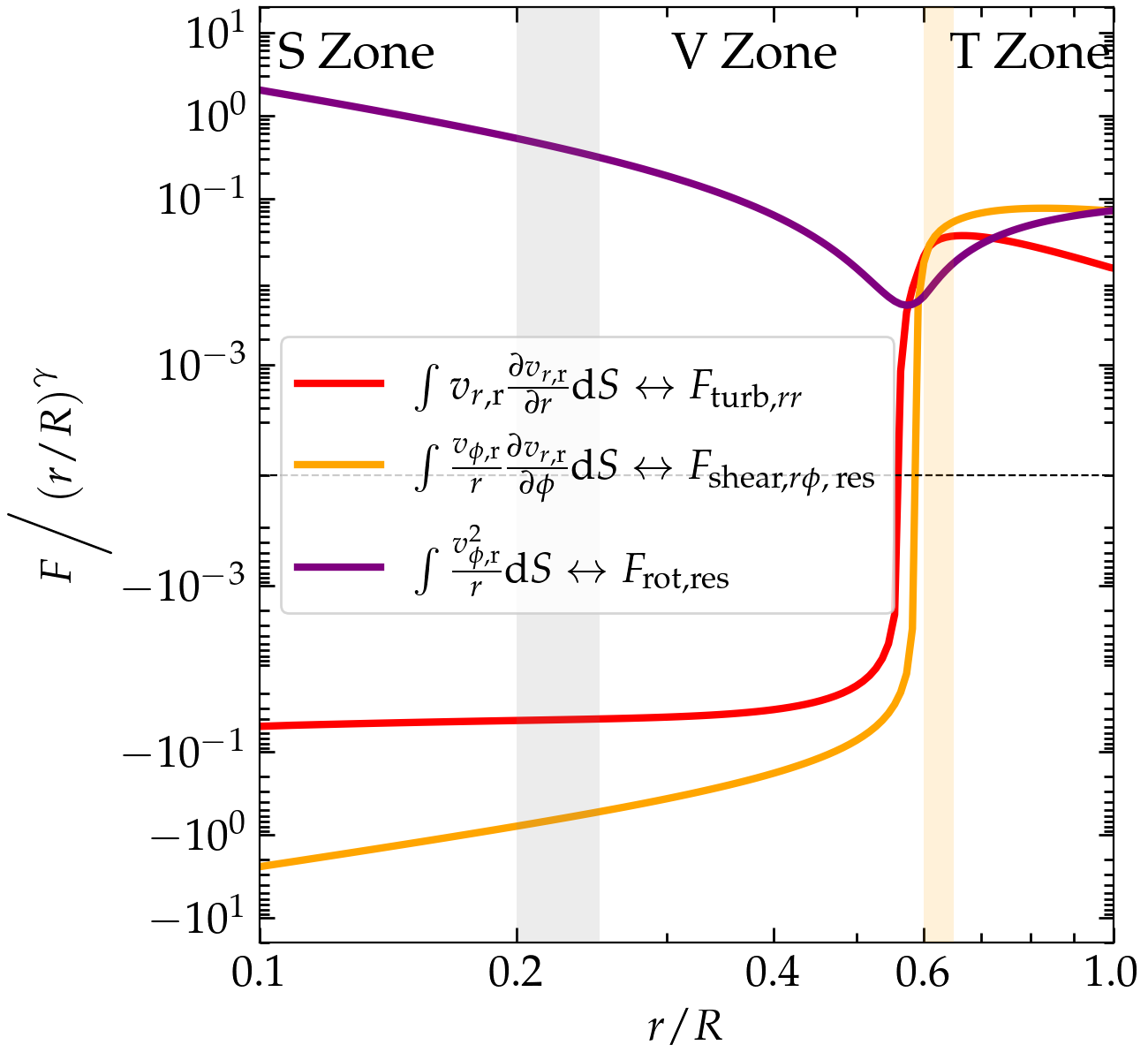}
    \caption{Inertial (fictitious) force terms resulting from our toy model of an ideal quadrupolar vortex structure (\fig{vortex velocity}). Since our model is completely symmetric, the mean parts of $v_r$ and $v_\phi$ are, by definition, $0$, and therefore only the residual parts contribute. Each force term of our model has been scaled by an additional $r^{\gamma}$ to account for the radial dependence of $F_{\rm g}$, the denominator in the ratios presented in \fig{summation method another}. Our simple model reproduces many of the trends seen in the simulation in the \textbf{V} zone, at $0.25\lsim r/r_{\rm shock}\lsim 0.6$ (\fig{summation method another}, middle panel, solid lines), where the ram pressure force has decreased and vortices dominate the dynamics of the filament (see the text for details).}
    \label{fig:ideal vortex results}
\end{figure}

\smallskip
The resulting force terms are plotted in \fig{ideal vortex results}. These reproduce many of the trends seen in the corresponding curves in the centre panel of \fig{summation method another}. At small radii ($r\lsim 0.5 R$), the residual rotation is mostly balanced by azimuthal shear, both of which appear to diverge as $r\rightarrow 0$ as $r^{-\gamma}$. The turbulent pressure also contributes to the balance of the rotation term, but saturates at a finite value as $r\rightarrow 0$. Even the sign-flip in this term at $r\lsim 0.6\,R$, near the singularity present in our model at the location of the vortices, resembles the sharp decrease in the amplitude of the corresponding term in \fig{summation method another} near the location of the vortices at $r\sim 0.6\,r_{\rm shock}$. The corresponding sign flip of the shear force at $r=a$ is not seen in the data. However, in the simulations, the inertial forces in the \textbf{T} zone at $r>0.6\,r_{\rm shock}$ are dominated by the ram pressure of the inflowing gas, which is not included in our ideal vortex model. Therefore, it is expected that it will only represent the data at $r<a=0.6\,R$. Furthermore, unlike our simple two-dimensional toy model, the actual filament dynamics is complex and three-dimensional, with vorticity that is not even truly aligned with the filament axis (\fig{vorticity}). Despite these caveats, our model confirms that the structure of the quadrupolar vorticity in intergalactic filaments produces fictitious force terms that resemble the inertial terms dominating the dynamics of the filament at $r\lsim 0.6\,r_{\rm shock}$. More detailed theoretical studies of cosmic velocity fields in filaments can be found in, e.g., \citet{Pichon.Bernardeau.99, Pichon.etal.11,Laigle.etal.15,hahn.etal.15}.

\section{Discussion}
\label{sec:disc}

\subsection{Comparison to Previous work}
\label{sec:comparison}

\subsubsection{Filament Thermal Structure and Size}

We compare our results with those of \citet{ramsoy2021rivers}, who used an AMR simulation to study the properties of an intergalactic filament that feeds the progenitor of a MW-mass halo ($\Mv\sim 10^{11.5}\msun$ at $z=0$). While they examined the filament in the redshift interval $z\sim (3.5-8)$, their main analysis focused on $z\sim 4$ as ours, and many of their results are similar to ours. They too find that the filament is surrounded by a cylindrical accretion shock and embedded in an intergalactic sheet with a planar accretion shock. The filament density in their simulation is well fit by the profile of an infinite, self-gravitating, isothermal cylinder, out to the radius where the sheet begins dominating the mass distribution. This was true for both gas and dark matter. Finally, similar to our results, their filaments were characterised by a quadrupolar vorticity structure with a weak radial dependence inside the filament shock radius. 

\smallskip
Despite these similarities, there are several important differences between our results. In their simulation, the density profiles of gas and dark matter were extremely similar. In particular, the characteristic scale radii, $r_0$, for the gas and dark matter were within a few tens of percent of each other, while these differed by a factor of $\gsim 3$ in our simulations. Similarly, the effective dark matter temperature in \citet{ramsoy2021rivers} was only a factor of $\lsim 2$ higher than the central gas temperature (see their figure 7), while it was $\sim 30$ times higher in our simulations (\se{DM_prop}). 
Furthermore, the shock surrounding the filament in their simulation was weaker and narrower than in ours. They measured an azimuthally-averaged gas-temperature increase of $\lsim 60\%$ between the filament core and the shock, 
while the temperature in our simulation increases by a factor of $\sim (20-30)$. Likewise, they found the shock to be nearly isothermal, with the gas temperature reaching its core value $\sim 3\kpc$ from the shock front, while in our case the gas remains hot outside the cooling radius of $\sim 0.5\,r_{\rm shock}\sim 25\kpc$ (\se{tcool}) and only reached the core temperature $\sim (30-40)\kpc$ from the shock front. Finally, while they find that thermal pressure alone can account for most of the filament support against gravity, we find this to be true only in the \textbf{T} zone near the shock radius, while the \textbf{V} and \textbf{S} zones are dominated by vortical, turbulent, and rotational motions driving the filament far from equilibrium. 

\smallskip
The primary reasons for these differences are likely the different environments and line-masses of the filaments studied in this work compared to that studied in \citet{ramsoy2021rivers}. 
While these authors studied a single, relatively low-mass and isolated filament feeding a single MW-progenitor halo, $\Mv\sim 10^{10.5}\msun$ at $z\sim 4$, we analyse ten slices of three filaments in a crowded region between three haloes with $\Mv\sim 10^{12}\msun$ at $z\sim 4$. In general, more massive filaments are predicted to have more prominent accretion shocks, similar to the transition from cold to hot CGM for haloes \citep{birnboim2016stability}. More quantitatively, the shock Mach number is predicted to scale as $\mathcal{M}\propto v_{\rm r}\propto v_{\rm vir} \propto \Lambda^{0.5}$ (\equnp{v_vir}, see also Appendix \ref{sec:Mach}). Based on Eq. (10) from \citet{mandelker2018cold}, the filament line-mass scales with the halo mass as $\Lambda \propto \Mv^{0.77}$, implying that our filament is $\lsim 15$ times more massive (per-unit-length) than theirs. For a filament in virial equilibrium per-unit-length (see \se{virial}), the effective DM temperature is expected to scale as $T\propto G\Lambda$, implying that the DM temperature in our filaments should be $\lsim 15$ times larger than in \citet{ramsoy2021rivers}. On the other hand, the central gas temperature is $\sim (2-3)\times 10^4\K$ in both cases, set by thermal equilibrium with UVB, as the central gas can efficiently cool. We thus expect the ratio of DM to gas temperature to be $\sim 15$ times larger in our simulations compared to \citet{ramsoy2021rivers}, consistent with the simulation results cited above. 
This demonstrates how filament properties are sensitive to the masses of the haloes they feed, and validates several scaling relations. 

\smallskip
Another important difference between our studies is the resolution in the filaments. In the simulations analysed by \citet{ramsoy2021rivers}, the filaments were resolved by $\sim 1.2\kpc$ with small patches surrounding haloes embedded in the filament resolved with $\sim 600\pc$. In our simulations, the typical cell size is $\sim (300-500)\pc$ in the \textbf{S} zone and $(0.5-1.0)\kpc$ in the \textbf{T} zone (\fig{bin sizes and cell sizes}). Combined with the fact that our filaments are $\sim 3$ times wider than the filament studied in \citet{ramsoy2021rivers}, we have nearly an order of magnitude more cells across the filament diameter in our simulation, allowing us to better resolve the turbulent cascade and reduce 
artificial dissipation of turbulent energy. Finally, it is possible that some of the differences are driven by differences in the hydro-solver (Eulerian AMR versus moving-mesh) and the different sub-grid models implemented in the simulation. More work with larger samples of filaments in different simulations using different codes will be required to break these degeneracies. 

\smallskip
Another interesting question is why the gas density profile is well-fit by an isothermal model with the correct core temperature, despite the temperature increasing by a factor of $\sim 20$ from the core to the shock. If we assume that the gas initially traced the DM and was thus roughly isothermal at the post-shock temperature, then the gas density is initially cored with a comparable radius to the DM core, as in \citet{ramsoy2021rivers}. This constant-density and constant-temperature core then begins to cool monolithically toward a temperature of $\sim 3\times 10^4$, set by thermal equilibrium with the UVB. If this cooling is isobaric, as suggested by the gas pressure profile (\fig{thermal profiles}), then the density in the core increases by a factor of $\sim (15-20)$ as the gas cools. From the mass conservation in the core, we thus expect the core radius to shrink by a factor of $\sim (15-20)^{1/2} \sim 4$, consistent with the ratio of the DM core radius to that of the gas. On the other hand, at large radii, near $r_{\rm shock}$, the gas density is much smaller. Cooling, therefore, is much less efficient, so we may assume that the density profile of the gas at these large distances continues to trace the DM, and thus asymptotically approaches the slope $\rho\propto r^{-4}$ of an isothermal cylinder irrespective of the central temperature. This will produce a density profile consistent with \fig{thermal profiles} despite the gas not being isothermal.

\smallskip
We can use these insights to derive a prediction for the size of the isothermal filament core (the \textbf{S} zone) as a function of redshift and the mass of the halo that the filament is feeding. From \citet{mandelker2018cold}, the line-mass and virial radius of filaments are given by (their equations 10 and 14)
\be 
\label{eq:Lambda_fil_M18}
\Lambda_{\rm fil}\simeq 2\times 10^9 \msun\,\kpc^{-1}\,M_{12}^{0.77}(1+z)_5^2f_{\rm s,3}\mathcal{M}_{\rm v}^{-1},
\ee 
\be 
\label{eq:Rvir_fil_M18}
R_{\rm v,fil} \simeq 55\kpc\,M_{12}^{0.38}(1+z)_5^{-0.5}f_{\rm s,3}^{0.5}\mathcal{M}_{\rm v}^{-0.5},
\ee 
{\no}where $M_{12}=\Mv/10^{12}\msun$, $(1+z)_5=(1+z)/5$, $\mathcal{M}_{\rm v}=V_{\rm s}/V_{\rm v}\sim 1$ is the inflow velocity of the filament towards the halo in units of the halo virial velocity, and $f_{\rm s,3}=f_{\rm s}/(1/3)\sim 1$ is the fraction of the total accretion onto the halo flowing along a given filament normalized by a fiducial value of $1/3$. These predicted values are only a few tens of percent larger than those found in our simulation, in \se{Line_Mass} and \se{virial} respectively. For a filament in virial equilibrium per-unit-length, \equ{virial_ideal} relates the line-mass to the kinetic energy per-unit-length. The latter is related to the virial temperature through \equ{DM_temp}, $\mathcal{K}\,\sim 1.5k_{\rm B}T_{\rm v} \Lambda_{\rm gas}/(\mu m_{\rm p})$, with $\Lambda_{\rm gas}\sim f_{\rm b}\Lambda_{\rm fil}$ the line-mass of gas within the filament. Taken together with \equ{Lambda_fil_M18}, we obtain an expression for the filament virial temperature, 
\be 
\label{eq:Tvir_fil_M18}
T_{\rm v,fil} \simeq 1.2\times 10^6 \K\,M_{12}^{0.77}(1+z)_5^2f_{\rm s,3}\mathcal{M}_{\rm v}^{-1}.
\ee 
{\no}Compared to the virial temperature of dark matter haloes \citep[e.g.,][]{Dekel.Birnboim.06}, $T_{\rm v}\sim 2.5\times 10^6\K \, M_{12}^{2/3}(1+z)_5$, the virial temperature of filaments increases more rapidly with redshift for a given halo mass. This is consistent with the results of \citet{birnboim2016stability}, who predicted more stable shocks in filaments feeding a given halo mass at higher redshifts. 

\smallskip
We now assume that the initial gas core is comparable to the DM core, which is comparable to $R_{\rm v,fil}$ (\se{DM_prop}), and that this core cools isobarically from $T_{\rm v,fil}$ to $T_{\rm s}\sim 2\times 10^4 \K$. The core density thus increases by a factor of $T_{\rm v,fil}/T_{\rm s}$, and thus mass conservation dictates that the core radius is roughly 
\be 
\label{eq:Rs_fil_M18}
R_{\rm s} \sim \left(T_{\rm v,fil}/T_{\rm s}\right)^{-0.5}R_{\rm v,fil} \simeq 7\kpc\,(1+z)_5^{-1.5},
\ee 
{\no}with no dependence on halo mass, $f_{\rm s}$ or $\mathcal{M}$. This is consistent with the fact that the core radius in our simulation of $R_{\rm s}\sim 10\kpc$ is similar to the core radius in \citet{ramsoy2021rivers} despite the factor $\sim 30$ difference in the halo mass. Associating the filament core (the \textbf{S} zone) with the cold stream that penetrates galaxy haloes, this model predicts a stream radius that scales differently from the predictions of both \citet{mandelker2018cold}, who assumed contraction from $R_{\rm v,fil}$ until full angular momentum support with a constant spin parameter and obtained $R_{\rm s}\propto M_{\rm v}^{0.38}(1+z)^{-0.5}$ (their equation 23), and \citet{mandelker2020LABs}, who assumed thermal pressure equilibrium between the cold stream and the hot CGM at the halo virial temperature and obtained $R_{\rm s}\propto (1+z)^{-1}$ independent of the halo mass (their equations 21 and 27). While all three of these models roughly agree on the radius of streams feeding haloes of $\Mv\sim 10^{12}\msun$ at $z\sim 4$, future studies of the properties of cold streams as a function of halo mass and redshift will be able to distinguish these models and shed further light on what determines the size of cold streams.

\begin{figure}
    \centering
    \includegraphics[trim={0.0cm 0.0cm 0.0cm 0.0cm}, clip, width =0.47 \textwidth]{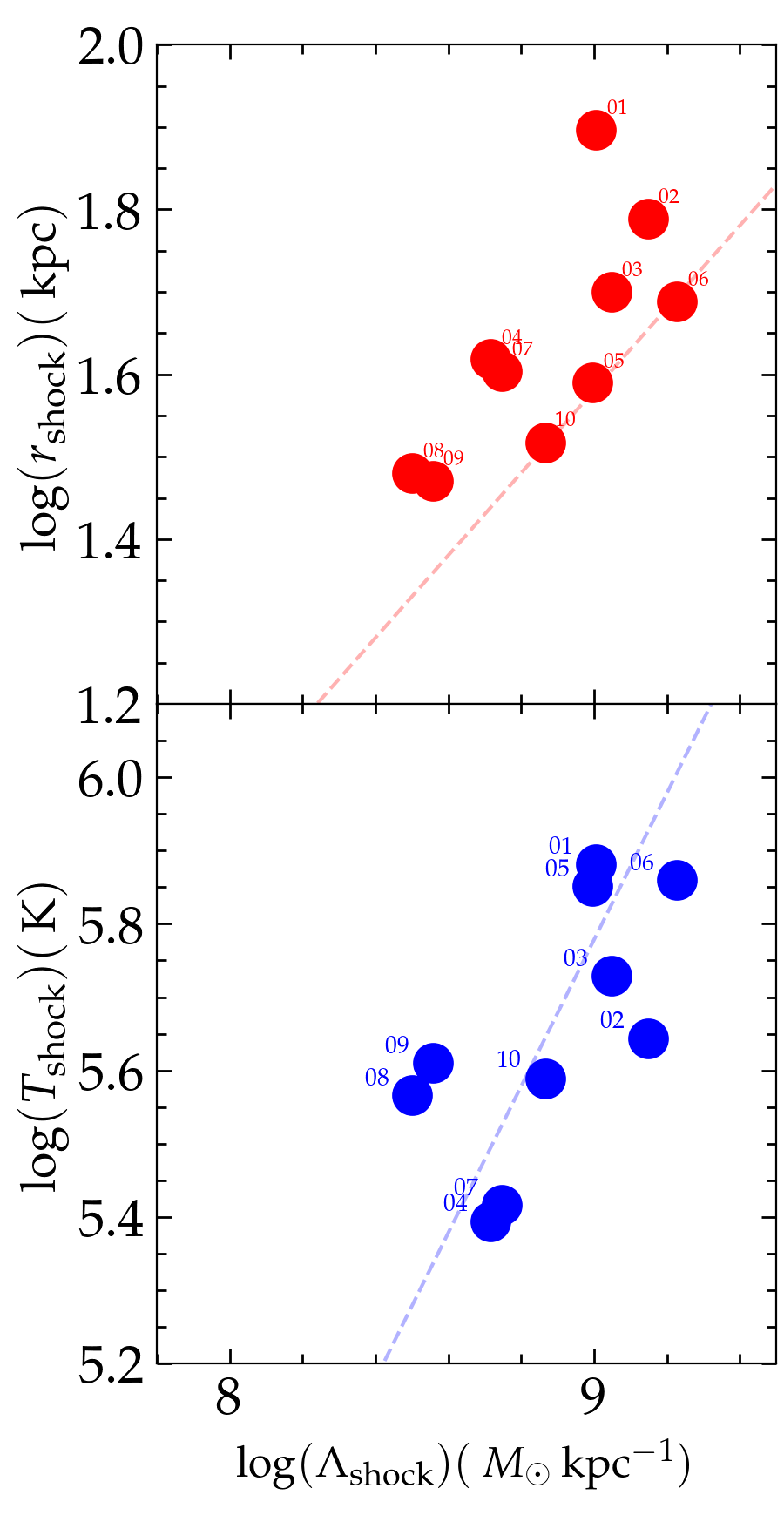}
    \caption{Correlations between the total line-mass within $r_{\rm shock}$, $\Lambda_{\rm shock}$, and the shock radius, $r_{\rm shock}$ (top) as well as the post-shock temperature, $T_{\rm shock}$ (bottom). Points mark the different slices, which are numbered in each figure (see also \fig{rho_0 and T_0 scatterings}). Dashed lines in each panel show our predictions from \equs{Rv_vs_Lambda} and \equm{Tv_vs_Lambda}, respectively, which are both reasonable matches to the simulation data.}
    \label{fig:Lambda_correlations}
\end{figure}

\smallskip
Combining \equs{Lambda_fil_M18}-\equm{Tvir_fil_M18} and setting $z=3.93$ (the redshift for the simulation we used), we obtain relations between the filament virial radius and temperature and the total filament line-mass,
\be
\label{eq:Rv_vs_Lambda}
    R_{\rm v, fil}\approx 40 \kpc \, \Lambda_9^{0.5},
\ee
\be
\label{eq:Tv_vs_Lambda}
    T_{\rm v, fil}\approx 0.6\times 10^6 \Kdegree \, \Lambda_9,
\ee

{\no} where $\Lambda_9\equiv \Lambda_{\rm fil}/10^9 {\rm M_\odot \, kpc^{-1}}$. Associating the filament virial radius with the shock radius (\fig{virial parameter}), we substitute $R_{\rm v, fil}\approx r_{\rm shock}$, $\Lambda_{\rm fil}\approx \Lambda_{\rm tot}(r_{\rm shock})\equiv \Lambda_{\rm shock}$ and $T_{\rm v, fil}\approx T(r_{\rm shock})\equiv T_{\rm shock}$. In \fig{Lambda_correlations}, we plot $r_{\rm shock}$ and $T_{\rm shock}$ versus $\Lambda_{\rm shock}$ for our ten slices. We find \equs{Rv_vs_Lambda}-\equm{Tv_vs_Lambda} to be reasonably good fits to the data, despite slice-to-slice variations and the fact that our measured shock radii seem to be on average $\sim 50\%$ higher than predicted.

\subsubsection{Filament Dynamics}

\smallskip
One of our main findings was the presence of a strong quadrupolar vorticity structure in the filaments, which dominates the filament dynamics and inertial forces in the \textbf{S} and \textbf{V} zones, at $r\lsim 0.6 r_{\rm shock}$ (\fig{summation method another} and \se{toy_model}). Similar quadrupolar vorticity structures in intergalactic filaments have been found in several previous works using a variety of simulation and analysis methods \citep[e.g.,][]{Pichon.Bernardeau.99,Pichon.etal.11,Codis.etal.12,Codis.etal.15b,Laigle.etal.15,ramsoy2021rivers,xia2021intergalactic}. This vortical structure is thought to result from anisotropic accretion onto filaments from surrounding sheets, 
either due to a non-zero impact parameter or to shearing flow through a curved shock. While this seems consistent with our simulations, we defer a more detailed study of the origin and evolution of vorticity to future work. 

\smallskip
Perhaps more intriguing, several recent works have claimed that filaments have a net coherent rotation or spin, giving rise to a helical flow along the filament, which can dominate over the quadrupolar vorticity structure discussed above. This has been claimed both using simulations \citep{xia2021intergalactic} and stacked observations at low-$z$ \citep{WangP.etal.21}. The filaments in our simulations show evidence for mean rotation which actually seems to be the primary support against gravity in the inner regions (\fig{summation method another}). However, the strength of the mean rotation in our simulations is much weaker than the ``residual'' rotation resulting from the quadrupolar vorticity field. Future work with larger samples of filaments of different masses and across different environments and redshifts will be required to study the evolution of filament spin.

\smallskip
Finally, it is worth commenting on the similarities and differences between our force analysis presented in \se{fil_dyn} and the recent force analysis of CGM gas in $\lsim 10^{12}\msun$ haloes at $z\sim 0$ from the FOGGIE simulations \citep{Lochhaas.etal.22}. While our mathematical formalism, based on previous works focusing on the ICM \citep[e.g.,][]{lau2013weighing}, derives directly from the Euler equation and is hence exact, the interpretation of some of the inertial terms can be rather abstract (see the discussion in \se{summation} and \se{sim_dynamics}). On the other hand, \citet{Lochhaas.etal.22} focused on five effective forces operating on the gas, which are all intuitive: gradients of thermal, turbulent, and ram pressure, centrifugal forces, and gravity. This results in a much more intuitive picture, as all the forces are well-defined and have clear physical meanings. However, while their thermal pressure gradients and gravitational forces have a one-to-one correspondence with ours, the other three forces are explicitly tied to a smoothing scale that separates mean from residual motions and serves only as approximations of the full shear tensor, which is accurately captured by the inertial forces in our method. 
Therefore, the exact force balance is not guaranteed in their method, even in the absence of temporal accelerations. Despite these differences and the very different systems studied (high-$z$ intergalactic filaments versus low-$z$ CGM), many of our results are similar. In particular, we both find that the outskirts of the system are primarily supported against gravity by thermal pressure gradients, while kinematic (inertial) forces dominate the inner regions and are not in a steady-state equilibrium. Moreover, we both find that locally, turbulent and ram pressure forces can act either outwards (opposing gravity) or inwards (aiding gravity). A more detailed comparison between these two mathematical methods of force decomposition is left for future work. This highlights the similarity between gaseous atmospheres in different cosmic-web elements and makes it tempting to describe the post-shock region in our filaments as a `\textit{circumfilamentary medium}' (CFM) surrounding the central cold stream within the potential well of the dark matter filament, much like the CGM surrounding the central galaxy within the potential well of the dark matter halo. 

\subsection{Implications for Cold Streams}
\label{sec:cold_streams}

\smallskip
The \textbf{S} zone of filaments is a dense, cold, isothermal core, where the density and temperature are roughly $n_{\rm H}\sim 10^{-2}\cmc$ and $T\sim 2\times 10^4\,{\rm K}$ (\se{thermal}, \figs{thermal profiles} - \figss{rho_0 and T_0 scatterings}). This core represents the `cold streams', predicted to be the main mode of gas accretion onto massive high-$z$ galaxies and to penetrate the virial accretion shocks around their dark matter haloes \citep{Dekel.Birnboim.06,dekel2009cold}. As described above, the radius, density, and temperature of this region are consistent with predictions from various analytic models for the properties of cold streams as a function of the halo mass and redshift. 

\smallskip
Several recent studies have attempted to model the evolution of cold streams as they travel towards massive central galaxies while interacting with the hot CGM, subject to various (magneto)hydrodynamic, thermal, and gravitational instabilities using analytic models and idealised numerical simulations \citep{mandelker.etal.2016,mandelker2018cold,mandelker.etal.2019,mandelker2020KHI4,mandelker2020LABs,padnos.etal.2018,Aung.etal.19,berlok.pfrommer.2019}. While these models are adding more and more physics and thus becoming more realistic, all of them have modelled the streams as initially laminar, hydrostatic systems supported predominantly by thermal pressure. While \citet{mandelker2018cold} did account for mild turbulence with Mach number $\mathcal{M}_{\rm turb}\lsim 1$ driven by accretion onto the filament from the surrounding sheet, this is still a fundamentally different picture from the highly turbulent environment of cold streams in our simulations, which is dominated by vorticity and accretion and is surrounded by a cylindrical accretion shock. These must be accounted for in future work studying the stability and evolution of intergalactic filaments and cold streams. In particular, future models should account for internal filament dynamics such as rotation and turbulence, and for the confining ram pressure from the accreting gas, along with turbulence in the confining medium. 

\smallskip
Finally, while the cold streams are dominated by non-thermal, turbulent, and vortical motions, they are in pressure equilibrium with the post-shock volume-filling CFM gas. This implies that as the streams penetrate the hot CGM of massive dark matter haloes, where both the densities and temperatures are larger, their confining pressure will increase by a factor of several. Indeed, we find in our simulations that the confining pressure of filaments jumps by a factor of $\sim (10-20)$ as they penetrate the virial shocks around $\gsim 10^{12}\msun$ haloes at $z\sim 4$ (Lu et al., in prep.). First, it is unclear what happens to the hot CFM gas as the cold stream penetrates the halo. Does it stay behind and accumulate at the virial radius, or does it partially penetrate as well, dragged along by the cold stream? Second, this can have important implications for the morphology of cold gas in the outer CGM of such haloes, since such a sudden increase in confining pressure may cause the streams to `shatter' into tiny fragments of order of the cooling length, $l_{\rm cool}\sim c_{\rm s}t_{\rm cool}$, with $c_{\rm s}$ and $t_{\rm cool}$ the sound speed and cooling time of gas with $T\gsim 10^4\K$ \citep[e.g.,][]{McCourt.etal.2018,Gronke.Oh.2020,Gronke.Oh.2022}. Similar phenomena have been predicted by other models of streams that penetrate virial shocks, although for different reasons \citep{Cornuault.etal.2018}. If the streams do shatter as they penetrate the virial shock, this may explain puzzling observations of large area covering fractions and small volume filling fractions of cold gas in the CGM of massive high-$z$ galaxies, which are consistent with a shattered mist-like collection of small cloudlets \citep[e.g.,][]{Borisova.etal.2016,Cantalupo.etal.2019,Pezzulli.etal.2019}. The process of how cold streams first penetrate the hot CGM around massive galaxies remains poorly understood and will be the subject of future work using idealised simulations of stream-shock interactions and shattering of filamentary systems (Yao, Mandelker, et al., in preparation).

\section{Summary and Conclusions}
\label{sec:conclusions}

\smallskip
Using a novel high-resolution cosmological simulation, \hypa{}, that zooms in on a large patch of the cosmic-web in a large-scale proto-group environment, we study the structure and internal dynamics of sections of $\sim \Mpc$-scale intergalactic filaments feeding three $\sim 10^{12}\msun$ haloes at $z\sim 4$. We select ten slices $30\kpc$ in length from three independent filaments that feed these three haloes and are embedded in the same cosmic sheet, so that no slice intersects any halo with $\Mv>10^{10}\msun$. We then study these slices independently and in stacked form to reach the following conclusions:

\begin{enumerate}

    \item \textbf{Radial zones:} The filaments can be broadly characterised by three radial zones (\fig{cartoon}). We summarise the key properties of these zones here, while further below we offer a more detailed summary of each property over the entire filament.
    
     \item[$\bullet$] In the outer ``\textit{thermal}'' (\textbf{T}) zone, a cylindrical accretion shock heats the gas to the virial temperature associated with the potential well of the dark-matter filament, and strong thermal pressure forces balance both inwardsgravitational and ram pressure forces resulting from the inflowing gas. Near $r_{\rm shock}$, excess thermal pressure forces cause the shock to expand, though the gas continues to flow with a roughly constant mass accretion rate throughout the \textbf{T} zone, both on and off the sheet, due to cooling of the gas behind the shock. The gas outside the sheet decelerates throughout this zone, while the sheet gas maintains a roughly constant inflow velocity. 
    
    \item[$\bullet$] In the intermediate ``\textit{vortex}'' (\textbf{V}) zone, the filament dynamics is dominated by a quadrupolar vortex structure in the velocity field, induced by the flow of gas through the sheet towards the central filament. 
    The outwards centrifugal force due to the global net rotation is already comparable to the gravitational force inward, but it is small compared to the centrifugal forces due to the residual rotations associated with the quadrupolar vortices. The latter are largely balanced by shear and turbulent forces. These vortices lead to net outflows outside the sheet, and the induced mixing causes the gas inflowing along the sheet to decelerate. 
    
    \item[$\bullet$] In the inner ``\textit{stream}'' (\textbf{S}) zone, a dense isothermal core forms from an isobaric cooling-flow resulting from post-shock cooling in a freefall time, and is associated with a decrease in outwards forces. The size of this region expands with time, though the gas within this zone can move both inwards and outwards due to the impact parameter of the sheet and the resulting vorticity. \textit{The \textbf{S} zone represents the cold streams that feed massive galaxies from the cosmic web}. 

    \item[] While this basic structure appears generic, the locations of the boundaries between these zones depend on filament line-mass (\se{comparison}). In our case the \textbf{T} zone is at $r\gsim 0.65r_{\rm shock}$ and the \textbf{S} zone is at $r\lsim 0.25r_{\rm shock}$.
    
    \item \textbf{Thermal structure:} The gas density profiles are well fit by the density profile of a self-gravitating isothermal filament with a temperature of $\sim 3\times 10^4\K$. However, the filaments are not isothermal, and the temperature increases by a factor of $\sim (20-30)$ from the core in the \textbf{S} zone to the cylindrical accretion shock outside the \textbf{T} zone, at $r_{\rm shock}\sim 50\kpc$ (\figs{thermal profiles}-\figss{rho_0 and T_0 scatterings}). The isothermal core in the \textbf{S} zone is characterised by densities $n_{\rm H}\sim 0.01\cmc$, temperatures $T\sim 3\times 10^4\K$, and sizes $\sim 0.25\,r_{\rm shock}\sim 10\kpc$, consistent with predictions for cold streams that feed massive high-$z$ galaxies from cosmic-web filaments \citep{mandelker2018cold,mandelker2020LABs}. The thermal pressure in the filament is roughly constant within the \textbf{S} and \textbf{V} zones, suggesting that the confining pressure of cold streams in the IGM is $\sim (10-20)$ times smaller than their confining pressure once they penetrate the CGM of massive haloes. Outside $r_{\rm shock}$, the density becomes dominated by the underlying sheet, which penetrates and feeds the filament much the same way that filaments penetrate and feed haloes.
    
    \item \textbf{Dark matter structure:} The gaseous filaments are embedded in dark matter filaments that set the gravitational potential wells. These are also well-fitted by models of self-gravitating isothermal cylinders, and dark matter is much more isothermal than gas with its effective temperature (i.e., kinetic energy or velocity dispersion squared) varying by a factor of $\lsim 2$ over the range $(0.1-1.0)r_{\rm shock}$ (\fig{DM prop}). The effective temperature of dark matter is very similar to the post-shock temperature of the gas, $T\sim 8\times 10^5\K$, yielding a core radius that is $\sim 4$ times larger than that of the gas (the \textbf{S} zone). However, this result is a function of the filament line-mass (\se{comparison}). The baryon fraction near $r_{\rm shock}$ approaches the universal value, while in the \textbf{S} zone $f_{\rm b}\gsim 0.5$ (\fig{fbar}). 

    \item \textbf{Line-mass:} The line-mass (mass per unit-length) of the dark matter filament is $<10\%$ of the maximal line-mass for hydrostatic equilibrium in a self-gravitating isothermal cylinder at the temperature of the dark matter (\fig{line mass}). The line-mass for the gas is $>30\%$ of this threshold, but still below it, which implies that intergalactic filaments are typically stable against gravitational fragmentation\footnote{Though they may still become unstable in the inner CGM \citep{mandelker2018cold,Aung.etal.19}.}. 
    
    \item \textbf{Thermal equilibrium:} The cooling time near $r_{\rm shock}$ is only slightly longer than the free fall time, $t_{\rm cool}\sim 3t_{\rm ff}$, implying that the post-shock state is unstable to thermal condensation and the formation of multiphase gas. This prevents the gas in the \textbf{T} zone from maintaining hydrostatic equilibrium and results in a cooling flow towards the cold-stream in the \textbf{S} zone, where $t_{\rm cool}<t_{\rm ff}$ (\fig{tcool_tff}).

    \item \textbf{Virial equilibrium:} The filament gas is in virial equilibrium \textit{per-unit-length}, once we account for surface-pressure terms and remove the gas that is rapidly inflowing toward the filament along the sheet, with $|v_r|>v_{\rm vir} = \sqrt{G\Lambda}$ (\figs{radial inflows}-\figss{virial parameter}). The virial radius of the filament is $r_{\rm vir}\sim r_{\rm shock}$. The associated virial temperature is approximately equal to the post-shock temperature, which is also roughly the dark matter effective temperature, suggesting that the dark matter is in virial equilibrium per unit-length as well. Throughout the \textbf{T} zone, the gas maintains the local virial temperature, with non-thermal motions offset by surface pressure. However, in the \textbf{V} and \textbf{S} zones these motions increase faster than the surface pressure with respect to the gravitational potential energy, resulting in net radial outflows induced by the vortices and the motion of the sheet. On the other hand, the filament is unbound along its axis, at all radii. 
    
    \item \textbf{Filament dynamics:} The component of vorticity along the filament axis has a characteristic quadrupolar structure, with two sets of counter-rotating vortices centred near the boundary between the \textbf{T} and \textbf{V} zones (\figs{streamplot}-\figss{vorticity}). However, despite a slight preference for the filament vorticity to align with its axis, there is substantial vorticity along the other axes as well (\fig{vorticity}, right). In the \textbf{T} zone, the filament dynamics is dominated by accretion onto the filament (\fig{streamplot}), with most of the mass flux flowing along the sheet (\fig{radial inflows}), whose impact parameter with respect to the filament centre seems related to the formation of the aforementioned vortices.

    \item \textbf{Dynamical equilibrium:} We quantify the various force terms acting on the filament gas by decomposing the Euler equation into gravitational forces, thermal pressure forces and inertial forces. The latter is composed of ram, turbulent pressure, and shear forces. We find that over a large radial range, these forces balance each other, and the velocity field within the filament is in a steady-state (\fig{summation method}). However, near the location of the vortices in the outer \textbf{V} zone, and near the location of the shock in the outer \textbf{T} zone, the radial velocity field accelerates outward. Similarly, in the \textbf{S} zone, the velocity field accelerates inwards, damping transient outflows caused by the impact parameter of the sheet. In the \textbf{T} zone, the outwards thermal pressure forces roughly balance the inwardsgravity and the ram pressure forces (\fig{summation method another}). In the \textbf{V} zone, where the dynamics is dominated by quadrupolar vortices, thermal pressure forces are comparable to centrifugal forces due to net average rotation, and these combine to provide support against gravity. At the same time, centrifugal forces due to residual rotation associated with the vortices are cancelledcanceled by shear forces and ram/turbulent pressure (\figs{summation method another}, \figss{ideal vortex results}). In the \textbf{S} zone, a drop in both centrifugal and thermal pressure forces, due to the development of a strong cooling flow, results in a net inwards acceleration of the velocity field. 
    
\end{enumerate}

\smallskip
While our analysis has been comprehensive and these results are enlightening, we stress that the properties of the filaments studied in this work certainly depend on the mass of the filament, the redshift, and the environment \citep[see, e.g.,][]{ramsoy2021rivers}. Future studies using similarly high-resolution simulations and employing similar analysis methods to study a much larger sample of filaments across different redshifts and environments will be necessary to elucidate the properties of these extremely important elements of the cosmic-web and their implications for galaxy formation.

\section*{Acknowledgements}

We thank the anonymous referee for their detailed reading of the paper and many constructive comments which improved our paper. We thank Han Aung, Yuval Birnboim, Dusan Keres, Mark Krumholz, Erwin Lau, Cassie Lochhaas, R\"udiger Pakmor, Mark Voit for helpful and insightful discussions. The simulations analysed in this work were funded by the Klauss Tschira Foundation, through the HITS-Yale Program in Astrophysics (HYPA). NM acknowledges support from ISF grant 3061/21 and from BSF grant 2020302, and partial support from the Gordon and Betty Moore Foundation through Grant GBMF7392 and from the National Science Foundation under Grant No. NSF PHY1748958. SPO is supported by NASA grant 19-ATP19-0205, and NSF grant AST-1911198. AD is supported by ISF grant 861/20. FvdB is supported by the National Aeronautics and Space Administration through Grant No. 19-ATP19-0059 issued as part of the Astrophysics Theory Program. DN is supported by NSF (AST-2206055) and NASA (80NSSC22K0821 \& TM3-24007X) grants. FvdV is supported by a Royal Society University Research Fellowship (URF\textbackslash R1\textbackslash191703).

\section*{Data Availability}

Simulation data will be shared upon reasonable request. The data supporting the plots within this article are available on reasonable request to the corresponding author.


\bibliographystyle{mnras}
\bibliography{references.bib} 

\begin{thebibliography}{}
\makeatletter
\relax
\def\mn@urlcharsother{\let\do\@makeother \do\$\do\&\do\#\do\^\do\_\do\%\do\~}
\def\mn@doi{\begingroup\mn@urlcharsother \@ifnextchar [ {\mn@doi@} {\mn@doi@[]}}
\def\mn@doi@[#1]#2{\def\@tempa{#1}\ifx\@tempa\@empty \href {http://dx.doi.org/#2} {doi:#2}\else \href {http://dx.doi.org/#2} {#1}\fi \endgroup}
\def\mn@eprint#1#2{\mn@eprint@#1:#2::\@nil}
\def\mn@eprint@arXiv#1{\href {http://arxiv.org/abs/#1} {{\tt arXiv:#1}}}
\def\mn@eprint@dblp#1{\href {http://dblp.uni-trier.de/rec/bibtex/#1.xml} {dblp:#1}}
\def\mn@eprint@#1:#2:#3:#4\@nil{\def\@tempa {#1}\def\@tempb {#2}\def\@tempc {#3}\ifx \@tempc \@empty \let \@tempc \@tempb \let \@tempb \@tempa \fi \ifx \@tempb \@empty \def\@tempb {arXiv}\fi \@ifundefined {mn@eprint@\@tempb}{\@tempb:\@tempc}{\expandafter \expandafter \csname mn@eprint@\@tempb\endcsname \expandafter{\@tempc}}}

\bibitem[\protect\citeauthoryear{{Aung}, {Mandelker}, {Nagai}, {Dekel}  \& {Birnboim}}{{Aung} et~al.}{2019}]{Aung.etal.19}
{Aung} H.,  {Mandelker} N.,  {Nagai} D.,  {Dekel} A.,   {Birnboim} Y.,  2019, \mn@doi [\mnras] {10.1093/mnras/stz1964}, \href {https://ui.adsabs.harvard.edu/abs/2019MNRAS.490..181A} {490, 181}

\bibitem[\protect\citeauthoryear{{Aung}, {Nagai}  \& {Lau}}{{Aung} et~al.}{2021}]{aung2021shock}
{Aung} H.,  {Nagai} D.,   {Lau} E.~T.,  2021, \mn@doi [\mnras] {10.1093/mnras/stab2598}, \href {https://ui.adsabs.harvard.edu/abs/2021MNRAS.508.2071A} {508, 2071}

\bibitem[\protect\citeauthoryear{{Bardeen}, {Bond}, {Kaiser}  \& {Szalay}}{{Bardeen} et~al.}{1986}]{bbks1986}
{Bardeen} J.~M.,  {Bond} J.~R.,  {Kaiser} N.,   {Szalay} A.~S.,  1986, \mn@doi [\apj] {10.1086/164143}, \href {https://ui.adsabs.harvard.edu/abs/1986ApJ...304...15B} {304, 15}

\bibitem[\protect\citeauthoryear{{Bennett} \& {Sijacki}}{{Bennett} \& {Sijacki}}{2020}]{Bennett.Sijacki.20}
{Bennett} J.~S.,  {Sijacki} D.,  2020, \mn@doi [\mnras] {10.1093/mnras/staa2835}, \href {https://ui.adsabs.harvard.edu/abs/2020MNRAS.tmp.2658B} {}

\bibitem[\protect\citeauthoryear{{Berlok} \& {Pfrommer}}{{Berlok} \& {Pfrommer}}{2019}]{berlok.pfrommer.2019}
{Berlok} T.,  {Pfrommer} C.,  2019, \mn@doi [\mnras] {10.1093/mnras/stz2347}, \href {https://ui.adsabs.harvard.edu/abs/2019MNRAS.489.3368B} {489, 3368}

\bibitem[\protect\citeauthoryear{{Bertschinger}}{{Bertschinger}}{1989}]{Bertschinger.1989}
{Bertschinger} E.,  1989, \mn@doi [\apj] {10.1086/167428}, \href {https://ui.adsabs.harvard.edu/abs/1989ApJ...340..666B} {340, 666}

\bibitem[\protect\citeauthoryear{{Birnboim} \& {Dekel}}{{Birnboim} \& {Dekel}}{2003}]{Birnboim2003shocks}
{Birnboim} Y.,  {Dekel} A.,  2003, \mn@doi [\mnras] {10.1046/j.1365-8711.2003.06955.x}, \href {https://ui.adsabs.harvard.edu/abs/2003MNRAS.345..349B} {345, 349}

\bibitem[\protect\citeauthoryear{{Birnboim}, {Padnos}  \& {Zinger}}{{Birnboim} et~al.}{2016}]{birnboim2016stability}
{Birnboim} Y.,  {Padnos} D.,   {Zinger} E.,  2016, \mn@doi [\apjl] {10.3847/2041-8205/832/1/L4}, \href {https://ui.adsabs.harvard.edu/abs/2016ApJ...832L...4B} {832, L4}

\bibitem[\protect\citeauthoryear{{Bond}, {Kofman}  \& {Pogosyan}}{{Bond} et~al.}{1996}]{Bond.etal.96}
{Bond} J.~R.,  {Kofman} L.,   {Pogosyan} D.,  1996, \mn@doi [\nat] {10.1038/380603a0}, \href {https://ui.adsabs.harvard.edu/abs/1996Natur.380..603B} {380, 603}

\bibitem[\protect\citeauthoryear{{Borisova} et~al.,}{{Borisova} et~al.}{2016}]{Borisova.etal.2016}
{Borisova} E.,  et~al., 2016, \mn@doi [\apj] {10.3847/0004-637X/831/1/39}, \href {https://ui.adsabs.harvard.edu/abs/2016ApJ...831...39B} {831, 39}

\bibitem[\protect\citeauthoryear{{Bouch{\'e}}, {Murphy}, {Kacprzak}, {P{\'e}roux}, {Contini}, {Martin}  \& {Dessauges-Zavadsky}}{{Bouch{\'e}} et~al.}{2013}]{Bouche.etal.13}
{Bouch{\'e}} N.,  {Murphy} M.~T.,  {Kacprzak} G.~G.,  {P{\'e}roux} C.,  {Contini} T.,  {Martin} C.~L.,   {Dessauges-Zavadsky} M.,  2013, \mn@doi [Science] {10.1126/science.1234209}, \href {http://adsabs.harvard.edu/abs/2013Sci...341...50B} {341, 50}

\bibitem[\protect\citeauthoryear{{Bouch{\'e}} et~al.,}{{Bouch{\'e}} et~al.}{2016}]{Bouche.etal.16}
{Bouch{\'e}} N.,  et~al., 2016, \mn@doi [\apj] {10.3847/0004-637X/820/2/121}, \href {http://adsabs.harvard.edu/abs/2016ApJ...820..121B} {820, 121}

\bibitem[\protect\citeauthoryear{{Bournaud} \& {Elmegreen}}{{Bournaud} \& {Elmegreen}}{2009}]{Bournaud.Elmegreen.09}
{Bournaud} F.,  {Elmegreen} B.~G.,  2009, \mn@doi [\apjl] {10.1088/0004-637X/694/2/L158}, \href {https://ui.adsabs.harvard.edu/abs/2009ApJ...694L.158B} {694, L158}

\bibitem[\protect\citeauthoryear{{Bryan} \& {Norman}}{{Bryan} \& {Norman}}{1998}]{Bryan98}
{Bryan} G.~L.,  {Norman} M.~L.,  1998, \mn@doi [\apj] {10.1086/305262}, \href {http://adsabs.harvard.edu/abs/1998ApJ...495...80B} {495, 80}

\bibitem[\protect\citeauthoryear{{Cai} et~al.,}{{Cai} et~al.}{2017}]{Cai.etal.2017}
{Cai} Z.,  et~al., 2017, \mn@doi [\apj] {10.3847/1538-4357/aa5d14}, \href {https://ui.adsabs.harvard.edu/abs/2017ApJ...837...71C} {837, 71}

\bibitem[\protect\citeauthoryear{{Cantalupo}, {Arrigoni-Battaia}, {Prochaska}, {Hennawi}  \& {Madau}}{{Cantalupo} et~al.}{2014}]{Cantalupo.etal.14}
{Cantalupo} S.,  {Arrigoni-Battaia} F.,  {Prochaska} J.~X.,  {Hennawi} J.~F.,   {Madau} P.,  2014, \mn@doi [\nat] {10.1038/nature12898}, \href {http://adsabs.harvard.edu/abs/2014Natur.506...63C} {506, 63}

\bibitem[\protect\citeauthoryear{{Cantalupo} et~al.,}{{Cantalupo} et~al.}{2019}]{Cantalupo.etal.2019}
{Cantalupo} S.,  et~al., 2019, \mn@doi [\mnras] {10.1093/mnras/sty3481}, \href {https://ui.adsabs.harvard.edu/abs/2019MNRAS.483.5188C} {483, 5188}

\bibitem[\protect\citeauthoryear{{Cautun}, {van de Weygaert}, {Jones}  \& {Frenk}}{{Cautun} et~al.}{2014}]{Cautun.etal.14}
{Cautun} M.,  {van de Weygaert} R.,  {Jones} B. J.~T.,   {Frenk} C.~S.,  2014, \mn@doi [\mnras] {10.1093/mnras/stu768}, \href {https://ui.adsabs.harvard.edu/abs/2014MNRAS.441.2923C} {441, 2923}

\bibitem[\protect\citeauthoryear{{Ceverino}, {Dekel}  \& {Bournaud}}{{Ceverino} et~al.}{2010}]{Ceverino.etal.10}
{Ceverino} D.,  {Dekel} A.,   {Bournaud} F.,  2010, \mn@doi [\mnras] {10.1111/j.1365-2966.2010.16433.x}, \href {https://ui.adsabs.harvard.edu/abs/2010MNRAS.404.2151C} {404, 2151}

\bibitem[\protect\citeauthoryear{{Chen} et~al.,}{{Chen} et~al.}{2015}]{Chen.etal.15}
{Chen} Y.-C.,  et~al., 2015, \mn@doi [\mnras] {10.1093/mnras/stv2260}, \href {https://ui.adsabs.harvard.edu/abs/2015MNRAS.454.3341C} {454, 3341}

\bibitem[\protect\citeauthoryear{{Codis}, {Pichon}, {Devriendt}, {Slyz}, {Pogosyan}, {Dubois}  \& {Sousbie}}{{Codis} et~al.}{2012}]{Codis.etal.12}
{Codis} S.,  {Pichon} C.,  {Devriendt} J.,  {Slyz} A.,  {Pogosyan} D.,  {Dubois} Y.,   {Sousbie} T.,  2012, \mn@doi [\mnras] {10.1111/j.1365-2966.2012.21636.x}, \href {https://ui.adsabs.harvard.edu/abs/2012MNRAS.427.3320C} {427, 3320}

\bibitem[\protect\citeauthoryear{{Codis} et~al.,}{{Codis} et~al.}{2015a}]{Codis.etal.15}
{Codis} S.,  et~al., 2015a, \mn@doi [\mnras] {10.1093/mnras/stv231}, \href {https://ui.adsabs.harvard.edu/abs/2015MNRAS.448.3391C} {448, 3391}

\bibitem[\protect\citeauthoryear{{Codis}, {Pichon}  \& {Pogosyan}}{{Codis} et~al.}{2015b}]{Codis.etal.15b}
{Codis} S.,  {Pichon} C.,   {Pogosyan} D.,  2015b, \mn@doi [\mnras] {10.1093/mnras/stv1570}, \href {https://ui.adsabs.harvard.edu/abs/2015MNRAS.452.3369C} {452, 3369}

\bibitem[\protect\citeauthoryear{{Colless} et~al.,}{{Colless} et~al.}{2001}]{Colless2001}
{Colless} M.,  et~al., 2001, \mn@doi [\mnras] {10.1046/j.1365-8711.2001.04902.x}, \href {https://ui.adsabs.harvard.edu/abs/2001MNRAS.328.1039C} {328, 1039}

\bibitem[\protect\citeauthoryear{{Cornuault}, {Lehnert}, {Boulanger}  \& {Guillard}}{{Cornuault} et~al.}{2018}]{Cornuault.etal.2018}
{Cornuault} N.,  {Lehnert} M.~D.,  {Boulanger} F.,   {Guillard} P.,  2018, \mn@doi [\aap] {10.1051/0004-6361/201629229}, \href {https://ui.adsabs.harvard.edu/abs/2018A&A...610A..75C} {610, A75}

\bibitem[\protect\citeauthoryear{{Daddi} et~al.,}{{Daddi} et~al.}{2021}]{Daddi.etal.20}
{Daddi} E.,  et~al., 2021, \mn@doi [\aap] {10.1051/0004-6361/202038700}, \href {https://ui.adsabs.harvard.edu/abs/2021A&A...649A..78D} {649, A78}

\bibitem[\protect\citeauthoryear{{Danovich}, {Dekel}, {Hahn}  \& {Teyssier}}{{Danovich} et~al.}{2012}]{Danovich.etal.12}
{Danovich} M.,  {Dekel} A.,  {Hahn} O.,   {Teyssier} R.,  2012, \mn@doi [\mnras] {10.1111/j.1365-2966.2012.20751.x}, \href {http://adsabs.harvard.edu/abs/2012MNRAS.422.1732D} {422, 1732}

\bibitem[\protect\citeauthoryear{{Danovich}, {Dekel}, {Hahn}, {Ceverino}  \& {Primack}}{{Danovich} et~al.}{2015}]{Danovich.etal.15}
{Danovich} M.,  {Dekel} A.,  {Hahn} O.,  {Ceverino} D.,   {Primack} J.,  2015, \mn@doi [\mnras] {10.1093/mnras/stv270}, \href {http://adsabs.harvard.edu/abs/2015MNRAS.449.2087D} {449, 2087}

\bibitem[\protect\citeauthoryear{{Dekel} \& {Birnboim}}{{Dekel} \& {Birnboim}}{2006}]{Dekel.Birnboim.06}
{Dekel} A.,  {Birnboim} Y.,  2006, \mn@doi [\mnras] {10.1111/j.1365-2966.2006.10145.x}, \href {https://ui.adsabs.harvard.edu/abs/2006MNRAS.368....2D} {368, 2}

\bibitem[\protect\citeauthoryear{Dekel et~al.,}{Dekel et~al.}{2009a}]{dekel2009cold}
Dekel A.,  et~al., 2009a, Nature, 457, 451

\bibitem[\protect\citeauthoryear{{Dekel}, {Sari}  \& {Ceverino}}{{Dekel} et~al.}{2009b}]{Dekel.etal.09b}
{Dekel} A.,  {Sari} R.,   {Ceverino} D.,  2009b, \mn@doi [\apj] {10.1088/0004-637X/703/1/785}, \href {https://ui.adsabs.harvard.edu/abs/2009ApJ...703..785D} {703, 785}

\bibitem[\protect\citeauthoryear{{Dekel}, {Zolotov}, {Tweed}, {Cacciato}, {Ceverino}  \& {Primack}}{{Dekel} et~al.}{2013}]{Dekel.etal.13}
{Dekel} A.,  {Zolotov} A.,  {Tweed} D.,  {Cacciato} M.,  {Ceverino} D.,   {Primack} J.~R.,  2013, \mn@doi [\mnras] {10.1093/mnras/stt1338}, \href {http://adsabs.harvard.edu/abs/2013MNRAS.435..999D} {435, 999}

\bibitem[\protect\citeauthoryear{{Dolag}, {Borgani}, {Murante}  \& {Springel}}{{Dolag} et~al.}{2009}]{Dolag09}
{Dolag} K.,  {Borgani} S.,  {Murante} G.,   {Springel} V.,  2009, \mn@doi [\mnras] {10.1111/j.1365-2966.2009.15034.x}, \href {https://ui.adsabs.harvard.edu/abs/2009MNRAS.399..497D} {399, 497}

\bibitem[\protect\citeauthoryear{{Eckert} et~al.,}{{Eckert} et~al.}{2015}]{Eckert.etal.2015}
{Eckert} D.,  et~al., 2015, \mn@doi [\nat] {10.1038/nature16058}, \href {https://ui.adsabs.harvard.edu/abs/2015Natur.528..105E} {528, 105}

\bibitem[\protect\citeauthoryear{{Elmegreen}, {Elmegreen}, {Ravindranath}  \& {Coe}}{{Elmegreen} et~al.}{2007}]{Elmegreen.etal.07}
{Elmegreen} D.~M.,  {Elmegreen} B.~G.,  {Ravindranath} S.,   {Coe} D.~A.,  2007, \mn@doi [\apj] {10.1086/511667}, \href {https://ui.adsabs.harvard.edu/abs/2007ApJ...658..763E} {658, 763}

\bibitem[\protect\citeauthoryear{{Emonts} et~al.,}{{Emonts} et~al.}{2023}]{Emonts.etal.2023}
{Emonts} B. H.~C.,  et~al., 2023, \mn@doi [arXiv e-prints] {10.48550/arXiv.2303.17484}, \href {https://ui.adsabs.harvard.edu/abs/2023arXiv230317484E} {p. arXiv:2303.17484}

\bibitem[\protect\citeauthoryear{{Farina} et~al.,}{{Farina} et~al.}{2017}]{Farina.etal.17}
{Farina} E.~P.,  et~al., 2017, \mn@doi [\apj] {10.3847/1538-4357/aa8df4}, \href {https://ui.adsabs.harvard.edu/abs/2017ApJ...848...78F} {848, 78}

\bibitem[\protect\citeauthoryear{{Faucher-Gigu{\`e}re}, {Lidz}, {Zaldarriaga}  \& {Hernquist}}{{Faucher-Gigu{\`e}re} et~al.}{2009}]{FG09}
{Faucher-Gigu{\`e}re} C.-A.,  {Lidz} A.,  {Zaldarriaga} M.,   {Hernquist} L.,  2009, \mn@doi [\apj] {10.1088/0004-637X/703/2/1416}, \href {https://ui.adsabs.harvard.edu/abs/2009ApJ...703.1416F} {703, 1416}

\bibitem[\protect\citeauthoryear{{Fielding}, {Quataert}, {McCourt}  \& {Thompson}}{{Fielding} et~al.}{2017}]{Fielding2017CGM}
{Fielding} D.,  {Quataert} E.,  {McCourt} M.,   {Thompson} T.~A.,  2017, \mn@doi [\mnras] {10.1093/mnras/stw3326}, \href {https://ui.adsabs.harvard.edu/abs/2017MNRAS.466.3810F} {466, 3810}

\bibitem[\protect\citeauthoryear{{Fillmore} \& {Goldreich}}{{Fillmore} \& {Goldreich}}{1984}]{FG84}
{Fillmore} J.~A.,  {Goldreich} P.,  1984, \mn@doi [\apj] {10.1086/162070}, \href {https://ui.adsabs.harvard.edu/abs/1984ApJ...281....1F} {281, 1}

\bibitem[\protect\citeauthoryear{{Forbes} et~al.,}{{Forbes} et~al.}{2022}]{Forbes.etal.22}
{Forbes} J.~C.,  et~al., 2022, arXiv e-prints, \href {https://ui.adsabs.harvard.edu/abs/2022arXiv220405344F} {p. arXiv:2204.05344}

\bibitem[\protect\citeauthoryear{{F{\"o}rster Schreiber} et~al.,}{{F{\"o}rster Schreiber} et~al.}{2006}]{ForsterSchreiber.etal.06}
{F{\"o}rster Schreiber} N.~M.,  et~al., 2006, \mn@doi [\apj] {10.1086/504403}, \href {https://ui.adsabs.harvard.edu/abs/2006ApJ...645.1062F} {645, 1062}

\bibitem[\protect\citeauthoryear{{F{\"o}rster Schreiber} et~al.,}{{F{\"o}rster Schreiber} et~al.}{2009}]{ForsterSchreiber.etal.09}
{F{\"o}rster Schreiber} N.~M.,  et~al., 2009, \mn@doi [\apj] {10.1088/0004-637X/706/2/1364}, \href {https://ui.adsabs.harvard.edu/abs/2009ApJ...706.1364F} {706, 1364}

\bibitem[\protect\citeauthoryear{{Fumagalli}, {Prochaska}, {Kasen}, {Dekel}, {Ceverino}  \& {Primack}}{{Fumagalli} et~al.}{2011}]{Fumagalli.etal.11}
{Fumagalli} M.,  {Prochaska} J.~X.,  {Kasen} D.,  {Dekel} A.,  {Ceverino} D.,   {Primack} J.~R.,  2011, \mn@doi [\mnras] {10.1111/j.1365-2966.2011.19599.x}, \href {http://adsabs.harvard.edu/abs/2011MNRAS.418.1796F} {418, 1796}

\bibitem[\protect\citeauthoryear{{Fumagalli} et~al.,}{{Fumagalli} et~al.}{2017}]{Fumagalli.etal.17}
{Fumagalli} M.,  et~al., 2017, \mn@doi [\mnras] {10.1093/mnras/stx1896}, \href {http://adsabs.harvard.edu/abs/2017MNRAS.471.3686F} {471, 3686}

\bibitem[\protect\citeauthoryear{{Gal{\'a}rraga-Espinosa}, {Aghanim}, {Langer}  \& {Tanimura}}{{Gal{\'a}rraga-Espinosa} et~al.}{2021}]{Galaraga2021}
{Gal{\'a}rraga-Espinosa} D.,  {Aghanim} N.,  {Langer} M.,   {Tanimura} H.,  2021, \mn@doi [\aap] {10.1051/0004-6361/202039781}, \href {https://ui.adsabs.harvard.edu/abs/2021A&A...649A.117G} {649, A117}

\bibitem[\protect\citeauthoryear{{Gal{\'a}rraga-Espinosa}, {Langer}  \& {Aghanim}}{{Gal{\'a}rraga-Espinosa} et~al.}{2022}]{Galaraga2022}
{Gal{\'a}rraga-Espinosa} D.,  {Langer} M.,   {Aghanim} N.,  2022, \mn@doi [\aap] {10.1051/0004-6361/202141974}, \href {https://ui.adsabs.harvard.edu/abs/2022A&A...661A.115G} {661, A115}

\bibitem[\protect\citeauthoryear{{Ganeshaiah Veena}, {Cautun}, {van de Weygaert}, {Tempel}, {Jones}, {Rieder}  \& {Frenk}}{{Ganeshaiah Veena} et~al.}{2018}]{Ganeshaiah.etal.18}
{Ganeshaiah Veena} P.,  {Cautun} M.,  {van de Weygaert} R.,  {Tempel} E.,  {Jones} B. J.~T.,  {Rieder} S.,   {Frenk} C.~S.,  2018, \mn@doi [\mnras] {10.1093/mnras/sty2270}, \href {https://ui.adsabs.harvard.edu/abs/2018MNRAS.481..414G} {481, 414}

\bibitem[\protect\citeauthoryear{{Ganeshaiah Veena}, {Cautun}, {Tempel}, {van de Weygaert}  \& {Frenk}}{{Ganeshaiah Veena} et~al.}{2019}]{Ganeshaiah.etal.19}
{Ganeshaiah Veena} P.,  {Cautun} M.,  {Tempel} E.,  {van de Weygaert} R.,   {Frenk} C.~S.,  2019, \mn@doi [\mnras] {10.1093/mnras/stz1343}, \href {https://ui.adsabs.harvard.edu/abs/2019MNRAS.487.1607G} {487, 1607}

\bibitem[\protect\citeauthoryear{{Ganeshaiah Veena}, {Cautun}, {van de Weygaert}, {Tempel}  \& {Frenk}}{{Ganeshaiah Veena} et~al.}{2021}]{Ganeshaiah.etal.20}
{Ganeshaiah Veena} P.,  {Cautun} M.,  {van de Weygaert} R.,  {Tempel} E.,   {Frenk} C.~S.,  2021, \mn@doi [\mnras] {10.1093/mnras/stab411}, \href {https://ui.adsabs.harvard.edu/abs/2021MNRAS.503.2280G} {503, 2280}

\bibitem[\protect\citeauthoryear{{Gaspari}, {Ruszkowski}  \& {Sharma}}{{Gaspari} et~al.}{2012}]{Gaspari2012multiphase}
{Gaspari} M.,  {Ruszkowski} M.,   {Sharma} P.,  2012, \mn@doi [\apj] {10.1088/0004-637X/746/1/94}, \href {https://ui.adsabs.harvard.edu/abs/2012ApJ...746...94G} {746, 94}

\bibitem[\protect\citeauthoryear{{Genel}, {Dekel}  \& {Cacciato}}{{Genel} et~al.}{2012}]{Genel.etal.12}
{Genel} S.,  {Dekel} A.,   {Cacciato} M.,  2012, \mn@doi [\mnras] {10.1111/j.1365-2966.2012.21652.x}, \href {https://ui.adsabs.harvard.edu/abs/2012MNRAS.425..788G} {425, 788}

\bibitem[\protect\citeauthoryear{{Genzel} et~al.,}{{Genzel} et~al.}{2006}]{Genzel.etal.06}
{Genzel} R.,  et~al., 2006, \mn@doi [\nat] {10.1038/nature05052}, \href {https://ui.adsabs.harvard.edu/abs/2006Natur.442..786G} {442, 786}

\bibitem[\protect\citeauthoryear{{Genzel} et~al.,}{{Genzel} et~al.}{2008}]{Genzel.etal.08}
{Genzel} R.,  et~al., 2008, \mn@doi [\apj] {10.1086/591840}, \href {https://ui.adsabs.harvard.edu/abs/2008ApJ...687...59G} {687, 59}

\bibitem[\protect\citeauthoryear{{Ginzburg}, {Dekel}, {Mandelker}  \& {Krumholz}}{{Ginzburg} et~al.}{2022}]{Ginzburg.etal.22}
{Ginzburg} O.,  {Dekel} A.,  {Mandelker} N.,   {Krumholz} M.~R.,  2022, \mn@doi [\mnras] {10.1093/mnras/stac1324}, \href {https://ui.adsabs.harvard.edu/abs/2022MNRAS.513.6177G} {513, 6177}

\bibitem[\protect\citeauthoryear{{Goerdt} \& {Ceverino}}{{Goerdt} \& {Ceverino}}{2015}]{Goerdt.etal.15}
{Goerdt} T.,  {Ceverino} D.,  2015, \mn@doi [\mnras] {10.1093/mnras/stv786}, \href {https://ui.adsabs.harvard.edu/abs/2015MNRAS.450.3359G} {450, 3359}

\bibitem[\protect\citeauthoryear{{Goerdt}, {Dekel}, {Sternberg}, {Ceverino}, {Teyssier}  \& {Primack}}{{Goerdt} et~al.}{2010}]{Goerdt.etal.10}
{Goerdt} T.,  {Dekel} A.,  {Sternberg} A.,  {Ceverino} D.,  {Teyssier} R.,   {Primack} J.~R.,  2010, \mn@doi [\mnras] {10.1111/j.1365-2966.2010.16941.x}, \href {http://adsabs.harvard.edu/abs/2010MNRAS.407..613G} {407, 613}

\bibitem[\protect\citeauthoryear{{Goerdt}, {Dekel}, {Sternberg}, {Gnat}  \& {Ceverino}}{{Goerdt} et~al.}{2012}]{Goerdt.etal.12}
{Goerdt} T.,  {Dekel} A.,  {Sternberg} A.,  {Gnat} O.,   {Ceverino} D.,  2012, \mn@doi [\mnras] {10.1111/j.1365-2966.2012.21397.x}, \href {https://ui.adsabs.harvard.edu/abs/2012MNRAS.424.2292G} {424, 2292}

\bibitem[\protect\citeauthoryear{{Gonz{\'a}lez}, {Prieto}, {Padilla}  \& {Jimenez}}{{Gonz{\'a}lez} et~al.}{2017}]{Gonzales.etal.17}
{Gonz{\'a}lez} R.~E.,  {Prieto} J.,  {Padilla} N.,   {Jimenez} R.,  2017, \mn@doi [\mnras] {10.1093/mnras/stw2715}, \href {https://ui.adsabs.harvard.edu/abs/2017MNRAS.464.4666G} {464, 4666}

\bibitem[\protect\citeauthoryear{{Gronke} \& {Oh}}{{Gronke} \& {Oh}}{2018}]{Gronke.Oh.2018}
{Gronke} M.,  {Oh} S.~P.,  2018, \mn@doi [\mnras] {10.1093/mnrasl/sly131}, \href {https://ui.adsabs.harvard.edu/abs/2018MNRAS.480L.111G} {480, L111}

\bibitem[\protect\citeauthoryear{{Gronke} \& {Oh}}{{Gronke} \& {Oh}}{2020}]{Gronke.Oh.2020}
{Gronke} M.,  {Oh} S.~P.,  2020, \mn@doi [\mnras] {10.1093/mnrasl/slaa033}, \href {https://ui.adsabs.harvard.edu/abs/2020MNRAS.494L..27G} {494, L27}

\bibitem[\protect\citeauthoryear{{Gronke} \& {Oh}}{{Gronke} \& {Oh}}{2022}]{Gronke.Oh.2022}
{Gronke} M.,  {Oh} S.~P.,  2022, \mn@doi [arXiv e-prints] {10.48550/arXiv.2209.00732}, \href {https://ui.adsabs.harvard.edu/abs/2022arXiv220900732G} {p. arXiv:2209.00732}

\bibitem[\protect\citeauthoryear{{Hahn}, {Angulo}  \& {Abel}}{{Hahn} et~al.}{2015}]{hahn.etal.15}
{Hahn} O.,  {Angulo} R.~E.,   {Abel} T.,  2015, \mn@doi [\mnras] {10.1093/mnras/stv2179}, \href {https://ui.adsabs.harvard.edu/abs/2015MNRAS.454.3920H} {454, 3920}

\bibitem[\protect\citeauthoryear{{Harford} \& {Hamilton}}{{Harford} \& {Hamilton}}{2011}]{Harford2011isothermal}
{Harford} A.~G.,  {Hamilton} A. J.~S.,  2011, \mn@doi [\mnras] {10.1111/j.1365-2966.2011.19220.x}, \href {https://ui.adsabs.harvard.edu/abs/2011MNRAS.416.2678H} {416, 2678}

\bibitem[\protect\citeauthoryear{{Hennawi}, {Prochaska}, {Cantalupo}  \& {Arrigoni-Battaia}}{{Hennawi} et~al.}{2015}]{Hennawi.etal.15}
{Hennawi} J.~F.,  {Prochaska} J.~X.,  {Cantalupo} S.,   {Arrigoni-Battaia} F.,  2015, \mn@doi [Science] {10.1126/science.aaa5397}, \href {https://ui.adsabs.harvard.edu/abs/2015Sci...348..779H} {348, 779}

\bibitem[\protect\citeauthoryear{{Huchra} et~al.,}{{Huchra} et~al.}{2005}]{Huchra05}
{Huchra} J.,  et~al., 2005, in {Fairall} A.~P.,  {Woudt} P.~A.,  eds,  Astronomical Society of the Pacific Conference Series Vol. 329, Nearby Large-Scale Structures and the Zone of Avoidance. p.~135

\bibitem[\protect\citeauthoryear{{Hummels} et~al.,}{{Hummels} et~al.}{2019}]{Hummels.etal.19}
{Hummels} C.~B.,  et~al., 2019, \mn@doi [\apj] {10.3847/1538-4357/ab378f}, \href {https://ui.adsabs.harvard.edu/abs/2019ApJ...882..156H} {882, 156}

\bibitem[\protect\citeauthoryear{{Katz}, {Weinberg}  \& {Hernquist}}{{Katz} et~al.}{1996}]{Katz96}
{Katz} N.,  {Weinberg} D.~H.,   {Hernquist} L.,  1996, \mn@doi [\apjs] {10.1086/192305}, \href {https://ui.adsabs.harvard.edu/abs/1996ApJS..105...19K} {105, 19}

\bibitem[\protect\citeauthoryear{{Kere{\v s}}, {Katz}, {Weinberg}  \& {Dav{\'e}}}{{Kere{\v s}} et~al.}{2005}]{Keres.etal.05}
{Kere{\v s}} D.,  {Katz} N.,  {Weinberg} D.~H.,   {Dav{\'e}} R.,  2005, \mn@doi [\mnras] {10.1111/j.1365-2966.2005.09451.x}, \href {http://adsabs.harvard.edu/abs/2005MNRAS.363....2K} {363, 2}

\bibitem[\protect\citeauthoryear{{Kimm}, {Devriendt}, {Slyz}, {Pichon}, {Kassin}  \& {Dubois}}{{Kimm} et~al.}{2011}]{Kimm.etal.11}
{Kimm} T.,  {Devriendt} J.,  {Slyz} A.,  {Pichon} C.,  {Kassin} S.~A.,   {Dubois} Y.,  2011, preprint, \href {http://adsabs.harvard.edu/abs/2011arXiv1106.0538K} {} (\mn@eprint {arXiv} {1106.0538})

\bibitem[\protect\citeauthoryear{{Klar} \& {M{\"u}cket}}{{Klar} \& {M{\"u}cket}}{2012}]{Klar.Mucket.2012}
{Klar} J.~S.,  {M{\"u}cket} J.~P.,  2012, \mn@doi [\mnras] {10.1111/j.1365-2966.2012.20877.x}, \href {https://ui.adsabs.harvard.edu/abs/2012MNRAS.423..304K} {423, 304}

\bibitem[\protect\citeauthoryear{Krumholz}{Krumholz}{2015}]{krumholz2015notes}
Krumholz M.~R.,  2015, arXiv preprint arXiv:1511.03457

\bibitem[\protect\citeauthoryear{{Laigle} et~al.,}{{Laigle} et~al.}{2015}]{Laigle.etal.15}
{Laigle} C.,  et~al., 2015, \mn@doi [\mnras] {10.1093/mnras/stu2289}, \href {https://ui.adsabs.harvard.edu/abs/2015MNRAS.446.2744L} {446, 2744}

\bibitem[\protect\citeauthoryear{{Lau}, {Nagai}  \& {Nelson}}{{Lau} et~al.}{2013}]{lau2013weighing}
{Lau} E.~T.,  {Nagai} D.,   {Nelson} K.,  2013, \mn@doi [\apj] {10.1088/0004-637X/777/2/151}, \href {https://ui.adsabs.harvard.edu/abs/2013ApJ...777..151L} {777, 151}

\bibitem[\protect\citeauthoryear{{Lau}, {Nagai}, {Avestruz}, {Nelson}  \& {Vikhlinin}}{{Lau} et~al.}{2015}]{lau2015gasprofile}
{Lau} E.~T.,  {Nagai} D.,  {Avestruz} C.,  {Nelson} K.,   {Vikhlinin} A.,  2015, \mn@doi [\apj] {10.1088/0004-637X/806/1/68}, \href {https://ui.adsabs.harvard.edu/abs/2015ApJ...806...68L} {806, 68}

\bibitem[\protect\citeauthoryear{{Ledos}, {Takasao}  \& {Nagamine}}{{Ledos} et~al.}{2023}]{Ledos.etal.2023}
{Ledos} N.,  {Takasao} S.,   {Nagamine} K.,  2023, \mn@doi [arXiv e-prints] {10.48550/arXiv.2308.05412}, \href {https://ui.adsabs.harvard.edu/abs/2023arXiv230805412L} {p. arXiv:2308.05412}

\bibitem[\protect\citeauthoryear{{Libeskind} et~al.,}{{Libeskind} et~al.}{2018}]{Libeskind.etal.18}
{Libeskind} N.~I.,  et~al., 2018, \mn@doi [\mnras] {10.1093/mnras/stx1976}, \href {https://ui.adsabs.harvard.edu/abs/2018MNRAS.473.1195L} {473, 1195}

\bibitem[\protect\citeauthoryear{{Lochhaas}, {Tumlinson}, {O'Shea}, {Peeples}, {Smith}, {Werk}, {Augustin}  \& {Simons}}{{Lochhaas} et~al.}{2021}]{Lochhaas.etal.21}
{Lochhaas} C.,  {Tumlinson} J.,  {O'Shea} B.~W.,  {Peeples} M.~S.,  {Smith} B.~D.,  {Werk} J.~K.,  {Augustin} R.,   {Simons} R.~C.,  2021, \mn@doi [\apj] {10.3847/1538-4357/ac2496}, \href {https://ui.adsabs.harvard.edu/abs/2021ApJ...922..121L} {922, 121}

\bibitem[\protect\citeauthoryear{{Lochhaas} et~al.,}{{Lochhaas} et~al.}{2023}]{Lochhaas.etal.22}
{Lochhaas} C.,  et~al., 2023, \mn@doi [\apj] {10.3847/1538-4357/acbb06}, \href {https://ui.adsabs.harvard.edu/abs/2023ApJ...948...43L} {948, 43}

\bibitem[\protect\citeauthoryear{{Lusso} et~al.,}{{Lusso} et~al.}{2019}]{Lusso.etal.19}
{Lusso} E.,  et~al., 2019, \mn@doi [\mnras] {10.1093/mnrasl/slz032}, \href {https://ui.adsabs.harvard.edu/abs/2019MNRAS.485L..62L} {485, L62}

\bibitem[\protect\citeauthoryear{{Mandelker}, {Padnos}, {Dekel}, {Birnboim}, {Burkert}, {Krumholz}  \& {Steinberg}}{{Mandelker} et~al.}{2016}]{mandelker.etal.2016}
{Mandelker} N.,  {Padnos} D.,  {Dekel} A.,  {Birnboim} Y.,  {Burkert} A.,  {Krumholz} M.~R.,   {Steinberg} E.,  2016, \mn@doi [\mnras] {10.1093/mnras/stw2267}, \href {https://ui.adsabs.harvard.edu/abs/2016MNRAS.463.3921M} {463, 3921}

\bibitem[\protect\citeauthoryear{{Mandelker}, {van Dokkum}, {Brodie}, {van den Bosch}  \& {Ceverino}}{{Mandelker} et~al.}{2018}]{mandelker2018cold}
{Mandelker} N.,  {van Dokkum} P.~G.,  {Brodie} J.~P.,  {van den Bosch} F.~C.,   {Ceverino} D.,  2018, \mn@doi [\apj] {10.3847/1538-4357/aaca98}, \href {https://ui.adsabs.harvard.edu/abs/2018ApJ...861..148M} {861, 148}

\bibitem[\protect\citeauthoryear{{Mandelker}, {Nagai}, {Aung}, {Dekel}, {Padnos}  \& {Birnboim}}{{Mandelker} et~al.}{2019a}]{mandelker.etal.2019}
{Mandelker} N.,  {Nagai} D.,  {Aung} H.,  {Dekel} A.,  {Padnos} D.,   {Birnboim} Y.,  2019a, \mn@doi [\mnras] {10.1093/mnras/stz012}, \href {https://ui.adsabs.harvard.edu/abs/2019MNRAS.484.1100M} {484, 1100}

\bibitem[\protect\citeauthoryear{{Mandelker}, {van den Bosch}, {Springel}  \& {van de Voort}}{{Mandelker} et~al.}{2019b}]{mandelker2019shattering}
{Mandelker} N.,  {van den Bosch} F.~C.,  {Springel} V.,   {van de Voort} F.,  2019b, \mn@doi [\apjl] {10.3847/2041-8213/ab30cb}, \href {https://ui.adsabs.harvard.edu/abs/2019ApJ...881L..20M} {881, L20}

\bibitem[\protect\citeauthoryear{{Mandelker}, {Nagai}, {Aung}, {Dekel}, {Birnboim}  \& {van den Bosch}}{{Mandelker} et~al.}{2020a}]{mandelker2020KHI4}
{Mandelker} N.,  {Nagai} D.,  {Aung} H.,  {Dekel} A.,  {Birnboim} Y.,   {van den Bosch} F.~C.,  2020a, \mn@doi [\mnras] {10.1093/mnras/staa812}, \href {https://ui.adsabs.harvard.edu/abs/2020MNRAS.494.2641M} {494, 2641}

\bibitem[\protect\citeauthoryear{{Mandelker}, {van den Bosch}, {Nagai}, {Dekel}, {Birnboim}  \& {Aung}}{{Mandelker} et~al.}{2020b}]{mandelker2020LABs}
{Mandelker} N.,  {van den Bosch} F.~C.,  {Nagai} D.,  {Dekel} A.,  {Birnboim} Y.,   {Aung} H.,  2020b, \mn@doi [\mnras] {10.1093/mnras/staa2421}, \href {https://ui.adsabs.harvard.edu/abs/2020MNRAS.498.2415M} {498, 2415}

\bibitem[\protect\citeauthoryear{{Mandelker}, {van den Bosch}, {Springel}, {van de Voort}, {Burchett}, {Butsky}, {Nagai}  \& {Oh}}{{Mandelker} et~al.}{2021}]{mandelker2021thermal}
{Mandelker} N.,  {van den Bosch} F.~C.,  {Springel} V.,  {van de Voort} F.,  {Burchett} J.~N.,  {Butsky} I.~S.,  {Nagai} D.,   {Oh} S.~P.,  2021, \mn@doi [\apj] {10.3847/1538-4357/ac2d29}, \href {https://ui.adsabs.harvard.edu/abs/2021ApJ...923..115M} {923, 115}

\bibitem[\protect\citeauthoryear{{Martin}, {Chang}, {Matuszewski}, {Morrissey}, {Rahman}, {Moore}  \& {Steidel}}{{Martin} et~al.}{2014a}]{Martin.etal.14a}
{Martin} D.~C.,  {Chang} D.,  {Matuszewski} M.,  {Morrissey} P.,  {Rahman} S.,  {Moore} A.,   {Steidel} C.~C.,  2014a, \mn@doi [\apj] {10.1088/0004-637X/786/2/106}, \href {http://adsabs.harvard.edu/abs/2014ApJ...786..106M} {786, 106}

\bibitem[\protect\citeauthoryear{{Martin}, {Chang}, {Matuszewski}, {Morrissey}, {Rahman}, {Moore}, {Steidel}  \& {Matsuda}}{{Martin} et~al.}{2014b}]{Martin.etal.14b}
{Martin} D.~C.,  {Chang} D.,  {Matuszewski} M.,  {Morrissey} P.,  {Rahman} S.,  {Moore} A.,  {Steidel} C.~C.,   {Matsuda} Y.,  2014b, \mn@doi [\apj] {10.1088/0004-637X/786/2/107}, \href {http://adsabs.harvard.edu/abs/2014ApJ...786..107M} {786, 107}

\bibitem[\protect\citeauthoryear{{Martin} et~al.,}{{Martin} et~al.}{2019}]{Martin.etal.19}
{Martin} D.~C.,  et~al., 2019, \mn@doi [Nature Astronomy] {10.1038/s41550-019-0791-2}, \href {https://ui.adsabs.harvard.edu/abs/2019NatAs.tmp..372M} {p.~372}

\bibitem[\protect\citeauthoryear{{Matsuda}, {Yamada}, {Hayashino}, {Yamauchi}  \& {Nakamura}}{{Matsuda} et~al.}{2006}]{Matsuda.etal.06}
{Matsuda} Y.,  {Yamada} T.,  {Hayashino} T.,  {Yamauchi} R.,   {Nakamura} Y.,  2006, \mn@doi [\apjl] {10.1086/503362}, \href {http://adsabs.harvard.edu/abs/2006ApJ...640L.123M} {640, L123}

\bibitem[\protect\citeauthoryear{{Matsuda} et~al.,}{{Matsuda} et~al.}{2011}]{Matsuda.etal.11}
{Matsuda} Y.,  et~al., 2011, \mn@doi [\mnras] {10.1111/j.1745-3933.2010.00969.x}, \href {http://adsabs.harvard.edu/abs/2011MNRAS.410L..13M} {410, L13}

\bibitem[\protect\citeauthoryear{{McCourt}, {Sharma}, {Quataert}  \& {Parrish}}{{McCourt} et~al.}{2012}]{mccourt2012thermal}
{McCourt} M.,  {Sharma} P.,  {Quataert} E.,   {Parrish} I.~J.,  2012, \mn@doi [\mnras] {10.1111/j.1365-2966.2011.19972.x}, \href {https://ui.adsabs.harvard.edu/abs/2012MNRAS.419.3319M} {419, 3319}

\bibitem[\protect\citeauthoryear{{McCourt}, {Oh}, {O'Leary}  \& {Madigan}}{{McCourt} et~al.}{2018}]{McCourt.etal.2018}
{McCourt} M.,  {Oh} S.~P.,  {O'Leary} R.,   {Madigan} A.-M.,  2018, \mn@doi [\mnras] {10.1093/mnras/stx2687}, \href {https://ui.adsabs.harvard.edu/abs/2018MNRAS.473.5407M} {473, 5407}

\bibitem[\protect\citeauthoryear{Nelson et~al.,}{Nelson et~al.}{2018}]{nelson2018first}
Nelson D.,  et~al., 2018, Monthly Notices of the Royal Astronomical Society, 475, 624

\bibitem[\protect\citeauthoryear{{O'Sullivan} et~al.,}{{O'Sullivan} et~al.}{2019}]{Sullivan.etal.19}
{O'Sullivan} S.~P.,  et~al., 2019, \mn@doi [\aap] {10.1051/0004-6361/201833832}, \href {https://ui.adsabs.harvard.edu/abs/2019A&A...622A..16O} {622, A16}

\bibitem[\protect\citeauthoryear{{Ocvirk}, {Pichon}  \& {Teyssier}}{{Ocvirk} et~al.}{2008}]{Ocvirk.etal.08}
{Ocvirk} P.,  {Pichon} C.,   {Teyssier} R.,  2008, \mn@doi [\mnras] {10.1111/j.1365-2966.2008.13763.x}, \href {https://ui.adsabs.harvard.edu/abs/2008MNRAS.390.1326O} {390, 1326}

\bibitem[\protect\citeauthoryear{{Ostriker}}{{Ostriker}}{1964}]{ostriker64}
{Ostriker} J.,  1964, \mn@doi [\apj] {10.1086/148057}, \href {https://ui.adsabs.harvard.edu/abs/1964ApJ...140.1529O} {140, 1529}

\bibitem[\protect\citeauthoryear{{Padnos}, {Mandelker}, {Birnboim}, {Dekel}, {Krumholz}  \& {Steinberg}}{{Padnos} et~al.}{2018}]{padnos.etal.2018}
{Padnos} D.,  {Mandelker} N.,  {Birnboim} Y.,  {Dekel} A.,  {Krumholz} M.~R.,   {Steinberg} E.,  2018, \mn@doi [\mnras] {10.1093/mnras/sty789}, \href {https://ui.adsabs.harvard.edu/abs/2018MNRAS.477.3293P} {477, 3293}

\bibitem[\protect\citeauthoryear{{Pandya} et~al.,}{{Pandya} et~al.}{2019}]{Pandya.etal.19}
{Pandya} V.,  et~al., 2019, \mn@doi [\mnras] {10.1093/mnras/stz2129}, \href {https://ui.adsabs.harvard.edu/abs/2019MNRAS.488.5580P} {488, 5580}

\bibitem[\protect\citeauthoryear{{Pasha}, {Mandelker}, {van den Bosch}, {Springel}  \& {van de Voort}}{{Pasha} et~al.}{2023}]{Pasha.etal.22}
{Pasha} I.,  {Mandelker} N.,  {van den Bosch} F.~C.,  {Springel} V.,   {van de Voort} F.,  2023, \mn@doi [\mnras] {10.1093/mnras/stac3776}, \href {https://ui.adsabs.harvard.edu/abs/2023MNRAS.520.2692P} {520, 2692}

\bibitem[\protect\citeauthoryear{{Peeples} et~al.,}{{Peeples} et~al.}{2019}]{Peeples.etal.19}
{Peeples} M.~S.,  et~al., 2019, \mn@doi [\apj] {10.3847/1538-4357/ab0654}, \href {https://ui.adsabs.harvard.edu/abs/2019ApJ...873..129P} {873, 129}

\bibitem[\protect\citeauthoryear{{Pezzulli} \& {Cantalupo}}{{Pezzulli} \& {Cantalupo}}{2019}]{Pezzulli.etal.2019}
{Pezzulli} G.,  {Cantalupo} S.,  2019, \mn@doi [\mnras] {10.1093/mnras/stz906}, \href {https://ui.adsabs.harvard.edu/abs/2019MNRAS.486.1489P} {486, 1489}

\bibitem[\protect\citeauthoryear{{Pichon} \& {Bernardeau}}{{Pichon} \& {Bernardeau}}{1999}]{Pichon.Bernardeau.99}
{Pichon} C.,  {Bernardeau} F.,  1999, \aap, \href {https://ui.adsabs.harvard.edu/abs/1999A&A...343..663P} {343, 663}

\bibitem[\protect\citeauthoryear{{Pichon}, {Pogosyan}, {Kimm}, {Slyz}, {Devriendt}  \& {Dubois}}{{Pichon} et~al.}{2011}]{Pichon.etal.11}
{Pichon} C.,  {Pogosyan} D.,  {Kimm} T.,  {Slyz} A.,  {Devriendt} J.,   {Dubois} Y.,  2011, \mn@doi [\mnras] {10.1111/j.1365-2966.2011.19640.x}, \href {http://adsabs.harvard.edu/abs/2011MNRAS.418.2493P} {418, 2493}

\bibitem[\protect\citeauthoryear{{Pillepich} et~al.,}{{Pillepich} et~al.}{2018a}]{pillepich2018Sim}
{Pillepich} A.,  et~al., 2018a, \mn@doi [\mnras] {10.1093/mnras/stx2656}, \href {http://adsabs.harvard.edu/abs/2018MNRAS.473.4077P} {473, 4077}

\bibitem[\protect\citeauthoryear{{Pillepich} et~al.,}{{Pillepich} et~al.}{2018b}]{pillepich2018first}
{Pillepich} A.,  et~al., 2018b, \mn@doi [\mnras] {10.1093/mnras/stx3112}, \href {https://ui.adsabs.harvard.edu/abs/2018MNRAS.475..648P} {475, 648}

\bibitem[\protect\citeauthoryear{{Planck Collaboration} et~al.,}{{Planck Collaboration} et~al.}{2016}]{Planck16}
{Planck Collaboration} et~al., 2016, \mn@doi [\aap] {10.1051/0004-6361/201525830}, \href {http://adsabs.harvard.edu/abs/2016A%26A...594A..13P} {594, A13}

\bibitem[\protect\citeauthoryear{{Powell}, {Slyz}  \& {Devriendt}}{{Powell} et~al.}{2011}]{Powell.etal.11}
{Powell} L.~C.,  {Slyz} A.,   {Devriendt} J.,  2011, \mn@doi [\mnras] {10.1111/j.1365-2966.2011.18668.x}, \href {https://ui.adsabs.harvard.edu/abs/2011MNRAS.414.3671P} {414, 3671}

\bibitem[\protect\citeauthoryear{{Prochaska}, {Lau}  \& {Hennawi}}{{Prochaska} et~al.}{2014}]{Prochaska.etal.14}
{Prochaska} J.~X.,  {Lau} M.~W.,   {Hennawi} J.~F.,  2014, \mn@doi [\apj] {10.1088/0004-637X/796/2/140}, \href {http://adsabs.harvard.edu/abs/2014ApJ...796..140P} {796, 140}

\bibitem[\protect\citeauthoryear{{Rahmati}, {Pawlik}, {Rai{\v{c}}evi{\'c}}  \& {Schaye}}{{Rahmati} et~al.}{2013}]{Rahmati13}
{Rahmati} A.,  {Pawlik} A.~H.,  {Rai{\v{c}}evi{\'c}} M.,   {Schaye} J.,  2013, \mn@doi [\mnras] {10.1093/mnras/stt066}, \href {https://ui.adsabs.harvard.edu/abs/2013MNRAS.430.2427R} {430, 2427}

\bibitem[\protect\citeauthoryear{{Rams{\o}y}, {Slyz}, {Devriendt}, {Laigle}  \& {Dubois}}{{Rams{\o}y} et~al.}{2021}]{ramsoy2021rivers}
{Rams{\o}y} M.,  {Slyz} A.,  {Devriendt} J.,  {Laigle} C.,   {Dubois} Y.,  2021, \mn@doi [\mnras] {10.1093/mnras/stab015}, \href {https://ui.adsabs.harvard.edu/abs/2021MNRAS.502..351R} {502, 351}

\bibitem[\protect\citeauthoryear{{Rees} \& {Ostriker}}{{Rees} \& {Ostriker}}{1977}]{Rees77}
{Rees} M.~J.,  {Ostriker} J.~P.,  1977, \mn@doi [\mnras] {10.1093/mnras/179.4.541}, \href {https://ui.adsabs.harvard.edu/abs/1977MNRAS.179..541R} {179, 541}

\bibitem[\protect\citeauthoryear{{Scannapieco} \& {Br{\"u}ggen}}{{Scannapieco} \& {Br{\"u}ggen}}{2015}]{Scannapieco.Bruggen.2015}
{Scannapieco} E.,  {Br{\"u}ggen} M.,  2015, \mn@doi [\apj] {10.1088/0004-637X/805/2/158}, \href {https://ui.adsabs.harvard.edu/abs/2015ApJ...805..158S} {805, 158}

\bibitem[\protect\citeauthoryear{{Shapiro} et~al.,}{{Shapiro} et~al.}{2008}]{Shapiro.etal.08}
{Shapiro} K.~L.,  et~al., 2008, \mn@doi [\apj] {10.1086/587133}, \href {https://ui.adsabs.harvard.edu/abs/2008ApJ...682..231S} {682, 231}

\bibitem[\protect\citeauthoryear{{Sharma}, {McCourt}, {Quataert}  \& {Parrish}}{{Sharma} et~al.}{2012}]{sharma2012thermal}
{Sharma} P.,  {McCourt} M.,  {Quataert} E.,   {Parrish} I.~J.,  2012, \mn@doi [\mnras] {10.1111/j.1365-2966.2011.20246.x}, \href {https://ui.adsabs.harvard.edu/abs/2012MNRAS.420.3174S} {420, 3174}

\bibitem[\protect\citeauthoryear{{Shaw}, {Weller}, {Ostriker}  \& {Bode}}{{Shaw} et~al.}{2006}]{shaw2006statistics}
{Shaw} L.~D.,  {Weller} J.,  {Ostriker} J.~P.,   {Bode} P.,  2006, \mn@doi [\apj] {10.1086/505016}, \href {https://ui.adsabs.harvard.edu/abs/2006ApJ...646..815S} {646, 815}

\bibitem[\protect\citeauthoryear{{Song} et~al.,}{{Song} et~al.}{2021}]{Song.etal.20}
{Song} H.,  et~al., 2021, \mn@doi [\mnras] {10.1093/mnras/staa3981}, \href {https://ui.adsabs.harvard.edu/abs/2021MNRAS.501.4635S} {501, 4635}

\bibitem[\protect\citeauthoryear{{Springel}}{{Springel}}{2010}]{springel2010pur}
{Springel} V.,  2010, \mn@doi [\mnras] {10.1111/j.1365-2966.2009.15715.x}, \href {https://ui.adsabs.harvard.edu/abs/2010MNRAS.401..791S} {401, 791}

\bibitem[\protect\citeauthoryear{{Springel} \& {Hernquist}}{{Springel} \& {Hernquist}}{2003}]{Springel03}
{Springel} V.,  {Hernquist} L.,  2003, \mn@doi [\mnras] {10.1046/j.1365-8711.2003.06206.x}, \href {https://ui.adsabs.harvard.edu/abs/2003MNRAS.339..289S} {339, 289}

\bibitem[\protect\citeauthoryear{{Springel}, {White}, {Tormen}  \& {Kauffmann}}{{Springel} et~al.}{2001}]{Springel01}
{Springel} V.,  {White} S. D.~M.,  {Tormen} G.,   {Kauffmann} G.,  2001, \mn@doi [\mnras] {10.1046/j.1365-8711.2001.04912.x}, \href {https://ui.adsabs.harvard.edu/abs/2001MNRAS.328..726S} {328, 726}

\bibitem[\protect\citeauthoryear{{Springel} et~al.,}{{Springel} et~al.}{2005}]{Springel.etal.05}
{Springel} V.,  et~al., 2005, \mn@doi [\nat] {10.1038/nature03597}, \href {http://adsabs.harvard.edu/abs/2005Natur.435..629S} {435, 629}

\bibitem[\protect\citeauthoryear{{Springel} et~al.,}{{Springel} et~al.}{2018}]{springel2018first}
{Springel} V.,  et~al., 2018, \mn@doi [\mnras] {10.1093/mnras/stx3304}, \href {https://ui.adsabs.harvard.edu/abs/2018MNRAS.475..676S} {475, 676}

\bibitem[\protect\citeauthoryear{{Stark}, {Swinbank}, {Ellis}, {Dye}, {Smail}  \& {Richard}}{{Stark} et~al.}{2008}]{Stark.etal.08}
{Stark} D.~P.,  {Swinbank} A.~M.,  {Ellis} R.~S.,  {Dye} S.,  {Smail} I.~R.,   {Richard} J.,  2008, \mn@doi [\nat] {10.1038/nature07294}, \href {https://ui.adsabs.harvard.edu/abs/2008Natur.455..775S} {455, 775}

\bibitem[\protect\citeauthoryear{{Steidel}, {Adelberger}, {Shapley}, {Pettini}, {Dickinson}  \& {Giavalisco}}{{Steidel} et~al.}{2000}]{Steidel.etal.00}
{Steidel} C.~C.,  {Adelberger} K.~L.,  {Shapley} A.~E.,  {Pettini} M.,  {Dickinson} M.,   {Giavalisco} M.,  2000, \mn@doi [\apj] {10.1086/308568}, \href {http://adsabs.harvard.edu/abs/2000ApJ...532..170S} {532, 170}

\bibitem[\protect\citeauthoryear{{Stern}, {Fielding}, {Faucher-Gigu{\`e}re}  \& {Quataert}}{{Stern} et~al.}{2020a}]{Stern2020maximum}
{Stern} J.,  {Fielding} D.,  {Faucher-Gigu{\`e}re} C.-A.,   {Quataert} E.,  2020a, \mn@doi [\mnras] {10.1093/mnras/staa198}, \href {https://ui.adsabs.harvard.edu/abs/2020MNRAS.492.6042S} {492, 6042}

\bibitem[\protect\citeauthoryear{{Stern}, {Fielding}, {Faucher-Gigu{\`e}re}  \& {Quataert}}{{Stern} et~al.}{2020b}]{stern.etal.2020}
{Stern} J.,  {Fielding} D.,  {Faucher-Gigu{\`e}re} C.-A.,   {Quataert} E.,  2020b, \mn@doi [\mnras] {10.1093/mnras/staa198}, \href {https://ui.adsabs.harvard.edu/abs/2020MNRAS.492.6042S} {492, 6042}

\bibitem[\protect\citeauthoryear{{Stern} et~al.,}{{Stern} et~al.}{2021a}]{Stern2021DLAs}
{Stern} J.,  et~al., 2021a, \mn@doi [\mnras] {10.1093/mnras/stab2240}, \href {https://ui.adsabs.harvard.edu/abs/2021MNRAS.507.2869S} {507, 2869}

\bibitem[\protect\citeauthoryear{{Stern} et~al.,}{{Stern} et~al.}{2021b}]{Stern2021virial}
{Stern} J.,  et~al., 2021b, \mn@doi [\apj] {10.3847/1538-4357/abd776}, \href {https://ui.adsabs.harvard.edu/abs/2021ApJ...911...88S} {911, 88}

\bibitem[\protect\citeauthoryear{{Stern}, {Fielding}, {Hafen}, {Su}, {Naor}, {Faucher-Gigu{\`e}re}, {Quataert}  \& {Bullock}}{{Stern} et~al.}{2023}]{stern.etal.2023}
{Stern} J.,  {Fielding} D.,  {Hafen} Z.,  {Su} K.-Y.,  {Naor} N.,  {Faucher-Gigu{\`e}re} C.-A.,  {Quataert} E.,   {Bullock} J.,  2023, \mn@doi [arXiv e-prints] {10.48550/arXiv.2306.00092}, \href {https://ui.adsabs.harvard.edu/abs/2023arXiv230600092S} {p. arXiv:2306.00092}

\bibitem[\protect\citeauthoryear{{Stewart}, {Kaufmann}, {Bullock}, {Barton}, {Maller}, {Diemand}  \& {Wadsley}}{{Stewart} et~al.}{2011}]{Stewart.etal.11}
{Stewart} K.~R.,  {Kaufmann} T.,  {Bullock} J.~S.,  {Barton} E.~J.,  {Maller} A.~H.,  {Diemand} J.,   {Wadsley} J.,  2011, \mn@doi [\apj] {10.1088/0004-637X/738/1/39}, \href {http://adsabs.harvard.edu/abs/2011ApJ...738...39S} {738, 39}

\bibitem[\protect\citeauthoryear{{Stewart}, {Brooks}, {Bullock}, {Maller}, {Diemand}, {Wadsley}  \& {Moustakas}}{{Stewart} et~al.}{2013}]{Stewart.etal.13}
{Stewart} K.~R.,  {Brooks} A.~M.,  {Bullock} J.~S.,  {Maller} A.~H.,  {Diemand} J.,  {Wadsley} J.,   {Moustakas} L.~A.,  2013, \mn@doi [\apj] {10.1088/0004-637X/769/1/74}, \href {http://adsabs.harvard.edu/abs/2013ApJ...769...74S} {769, 74}

\bibitem[\protect\citeauthoryear{{Suresh}, {Nelson}, {Genel}, {Rubin}  \& {Hernquist}}{{Suresh} et~al.}{2019}]{Suresh.etal.2019}
{Suresh} J.,  {Nelson} D.,  {Genel} S.,  {Rubin} K. H.~R.,   {Hernquist} L.,  2019, \mn@doi [\mnras] {10.1093/mnras/sty3402}, \href {https://ui.adsabs.harvard.edu/abs/2019MNRAS.483.4040S} {483, 4040}

\bibitem[\protect\citeauthoryear{{Tegmark} et~al.,}{{Tegmark} et~al.}{2004}]{tegmark04}
{Tegmark} M.,  et~al., 2004, \mn@doi [\prd] {10.1103/PhysRevD.69.103501}, \href {http://adsabs.harvard.edu/abs/2004PhRvD..69j3501T} {69, 103501}

\bibitem[\protect\citeauthoryear{{Teyssier}}{{Teyssier}}{2002}]{Teyssier.02}
{Teyssier} R.,  2002, \mn@doi [\aap] {10.1051/0004-6361:20011817}, \href {https://ui.adsabs.harvard.edu/abs/2002A&A...385..337T} {385, 337}

\bibitem[\protect\citeauthoryear{{Tillson}, {Devriendt}, {Slyz}, {Miller}  \& {Pichon}}{{Tillson} et~al.}{2015}]{Tillson.etal.15}
{Tillson} H.,  {Devriendt} J.,  {Slyz} A.,  {Miller} L.,   {Pichon} C.,  2015, \mn@doi [\mnras] {10.1093/mnras/stv557}, \href {https://ui.adsabs.harvard.edu/abs/2015MNRAS.449.4363T} {449, 4363}

\bibitem[\protect\citeauthoryear{{Tomassetti} et~al.,}{{Tomassetti} et~al.}{2016}]{Tomassetti.etal.16}
{Tomassetti} M.,  et~al., 2016, \mn@doi [\mnras] {10.1093/mnras/stw606}, \href {https://ui.adsabs.harvard.edu/abs/2016MNRAS.458.4477T} {458, 4477}

\bibitem[\protect\citeauthoryear{{Uhlemann}, {Friedrich}, {Villaescusa-Navarro}, {Banerjee}  \& {Codis}}{{Uhlemann} et~al.}{2020}]{Uhlemann.etal.20}
{Uhlemann} C.,  {Friedrich} O.,  {Villaescusa-Navarro} F.,  {Banerjee} A.,   {Codis} S.~r.,  2020, \mn@doi [\mnras] {10.1093/mnras/staa1155}, \href {https://ui.adsabs.harvard.edu/abs/2020MNRAS.495.4006U} {495, 4006}

\bibitem[\protect\citeauthoryear{{Umehata} et~al.,}{{Umehata} et~al.}{2019}]{Umehata.etal.19}
{Umehata} H.,  et~al., 2019, \mn@doi [Science] {10.1126/science.aaw5949}, \href {https://ui.adsabs.harvard.edu/abs/2019Sci...366...97U} {366, 97}

\bibitem[\protect\citeauthoryear{{Vernstrom}, {Heald}, {Vazza}, {Galvin}, {West}, {Locatelli}, {Fornengo}  \& {Pinetti}}{{Vernstrom} et~al.}{2021}]{Vernstrom.etal.21}
{Vernstrom} T.,  {Heald} G.,  {Vazza} F.,  {Galvin} T.~J.,  {West} J.~L.,  {Locatelli} N.,  {Fornengo} N.,   {Pinetti} E.,  2021, \mn@doi [\mnras] {10.1093/mnras/stab1301}, \href {https://ui.adsabs.harvard.edu/abs/2021MNRAS.505.4178V} {505, 4178}

\bibitem[\protect\citeauthoryear{{Vogelsberger}, {Genel}, {Sijacki}, {Torrey}, {Springel}  \& {Hernquist}}{{Vogelsberger} et~al.}{2013}]{Vogelsberger13}
{Vogelsberger} M.,  {Genel} S.,  {Sijacki} D.,  {Torrey} P.,  {Springel} V.,   {Hernquist} L.,  2013, \mn@doi [\mnras] {10.1093/mnras/stt1789}, \href {https://ui.adsabs.harvard.edu/abs/2013MNRAS.436.3031V} {436, 3031}

\bibitem[\protect\citeauthoryear{{Voit} \& {Donahue}}{{Voit} \& {Donahue}}{2015}]{Voit2015a}
{Voit} G.~M.,  {Donahue} M.,  2015, \mn@doi [\apjl] {10.1088/2041-8205/799/1/L1}, \href {https://ui.adsabs.harvard.edu/abs/2015ApJ...799L...1V} {799, L1}

\bibitem[\protect\citeauthoryear{{Voit}, {Donahue}, {Bryan}  \& {McDonald}}{{Voit} et~al.}{2015a}]{Voit2015b}
{Voit} G.~M.,  {Donahue} M.,  {Bryan} G.~L.,   {McDonald} M.,  2015a, \mn@doi [\nat] {10.1038/nature14167}, \href {https://ui.adsabs.harvard.edu/abs/2015Natur.519..203V} {519, 203}

\bibitem[\protect\citeauthoryear{{Voit}, {Bryan}, {O'Shea}  \& {Donahue}}{{Voit} et~al.}{2015b}]{Voit2015c}
{Voit} G.~M.,  {Bryan} G.~L.,  {O'Shea} B.~W.,   {Donahue} M.,  2015b, \mn@doi [\apjl] {10.1088/2041-8205/808/1/L30}, \href {https://ui.adsabs.harvard.edu/abs/2015ApJ...808L..30V} {808, L30}

\bibitem[\protect\citeauthoryear{{Wang}, {Libeskind}, {Tempel}, {Kang}  \& {Guo}}{{Wang} et~al.}{2021}]{WangP.etal.21}
{Wang} P.,  {Libeskind} N.~I.,  {Tempel} E.,  {Kang} X.,   {Guo} Q.,  2021, \mn@doi [Nature Astronomy] {10.1038/s41550-021-01380-6}, \href {https://ui.adsabs.harvard.edu/abs/2021NatAs...5..839W} {5, 839}

\bibitem[\protect\citeauthoryear{{Weinberger} et~al.,}{{Weinberger} et~al.}{2017}]{Weinberger2017Sim}
{Weinberger} R.,  et~al., 2017, \mn@doi [\mnras] {10.1093/mnras/stw2944}, \href {http://adsabs.harvard.edu/abs/2017MNRAS.465.3291W} {465, 3291}

\bibitem[\protect\citeauthoryear{{White} \& {Rees}}{{White} \& {Rees}}{1978}]{White78}
{White} S.~D.~M.,  {Rees} M.~J.,  1978, \mn@doi [\mnras] {10.1093/mnras/183.3.341}, \href {https://ui.adsabs.harvard.edu/abs/1978MNRAS.183..341W} {183, 341}

\bibitem[\protect\citeauthoryear{{Wiersma}, {Schaye}  \& {Smith}}{{Wiersma} et~al.}{2009}]{Wiersma09}
{Wiersma} R. P.~C.,  {Schaye} J.,   {Smith} B.~D.,  2009, \mn@doi [\mnras] {10.1111/j.1365-2966.2008.14191.x}, \href {https://ui.adsabs.harvard.edu/abs/2009MNRAS.393...99W} {393, 99}

\bibitem[\protect\citeauthoryear{{Wisnioski} et~al.,}{{Wisnioski} et~al.}{2015}]{Wisnioski.etal.15}
{Wisnioski} E.,  et~al., 2015, \mn@doi [\apj] {10.1088/0004-637X/799/2/209}, \href {http://adsabs.harvard.edu/abs/2015ApJ...799..209W} {799, 209}

\bibitem[\protect\citeauthoryear{{Xia}, {Neyrinck}, {Cai}  \& {Arag{\'o}n-Calvo}}{{Xia} et~al.}{2021}]{xia2021intergalactic}
{Xia} Q.,  {Neyrinck} M.~C.,  {Cai} Y.-C.,   {Arag{\'o}n-Calvo} M.~A.,  2021, \mn@doi [\mnras] {10.1093/mnras/stab1713}, \href {https://ui.adsabs.harvard.edu/abs/2021MNRAS.506.1059X} {506, 1059}

\bibitem[\protect\citeauthoryear{{Zel'dovich}}{{Zel'dovich}}{1970}]{Zeldovich1970}
{Zel'dovich} Y.~B.,  1970, \aap, \href {https://ui.adsabs.harvard.edu/abs/1970A&A.....5...84Z} {5, 84}

\bibitem[\protect\citeauthoryear{{Zhang} et~al.,}{{Zhang} et~al.}{2023}]{zhang.etal.2023}
{Zhang} S.,  et~al., 2023, \mn@doi [Science] {10.1126/science.abj9192}, \href {https://ui.adsabs.harvard.edu/abs/2023Sci...380..494Z} {380, 494}

\bibitem[\protect\citeauthoryear{{Zinger}, {Dekel}, {Kravtsov}  \& {Nagai}}{{Zinger} et~al.}{2018}]{zinger2018quenching}
{Zinger} E.,  {Dekel} A.,  {Kravtsov} A.~V.,   {Nagai} D.,  2018, \mn@doi [\mnras] {10.1093/mnras/stx3329}, \href {https://ui.adsabs.harvard.edu/abs/2018MNRAS.475.3654Z} {475, 3654}

\bibitem[\protect\citeauthoryear{{van de Voort}, {Schaye}, {Booth}, {Haas}  \& {Dalla Vecchia}}{{van de Voort} et~al.}{2011}]{vandeVoort.etal.11}
{van de Voort} F.,  {Schaye} J.,  {Booth} C.~M.,  {Haas} M.~R.,   {Dalla Vecchia} C.,  2011, \mn@doi [\mnras] {10.1111/j.1365-2966.2011.18565.x}, \href {https://ui.adsabs.harvard.edu/abs/2011MNRAS.414.2458V} {414, 2458}

\bibitem[\protect\citeauthoryear{{van de Voort}, {Schaye}, {Altay}  \& {Theuns}}{{van de Voort} et~al.}{2012}]{vandeVoort.etal.12}
{van de Voort} F.,  {Schaye} J.,  {Altay} G.,   {Theuns} T.,  2012, \mn@doi [\mnras] {10.1111/j.1365-2966.2012.20487.x}, \href {https://ui.adsabs.harvard.edu/abs/2012MNRAS.421.2809V} {421, 2809}

\bibitem[\protect\citeauthoryear{{van de Voort}, {Springel}, {Mandelker}, {van den Bosch}  \& {Pakmor}}{{van de Voort} et~al.}{2019}]{vandeVoort.etal.19}
{van de Voort} F.,  {Springel} V.,  {Mandelker} N.,  {van den Bosch} F.~C.,   {Pakmor} R.,  2019, \mn@doi [\mnras] {10.1093/mnrasl/sly190}, \href {https://ui.adsabs.harvard.edu/abs/2019MNRAS.482L..85V} {482, L85}

\makeatother
\end{thebibliography}



\appendix

\section{Mach Number}\label{sec:Mach}

\begin{figure}
    \centering
    \includegraphics[trim={0.0cm 0.0cm 0.0cm 0.0cm}, clip, width =0.48 \textwidth]{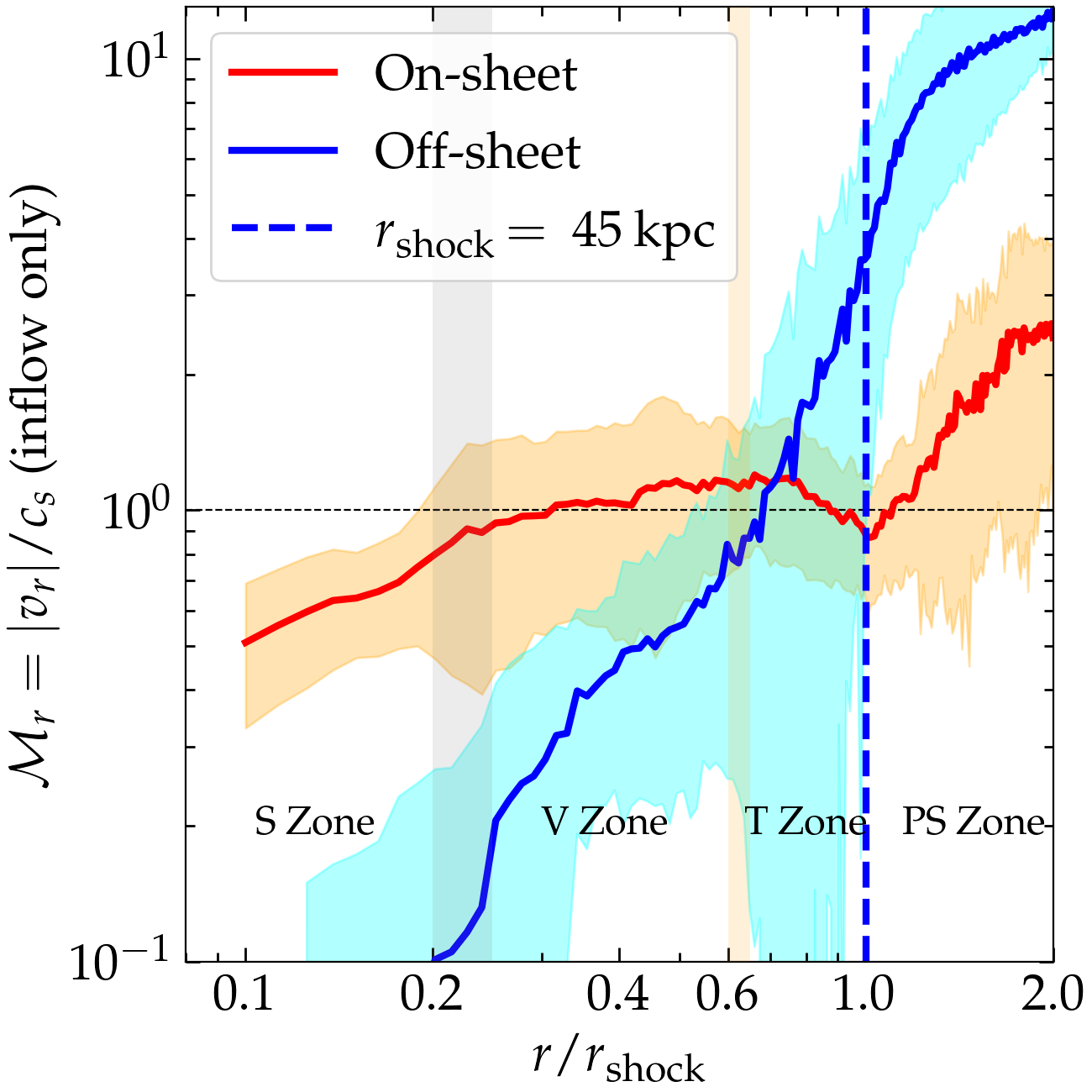}
    \caption{The radial Mach number of inflowing gas, $\mathcal{M}_r$, for the on-sheet (red) and off-sheet (blue) components, stacked among our ten filament slices. The cyan and orange shaded regions show the $1-\sigma$ scatter for the two components, respectively. The off-sheet $\mathcal{M}_r$ changes much more drastically than the on-sheet $\mathcal{M}_r$, indicating that the on-sheet gas has undergone a much weaker and more diffusive shock compared the off-sheet gas.}
    \label{fig:Mach}
\end{figure}

To add additional context to the strength of the filamentary accretion shock (\se{thermal}) and the radial velocity profiles (\se{inflows}), we here present an analysis of the radial Mach number of inflowing gas, $\mathcal{M}_r$, defined as 
\begin{equation}
    \mathcal{M}_r\equiv \frac{\left|v_r\right|}{c_{\rm s}},
\end{equation}
{\no}where $v_r$ is the radial velocity, and $c_{\rm s}=\sqrt{\gamma P/\rho}=\sqrt{\gamma k_{\rm B}T/(\mu m_{\rm p})}$ is the adiabatic (constant entropy) sound speed. In our case, $\gamma=5/3$. We compute $\mathcal{M}_r$ only for radially inflowing gas cells, and then take the volume-weighted average over each radial bin, as in the profiles presented throughout the paper. However, we distinguish the on-sheet and off-sheet components of gas when taking the average, as in our analysis of radial velocity profiles in \se{inflows}.

\smallskip
The radial profiles of $\mathcal{M}_{r}$ for the on-sheet and off-sheet components, stacked among our ten filament slices, are shown in \fig{Mach}. Outside the shock, at $r\lsim 2 \, r_{\rm shock}$, the off-sheet gas is inflowing with $\mathcal{M}_r\sim 10$ on average\footnote{We note that Slice 01 (\fig{slices}) has the highest Mach number among our slices, reaching $\mathcal{M}_r\gsim 15$ at $r_{\rm shock}$, and it also has the largest $r_{\rm shock}$ (\fig{rho_0 and T_0 scatterings}). This is due to the fact that Slice 01 is close to a massive halo, and the circum-filamentary shock gets mixed up with the halo accretion shock.}while the on-sheet gas has $\mathcal{M}_r\sim 3$. This is consistent with what we found in \fig{radial inflows}: the off-sheet gas inflows much faster than the on-sheet gas at large distances from the shock. Close to the shock, at $r\sim 1.2 \, r_{\rm shock}$, the off-sheet $\mathcal{M}_r$ begins to rapidly decline, reaching $\mathcal{M}_r\sim 1$ just outside the \textbf{V} zone, and continuing to decline towards very subsonic values at smaller radii. This suggests that the radial extent of the \textbf{T} zone, over which the Mach number declines towards unity converting bulk kinetic energy into thermal energy, is representative of the shock thickness. On the other hand, the on-sheet $\mathcal{M}_r$ has a much shallower profile, retaining $\mathcal{M}_r\sim 1$ throughout the \textbf{T} and \textbf{V} zones, and declining to $\mathcal{M}_r\sim 0.5$ in the \textbf{S} zone.

This suggests that the effect of the shock is much less significant on the sheet compared to off the sheet. The on-sheet gas might partially bypass the shock and therefore avoid interacting with it. This is reminiscent of speculations concerning how cold streams penetrate the virial shocks around the CGM in massive halos as seen in simulations \citep{Dekel.Birnboim.06, dekel2009cold,Bennett.Sijacki.20}. However, a comprehensive discussion of this topic is beyond the scope of this paper.


\section{Magnetic Fields and Hubble Drag}\label{sec:simplification}

\begin{figure}
    \centering
    \includegraphics[trim={0.0cm 0.0cm 0.0cm 0.0cm}, clip, width =0.48 \textwidth]{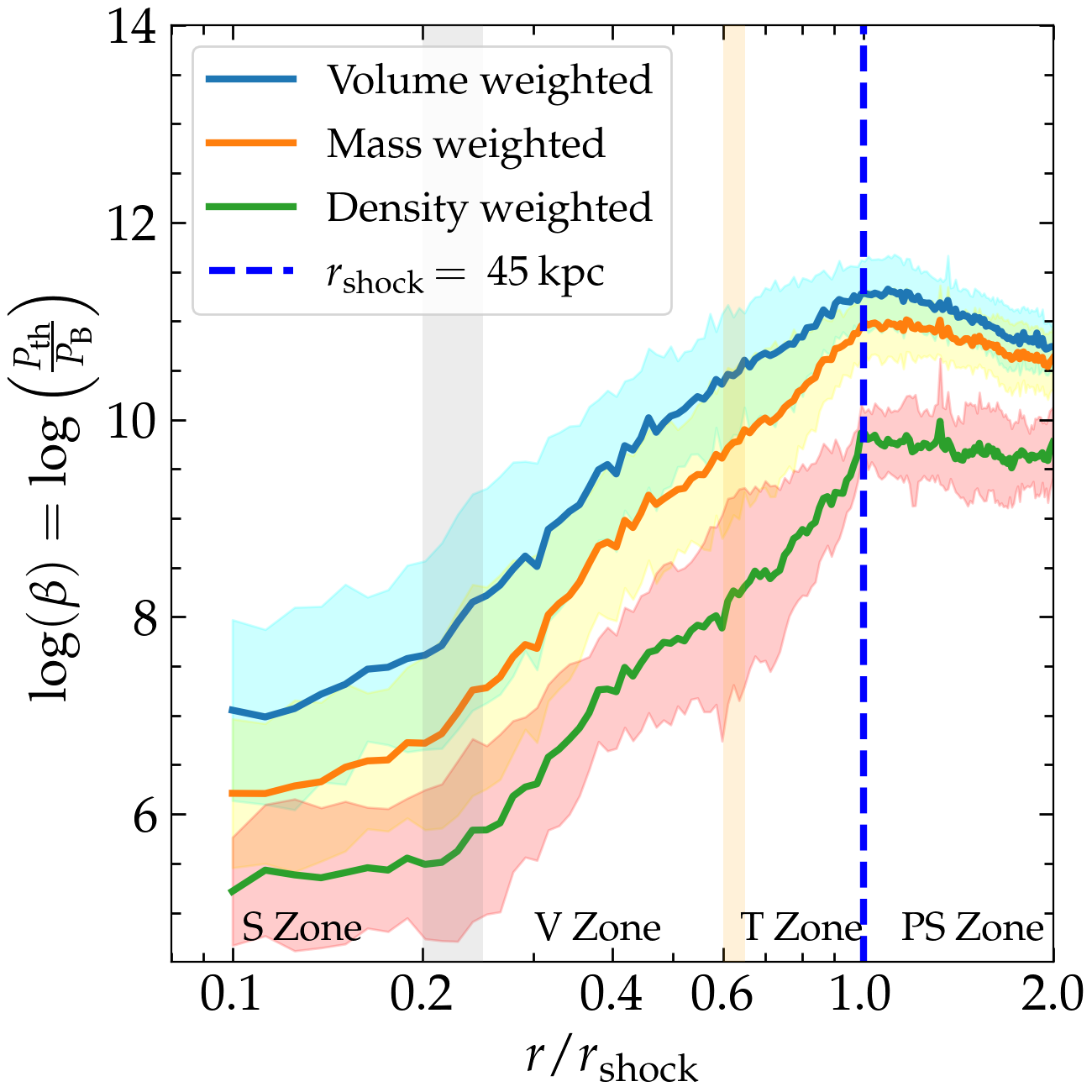}
    \caption{Radial profile of the plasma $\beta$ parameter, the ratio of thermal to magnetic pressure, stacked among our filament slices. We compute the value of $\beta$ for each individual gas cell, and then average them over each radial bin weighted by volume (blue line, cyan shaded region), mass (orange line, yellow shaded region), and density (green line, pink shaded region). The three solid lines show the results after stacking the ten slices in log space, while the shaded regions show the $1-\sigma$ scatter. $\beta$ evolves from $\sim (10^5-10^6)$ at $r\sim 0.1\,r_{\rm shock}$ to $\sim (10^{10}-10^{11})$ at $r\sim r_{\rm shock}$, implying that magnetic fields are not dynamically important in the filaments.}
    \label{fig:plasma_beta}
\end{figure}

In \se{summation}, we analysed the gas dynamics assuming that this obeyed the Euler equation (\equnp{euler}). However, this was just an approximation, as it neglects the effect of magnetic fields and the expansion of the Universe. In this section, we justify both of these approximations.

\smallskip
The gas dynamics in the simulation follow the ideal magnetohydrodynamics (MHD) equations. In particular, the Eulerian momentum equation reads
\be 
\label{eq:MHD}
\pdv{\vec{v}}{t}+\left(\vec{v}\cdot \vec{\nabla} \right) \vec{v} = -\frac{1}{\rho} \vec{\nabla} \left(P+\frac{B^2}{8\pi}\right) + \frac{1}{4\pi\rho}\left(\vec{B}\cdot\vec{\nabla}\right)\vec{B} - \vec{\nabla} \Phi,
\ee
{\no}where $\vec{v}$ is the fluid velocity, $\rho$ is its density, $P$ its thermal pressure, $\Phi$ the gravitational potential, and $\vec{B}$ the magnetic field. Here, $\vec{v}$ represents the physical velocity, including the Hubble flow in an expanding Universe. In practice, the simulation stores the peculiar velocity, $\vec{v}_{\rm pec}=\vec{v}-H(z)\vec{r}$ with $H(z)$ the Hubble parameter at redshift $z$, for which an additional term of $(-H(z)\vec{v})$ representing the so-called ``Hubble drag'' must be added to the right-hand side of \equ{MHD}. However, over the scales of interest, $r\lsim r_{\rm shock}\sim 50\kpc$, the Hubble flow $H(z)r\lsim 10\kms$ is negligible compared to the peculiar velocity. Likewise, the Hubble drag is negligible compared to the other forces discussed in \se{fil_dyn}, at the level of a few percent at most, and has no effect on our results. We therefore have neglected this subtlety, and treat peculiar and physical velocities interchangeably.

\smallskip 

The dynamical importance of the magnetic field is often quantified by the plasma $\beta$ parameter, which is the ratio of thermal to magnetic pressure,
\be 
\label{eq:beta}
\beta \equiv \frac{P_{\rm th}}{P_{\rm B}} = \frac{8\pi P}{B^2},
\ee
{\no}where $P_{\rm B}=B^2/(8\pi)$ is the magnetic pressure. 
In \fig{plasma_beta}, we show profiles of $\beta$ stacked among our filament slices. We compute the value of $\beta$ for each cell in the simulation and compute the mean value at each radius $r$ weighted by volume (blue line, cyan shading), mass (orange line, yellow shading), and density (green line, pink shading). In all cases, $\beta$ increases monotonically from the \textbf{S} zone to the \textbf{T} zone, consistent with magnetic field amplification due to gas condensation as it cools together with a turbulent dynamo effect. The volume-weighting, which highlights the hot and diffuse volume-filling gas, presents the highest values of $\beta$, i.e. the lowest values of the magnetic fields, with $\beta\sim 10^7$ in the \textbf{S} zone and $\beta\sim 10^{11}$ in the \textbf{T} zone. The density-weighting, which highlights cold and dense gas, presents the strongest magnetic fields, but even here $\beta\sim 10^5$ in the \textbf{S} zone and $\beta\sim 10^{10}$ in the \textbf{T} zone. The magnetic pressure is thus negligible compared to the thermal pressure. 

\smallskip
Furthermore, magnetic pressure gradients are negligible compared to thermal pressure gradients. Even in the extreme case where the thermal pressure varies over a characteristic length-scale of order $r_{\rm shock}\lsim 50\kpc$, and the magnetic pressure varies over a characteristic length-scale of order the smallest cell size, $\Delta\gsim 100\pc$, magnetic pressure gradients would still be at least 100 times smaller than thermal pressure gradients. We conclude that magnetic fields are not dynamically important in our filament sample, and do not contribute to filament support against self-gravity. We, therefore, neglect them in our further analysis, simplifying \equ{MHD} to the Euler equation (\equnp{euler}).

\smallskip
Observations of low-$z$ filaments in the vicinity of galaxy clusters do reveal the presence of magnetic fields \citep[e.g.,][]{Sullivan.etal.19, Vernstrom.etal.21}, though to our knowledge no such observational constraints exist for high-$z$ filaments. However, we note that values of $\beta\sim 10^5$, as found for cold and dense gas in the \textbf{S} zone, have been assumed for cold streams feeding massive high-z galaxies in idealized studies of stream evolution in the CGM \citep{Ledos.etal.2023}. A study of the growth of magnetic fields in intergalactic filaments over cosmic time is beyond the scope of this paper and is left for future work.

\section{Numerical Details of the Force Decomposition}
\label{sec:sim_numerics}

\smallskip
In this section, we provide further details regarding our numerical method for analysing the forces (\se{summation}-\se{sim_dynamics}) and present several validation tests. 
Using our simulations, we numerically evaluate \equs{accel}-\equm{Mgrav_a} as a function of radius, $r$, for each filament slice. We begin by generating a uniform cylindrical grid in each slice, with $N_r$ radial bins in the range $r=(0.05-1.1)r_{\rm shock}$, $N_\phi$ azimuthal bins in the range $\phi=(0-2\pi)$, and $N_z$ axial bins from $z=-15\kpc$ to $z=15\kpc$. We assign to each bin the density, velocity, and pressure of the gas cell whose centre is closest to the bin centre. While this does not conserve the total mass, momentum, or thermal energy in the slice, it accurately represents the fluid properties on the grid points, provided the bins are not much larger than the gas cells in the simulation. On the other hand, if the bins are much smaller than the gas cells, this can result in artificially null-gradients when using the grid to compute the terms in the integrals of \equs{accel}-\equm{Mgrav_a}.
We, therefore, wish to use bin sizes as close as possible to the typical sizes of gas cells in the simulation. 

\begin{figure}
    \centering
    \includegraphics[trim={0.0cm 0.0cm 0.0cm 0.0cm}, clip, width =0.48 \textwidth]{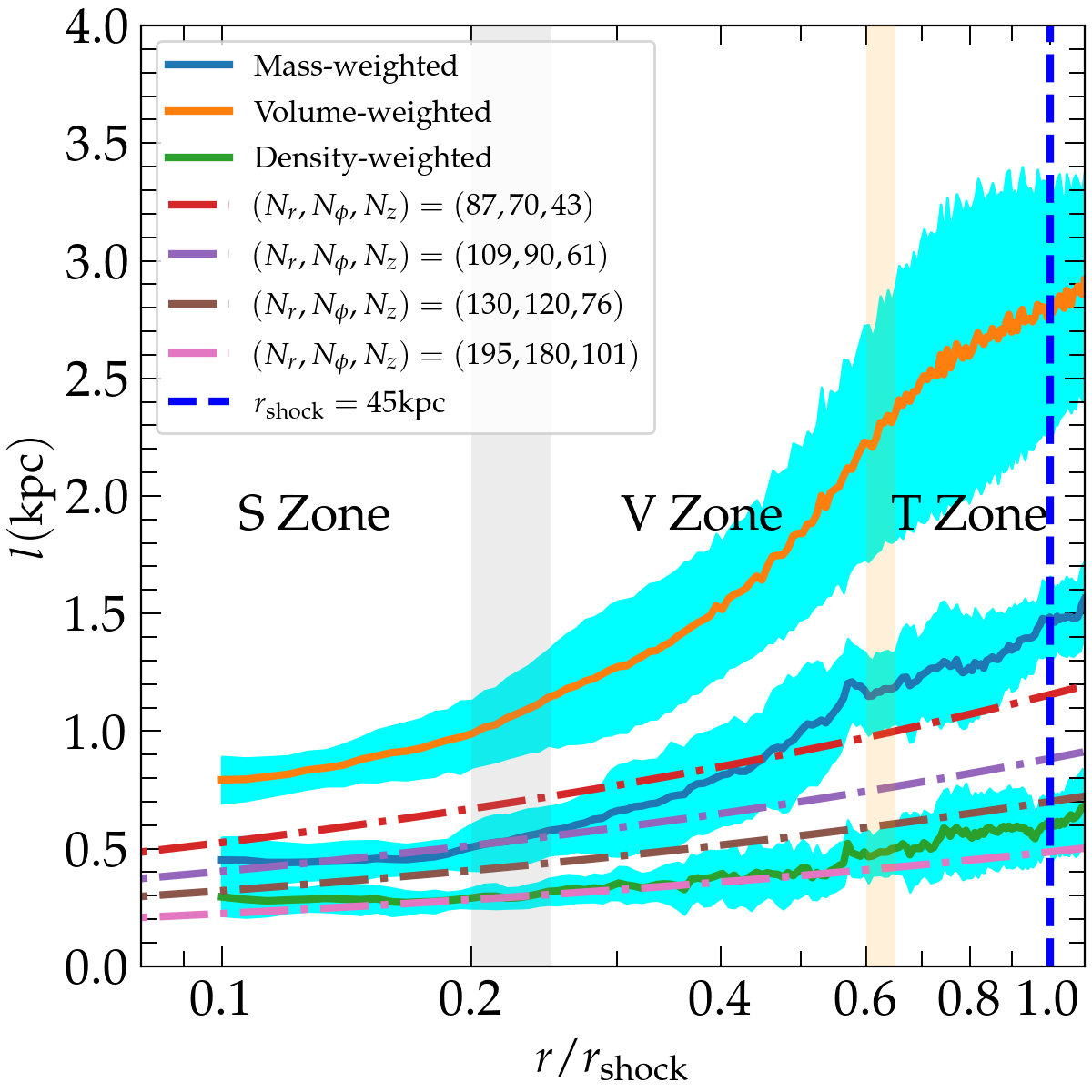}
    \caption{Cell-sizes in the simulation compared to bin-sizes of the uniform cylindrical grid we use to evaluate \equs{accel}-\equm{Mgrav_a}. Solid lines show the average cell size as a function of radius, weighted by volume (orange), mass (blue), or density (green), and stacked among our 10 filament slices. Dot-dashed lines show the bin sizes of our cylindrical grids, with different numbers of bins as indicated in the legend. At all radii, the bin size in our fiducial grid is comparable to the density-weighted average cell-size, representing the smallest and densest cells in the simulation.}
    \label{fig:bin sizes and cell sizes}
\end{figure}

\smallskip
In \fig{bin sizes and cell sizes}, the solid lines show the radial profiles of the average cell size, defined as the cubic root of the cell volume, $l={\rm Vol}^{1/3}$, as a function of $r/r_{\rm shock}$ stacked among our 10 filament slices. We show profiles for the volume-weighted average cell size (orange), the mass-weighted average cell size (blue), and the density-weighted average cell size (green). Since cells have the same mass to within a factor of $\sim 2$, these averages roughly represent the largest, median, and smallest cells at a given radius, respectively. As in previous figures, the cyan-shaded regions represent the $1-\sigma$ standard deviations among the slices. The median cell size increases from $\sim 0.5\kpc$ in $r\sim 0.1r_{\rm shock}$, to $\sim 1.5\kpc$ at $r\gsim r_{\rm shock}$, while the smallest densest cells grow from $\sim (0.3-0.7)\kpc$ in the same range. We also show as dot-dashed lines in \fig{bin sizes and cell sizes} the bin sizes, defined as the cubic root of the bin volume, for cylindrical grids with different numbers of radial, azimuthal, and axial bins as indicated in the legend. Note that in each case, the bin size scales as $r^{1/3}$, since the bin volume is $\sim r\,\Delta r \Delta \phi \Delta z$. For our fiducial grid we use $(N_r, N_{\phi}, N_z)=(195, 180, 101)$, the highest-resolution grid shown in \fig{bin sizes and cell sizes}, corresponding to $\Delta r\sim 0.005\, r_{\rm shock}\sim 0.23 \kpc$, $\Delta \phi \sim \pi/90$, and $\Delta z\sim 0.3\kpc$. These dimensions are very similar to the density-weighted average cell size at all radii. 
However, all of our results are very similar when varying the number of bins as in \fig{bin sizes and cell sizes} (see \figs{cumulative gas mass}-\figss{Grad Phi together} below) 

\smallskip
To verify the validity of our method of depositing gas properties in the grid, which is not strictly conservative, we test the degree to which various gas quantities are conserved. In \fig{cumulative gas mass}, we show the ratio of the enclosed gas mass profile inferred from our grid to that of the actual simulation cells. We see that for all but our lowest resolution grid, with $(N_r,N_\phi, N_z)=(87, 70, 43)$, the error on the enclosed mass is $\lsim 5\%$ at all radii and approaches $0$ at $r_{\rm shock}$. For our highest resolution grid, $(N_r,N_\phi, N_z)=(190, 180, 101)$, the error on both the total mass interior to $r$ and the mass contained between $r$ and $r+\Delta r$ (not shown) is $\lsim 2\%$ at all radii. The ratio of enclosed momenta and thermal energies is similar. We conclude that the fluid properties are sufficiently conserved when generating our highest-resolution grid.

\smallskip
We also experimented with other methods for depositing the fluid properties into the cylindrical grids. Specifically, we deposited the mass, momentum, and thermal energy of each gas cell onto the grid using either a top-hat or a cubic-spline smoothing kernel, with the kernel size set to the cell size (not shown). While this conserves the total mass, momentum, and thermal energy of gas within each slice, the unstructured nature of the mesh and the non-uniform sizes of gas cells result in many ``empty'' bins, into which no gas cell is deposited, unless we use a kernel size much larger than the cell size. For this reason, we prefer the deposition method described above, which guarantees that each bin has a well-defined non-zero value. However, we note that all three methods for depositing the gas data onto the cylindrical grid yield qualitatively similar results. 

\begin{figure}
    \centering
    \includegraphics[trim={0.0cm 0.0cm 0.0cm 0.0cm}, clip, width =0.48 \textwidth]{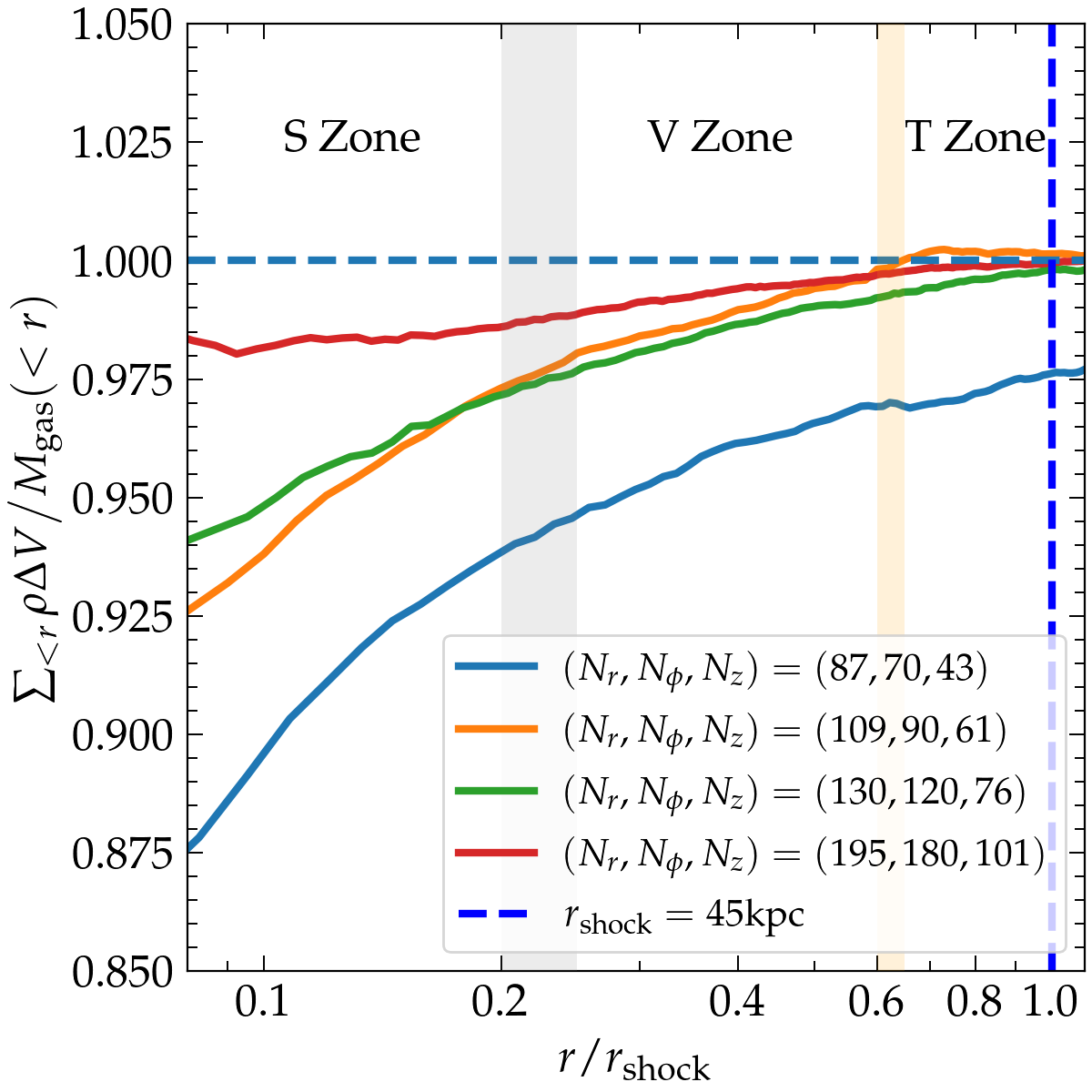}
    \caption{Mass conservation test of our method for depositing gas cells into uniform cylindrical grids. We show the ratio of the enclosed gas mass inferred from our grid to the actual enclosed gas mass from the simulation. At all but our lowest resolution (blue line), the cumulative error in the gas mass in the grid is $<5\%$ everywhere and approaches $0$ at $r_{\rm shock}$. At our highest resolution (red line), both the cumulative and the \textit{local} error in gas mass (not shown) are $<2\%$ at all radii.}
    \label{fig:cumulative gas mass}
\end{figure}

\begin{figure*}
    \centering
    \includegraphics[trim={0.0cm 0.0cm 0.0cm 0.0cm}, clip, width =0.98 \textwidth]{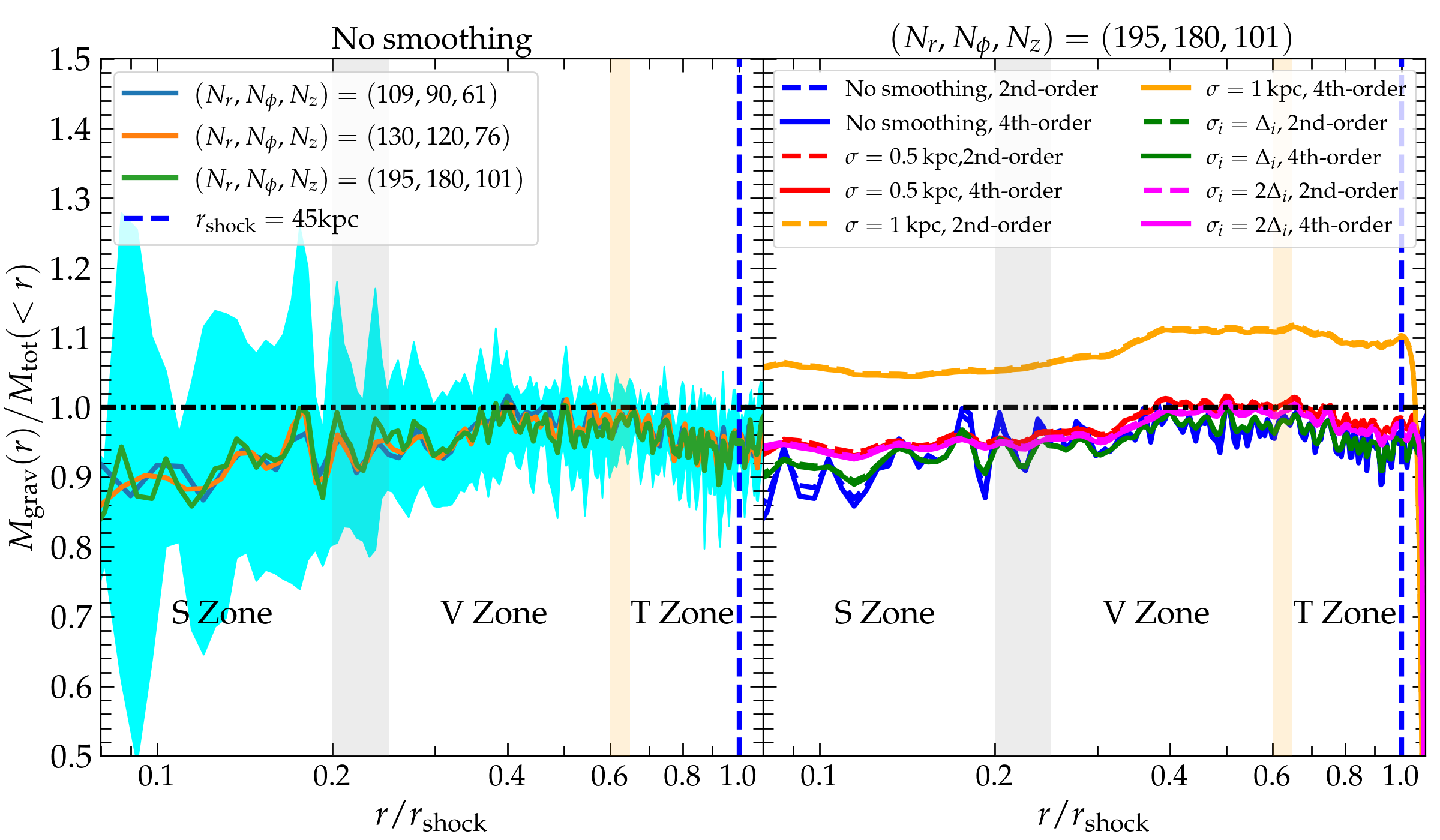}
    \caption{ Testing our numerical methods for evaluating gradients and surface integrals on our cylindrical grids, using Gauss's law. We show the ratio of the gravitational mass at radius $r$, $M_{\rm grav}$ from \equ{gauss}, and the total mass interior to radius $r$. Gauss's law states that this ratio must be unity everywhere. 
    \textbf{\textit{On the left,}} we show the results using fourth-order differentiation (\equnp{derivative_4}) with no smoothing applied. Different colour solid lines represent the mean ratio among the ten filament slices for different grid resolutions as shown in the legend. The cyan-shaded region represents the $1-\sigma$ standard deviation among the filament slices for the highest resolution. All resolutions yield similar results, with $M_{\rm grav}$ systematically underpredicting the true enclosed mass by $\sim 5\%$. \textbf{\textit{On the right,}} we show results for our fiducial (highest) resolution grid, but using different methods for smoothing the data and computing the gradients, as indicated in the legend. Using second or fourth-order derivatives has no effect on the results. The best results are obtained when smoothing the data with a Gaussian of $\sigma=0.5\kpc$ prior to computing the derivatives. In this case, there is effectively no error on the enclosed mass at $0.3\lsim r/r_{\rm shock}\lsim 0.8$, and $\lsim 5\%$ error elsewhere. The worst results are obtained with $1.0\kpc$ smoothing, but even here the errors are only $\sim (5-10)\%$. This confirms the validity and robustness of our methodology. 
    } 
    \label{fig:Grad Phi together}
\end{figure*}

\smallskip
We compute the gradients, $\partial/\partial r$, $\partial/\partial \phi$, $\partial/\partial z$, on the uniform cylindrical grid using the standard fourth-order centred finite-difference approximation for evenly-spaced data:

\begin{equation}
\label{eq:derivative_4}
    \left(\frac{\text{d} f}{\text{d} x}\right)_i=\frac{f(x_{i-2})-8f(x_{i-1})+8f(x_{i+1})-f(x_{i+2})}{12h} + \mathcal{O}(h^4).
\end{equation}
{\no}We also experimented with second-order centred finite-difference derivatives:
\begin{equation}
\label{eq:derivative_2}
    \left(\frac{\text{d} f}{\text{d} x}\right)_i=\frac{f(x_{i+1})-f(x_{i+1})}{2h} + \mathcal{O}(h^2),
\end{equation}
and obtained very similar results with no qualitative differences (see \Fig{Grad Phi together} below). Note that in order to compute these derivatives at the grid boundaries self-consistently, we padded our grid with extra bins in both the radial and axial directions, with fluid properties deposited in the same way as described above. This is especially important for the $z$ direction, since the axial terms (\equsnp{accel_2}, \equmnp{Mtherm_a}, \equmnp{axial inertial term}, and \equmnp{Mgrav_a}) are evaluated only at the boundaries of $z=\pm 15\kpc$. However, all of our results are presented only for the main grid, without the extra padding cells. 

\smallskip
Prior to computing each spatial derivative, we smooth the grid with a one-dimensional Gaussian kernel along the direction of the derivative. For example, prior to computing $\partial P/\partial r$, we smooth the pressure values on the grid using a smoothing kernel $\propto \exp[-r^2/(2\sigma_r^2)]$, while prior to computing $\partial P/\partial z$, we smooth the pressure values on the grid using a smoothing kernel $\propto \exp[-z^2/(2\sigma_z^2)]$. This is similar to the smoothing method employed by \citet{lau2013weighing}, though they used one-dimensional Savitzky-Golay filters rather than Gaussians. Our fiducial results use $\sigma_r=\sigma_{\phi}=\sigma_z=0.5\kpc$, $\sim (1-2)$ times the bin size in our highest resolution grid. We also present results for no-smoothing and for smoothing with $\sigma_r=\sigma_{\phi}=\sigma_z=1.0\kpc$ in \fig{Grad Phi together} below, and find overall good agreement between these different methods. In the same figure, we also present results for $\sigma_i=a\Delta i$, with $i=r,\phi,z$ and $a=1,2$, which yield very similar results as well. 

\smallskip
Using the above procedures, we obtain all the terms in the integrands of \equs{Mtherm_r}-\equm{Mgrav_a} (we address the temporal derivatives of \equs{accel}-\equm{accel_2} below). We then perform the integrations using the ``trapezoid-method'', the standard second-order centred approximation for numerical integration of evenly-spaced data,
\begin{equation}
\label{eq:integration}
    \int f(x) {\rm d}x = \sum_{i=1}^{n-1}h\frac{f_i+f_{i+1}}{2} + \mathcal{O}(h^2).
\end{equation}
{\no}We also experimented with a first-order, ``rectangular'', approximation, which yielded a slightly poorer fit to Gauss's theorem. We did not experiment with higher-order approximations involving higher-order interpolations between the grid points, as we found the trapezoidal method to be satisfactory and numerical integration is inherently less sensitive to noise than numerical differentiation. 

\smallskip
We evaluated the robustness and validity of our methodology by directly computing $\vec{\nabla}\Phi$ on our grid and testing Gauss's law. The gravitational potential at the location of each gas cell in the simulation is provided in the simulation output and deposited into our uniform cylindrical grids in the same manner as the gas density, pressure, and velocity. We then compute $M_{\rm grav}(r)$ following \equ{gauss}, and compare this to  $M_{\rm tot}(<r)$, the total mass of gas, stars, and dark matter enclosed within the cylinder of radius $r$. 
The results of this test are shown in \fig{Grad Phi together}. The $y$-axes show the ratio of 
$M_{\rm grav}(r)/M_{\rm tot}(<r)$, which should be unity everywhere according to Gauss's law. The $x$-axes show the radius normalized to $r_{\rm shock}$. We computed the radial profile of this ratio for each slice and then stacked the slices by computing the mean and standard deviation of the ratio at each radius $r$. 
In the left panel, we show results for different grid resolutions, focusing on those where the error on the enclosed mass is $\lsim 5\%$ (\fig{cumulative gas mass}). In each case, we used fourth-order derivatives and applied no smoothing prior to computing the derivatives. The three solid curves represent the mean ratio among our ten slices, while the cyan-shaded region represents the $1-\sigma$ standard deviation of the highest resolution grid. The three different grid resolutions yield extremely similar results, with no systematic differences. Overall, our method for evaluating the enclosed mass using Gauss's law seems to systematically underpredict the true mass, with an average error of $\sim 5\%$ at $r>0.3r_{\rm shock}$, and $\lsim 10\%$ at smaller radii. The scatter, however, is more symmetric about an error of $0$ and is still limited to $\lsim 10\%$ in most of the radial range. The scatter in the lower resolution versions (not shown) is somewhat larger, but qualitatively very similar with errors in the range of $\sim (10-20)\%$. In the right panel, we focus on the highest resolution grid and examine the effect of different smoothing widths (no smoothing, constant $\sigma=0.5$ or $1.0\kpc$, and coordinate-dependent $\sigma_i=a\Delta i$, where $\Delta_i$ is the bin width along the $i$ direction with $i=r,\phi,z$ and $a=1,2$) and differentiation methods (second-order and fourth-order), as noted in the legend. At this resolution, the order of the numerical derivatives has hardly any impact on the results for each smoothing method. The version smoothed with $\sigma=0.5\kpc$ seems to yield the best results, with effectively no error on the enclosed mass at $0.3\lsim r/r_{\rm shock}\lsim 0.8$, and errors of $<5\%$ elsewhere. The version smoothed with $\sigma=1.0\kpc$ yields the largest errors, of $(5-10)\%$. Also, interestingly, unlike the no-smoothing and $\sigma=0.5\kpc$ versions, which both systematically underestimate the enclosed mass, the $\sigma=1.0\kpc$ version systematically overestimates the enclosed mass. As expected, given the bin sizes at this grid resolution, $\sigma_i=\Delta_i$ is very similar to no smoothing while $\sigma_i=2\Delta_i$ is very similar to $0.5\kpc$ smoothing. We note that similar errors of $\sim 5\%$ on estimates of the enclosed mass using Gauss's law were also reported in \citet{lau2013weighing} (the bottom panel of Fig. 1), which formed the inspiration for our methodology, but whose simulations were run on a much simpler Cartesian mesh. 

\smallskip
We adopt a constant $\sigma=0.5\kpc$ as our fiducial smoothing method, along with fourth-order centred finite differences and our highest resolution grid. The results presented in \figs{cumulative gas mass} and \figss{Grad Phi together} suggest that this method produces reliable and robust results, both in terms of the conservation of fluid properties and in terms of accurately evaluating the necessary surface integrals in \equs{accel}-\equm{Mgrav_a}.

\smallskip
Finally, we address the computation of the two acceleration terms in \equs{accel}-\equm{accel_2}, which represent a temporal change in velocity within a fixed volume. We attempted to compute these directly using multiple simulation snapshots, experimenting with first, second, and fourth-order finite-difference approximations for the time derivatives. When doing so, we attempted in each snapshot to centre ourselves on the same ten filament slices as in our fiducial snapshot. However, since the filaments themselves are all moving with respect to the simulation box, since the first stage of our filament selection process (\se{Fil_selec}) is done by-eye. Since there is some inherent uncertainty associated with defining the filament centre and rest-frame (as discussed in \se{Fil_selec}), this process is prone to errors and uncertainties. Furthermore, the time in between consecutive snapshots of the simulation, $\Delta t\sim 55\Myr$ near $z\sim 4$, is within a factor of $\lsim 2$ of both the cooling time and the eddy-crossing time for gas within the \textbf{V} and \textbf{S} zones, $r\lsim 0.6r_{\rm shock}$. All of this makes the direct computation of the acceleration terms highly uncertain. As a result, in none of our experiments did the acceleration terms balance the other terms in \equ{summation_radial}. Since we have shown that our method reliably reproduces $M_{\rm grav}(r)$ (\fig{Grad Phi together}), and since $M_{\rm therm}(r)$ and $M_{\rm inertial}(r)$ are computed similarly, we can assume that these are also computed reliably. We, therefore, use \equ{summation_radial} and the corresponding axial equation to infer the acceleration terms from the difference between the gravitational, thermal, and inertial terms.

\smallskip
After computing the radial profiles of each mass-term from \equs{Mtherm_r}-\equm{Mgrav_a} for each filament slice, we normalise these by $M_{\rm tot}(<r)$ and stack the ten filament slices by taking the average of each normalised mass term in each radial bin. While it is the ratio of each mass term to $M_{\rm grav}(r)$ that tells us the contribution of the respective 
force term to the support of the filament against gravity at that radius, we nonetheless opt to normalise the mass terms by $M_{\rm tot}(<r)$. As shown in \fig{Grad Phi together}, $M_{\rm tot}(<r) \simeq M_{\rm grav}(r)$ to within $\sim 5\%$. However, while $M_{\rm grav}(r)$ can fluctuate, especially for individual slices, $M_{\rm tot}(<r)$ is a smooth, monotonically increasing function. We have verified that normalising by $M_{\rm grav}(r)$ instead does not have a noticeable impact on our results. In \se{sim_dynamics}, we use the notations $M_{...}/M_{\rm tot}$, $M_{...}/M_{\rm grav}$, and $F_{...}/F_{\rm grav}$ interchangeably, and think of these as ratios of forces/accelerations rather than ratios of masses.


\bsp	
\label{lastpage}
\end{document}